\documentclass[11pt]{article}
\pdfoutput=1
\usepackage[utf8]{inputenc}
\usepackage{amsfonts}
\usepackage{amsmath,amssymb,amsthm}
\usepackage[margin=2.5cm]{geometry}
\usepackage{physics}
\usepackage{ulem}
\usepackage[dvipsnames]{xcolor}
\definecolor{darkgreen}{rgb}{0,0.5,0}
\definecolor{darkblue}{rgb}{0,0,0.6}
\definecolor{purple}{rgb}{0.4,.2,0.7}
\definecolor{contourred}{rgb}{0.816,0.008,0.106}
\definecolor{saddleblue}{rgb}{0.290,0.565,0.886}
\definecolor{bromwichgreen}{rgb}{0.494,0.827,0.129}
\usepackage{graphicx}
\usepackage{float}
\usepackage{enumerate}
\usepackage{pgfplots}
\pgfplotsset{compat=1.18}
\usetikzlibrary{decorations.pathmorphing, decorations.markings}
\usepackage{cite}
\usepackage[hyperfootnotes=false,colorlinks=true,linkcolor=darkblue,citecolor=purple]{hyperref}

\newcommand{\be}{\begin{equation}}
\newcommand{\ee}{\end{equation}}
\newcommand{\bea}{\begin{equation}\begin{aligned}}

\newcommand{\eea}{\end{aligned}\end{equation}}
\def\la{\label}

\numberwithin{equation}{section}

\begin{document}

\thispagestyle{empty}
\begin{center}
    ~\vspace{5mm}

  \vskip 2cm

   {\LARGE \bf
       Inner Horizon Saddles and a Spectral KSW Criterion

   }

   \vspace{0.5in}

  {\bf Jacqueline Caminiti$^{1,2}$ and Aidan Herderschee$^3$
   }

    \vspace{0.5in}

  $^1$
  Perimeter Institute for Theoretical Physics, Waterloo, ON N2L 2Y5, Canada
  \\
   ~
   \\
   $^2$
 Department of Physics \& Astronomy, University of Waterloo, Waterloo, ON N2L 3G1, Canada
   \\
   ~
   \\
  $^3$
  School of Natural Sciences, Institute for Advanced Study, Princeton, NJ 08540, USA

    \vspace{0.5in}

    \vspace{0.5in}

\end{center}


\begin{abstract}

\noindent
The Bekenstein-Hawking entropy formula $\rho = e^{A/4G}$ receives significant corrections for charged black holes near extremality.
Using standard results in JT gravity, the correction term can semiclassically be expressed as minus the exponential of the inner horizon area, $e^{A_{\text{inner}}/4G}$, and the cancellation between these two exponentials enforces a vanishing density of states towards extremality, when the two horizons collide. 
Building on Ref.~\cite{Kruthoff:2024gxc}, we argue that the correction term corresponds to a complex saddle geometry of the bulk gravitational path integral.
The proposed geometry has a negative boundary length and caps off at the inner horizon; we refer to it as the inner horizon saddle. 
We discuss how the saddle, and its accompanying minus sign, contribute to the density of states through a Picard-Lefschetz analysis of the inverse Laplace contour, together with a stability analysis of the saddle.
We also address the inner horizon saddle's violation of the  Kontsevich-Segal-Witten (KSW) allowability criterion for the inclusion of complex metrics. 
Despite this violation, which is believed to cause unphysical divergences in path integral computations, one can describe one-loop effects on the inner horizon saddle by carefully treating wrong-sign modes.
Motivated by this observation, we propose a weaker version of the KSW criterion, which we call the spectral KSW criterion. Its purpose is to characterize when one-loop corrections around complex gravitational saddles are well defined.

\end{abstract}

\vspace{1in}

\pagebreak

\setcounter{tocdepth}{3}
{\hypersetup{linkcolor=black}\tableofcontents}

\pagebreak

\section{Introduction}
\label{sec:intro}

For over half a century, the gravitational path integral (GPI) has provided a powerful window into the microscopic nature of black holes.
First, the GPI taught us that black holes mimic thermodynamic systems: they can be assigned an entropy and temperature~\cite{Gibbons:1976ue}, and they exhibit phase transitions~\cite{Hawking:1982dh}.
More recently, the inclusion of replica wormhole saddles in the gravitational path integral has led to a derivation of the Page curve, providing a semiclassical ``smoking gun'' of microscopic unitarity~\cite{Penington:2019npb,Almheiri:2019psf,Penington:2019kki}.
These developments, and many more, have sharpened the gravitational path integral into a predictive framework for probing nonperturbative aspects of black hole physics.

In the semiclassical regime, the GPI is computed by summing over saddle point geometries with appropriate perturbative corrections. Surprisingly, in many cases, a complete understanding of the physics requires studying subleading saddles in addition to the leading ones.
In particular, non-perturbatively small corrections from spacetime wormholes to the gravitational inner product enforce that the dimension of the black hole Hilbert space is of the correct order~\cite{ Penington:2019kki,Marolf:2020xie,Balasubramanian:2022gmo,Boruch:2023trc,Iliesiu:2024cnh,Boruch:2024kvv}.
In another context, subleading giant graviton saddles in the half-BPS partition function enforce finite-$N$ trace relations \cite{Lee:2024hef,Lee:2023iil,Gaiotto:2021xce,Eleftheriou:2023jxr} without which the index overcounts states that are interdependent at finite $N$. 
In both cases, although these contributions are invisible order by order in perturbation theory, they impose essential constraints needed to recover the physical behavior of the system.

In this paper, we consider another example in which subleading saddles are structurally essential: an ``inner horizon'' saddle that contributes to the density of states of near-extremal black holes.
Famously, the near-horizon geometry of an extremal charged black hole is AdS$_2 \times S^{d-2}$, and corrections away from extremality can be studied by dimensionally reducing to AdS$_2$ dilaton gravity, with the dilaton $\Phi$ encoding the area of the transverse sphere~\cite{Iliesiu:2020qvm,Turiaci:2024cad}.
In particular, at linear order in the dilaton perturbation $\phi\equiv \Phi-\Phi_0$, the theory is simply JT gravity \cite{Jackiw:1984je}, and the density of states at energy $E$ above extremality reads
\be\label{eq:rho-JT}
\rho(E) = \frac{e^{S_0}\,\phi_b}{2\,\pi^2}\,\sinh\!\bigl(2\pi\sqrt{2\phi_b E}\bigr)\,,
\ee
at leading order in $e^{S_0}$.
While not apparent at first glance, this formula represents a correction to the Bekenstein--Hawking entropy coming from the inner horizon area.
In particular, the above density of states can be expressed as a difference of exponentials
\be\label{eq:rho-two-areas}
\rho(E) \sim e^{A_{\text{out}}/4G} -  e^{A_{\text{in}}/4G}\,,
\ee
where we have identified
\begin{equation}
\frac{A_{\text{out/in}}}{4G} = S_0 \pm 2\pi\sqrt{2\phi_b E}
\end{equation}
as the areas over $4G$ of the two horizons of the original near-extremal black hole, in the semiclassical regime. 
Away from extremality, the outer horizon term dominates exponentially, recovering the standard Bekenstein-Hawking relation.
On the other hand, towards extremality ($E \to 0$), the two horizons approach one another ($A_{\text{out}} \to A_{\text{in}}$) and the density of states approaches zero. 

\begin{figure}[t]
\centering
\includegraphics[width=.48\linewidth]{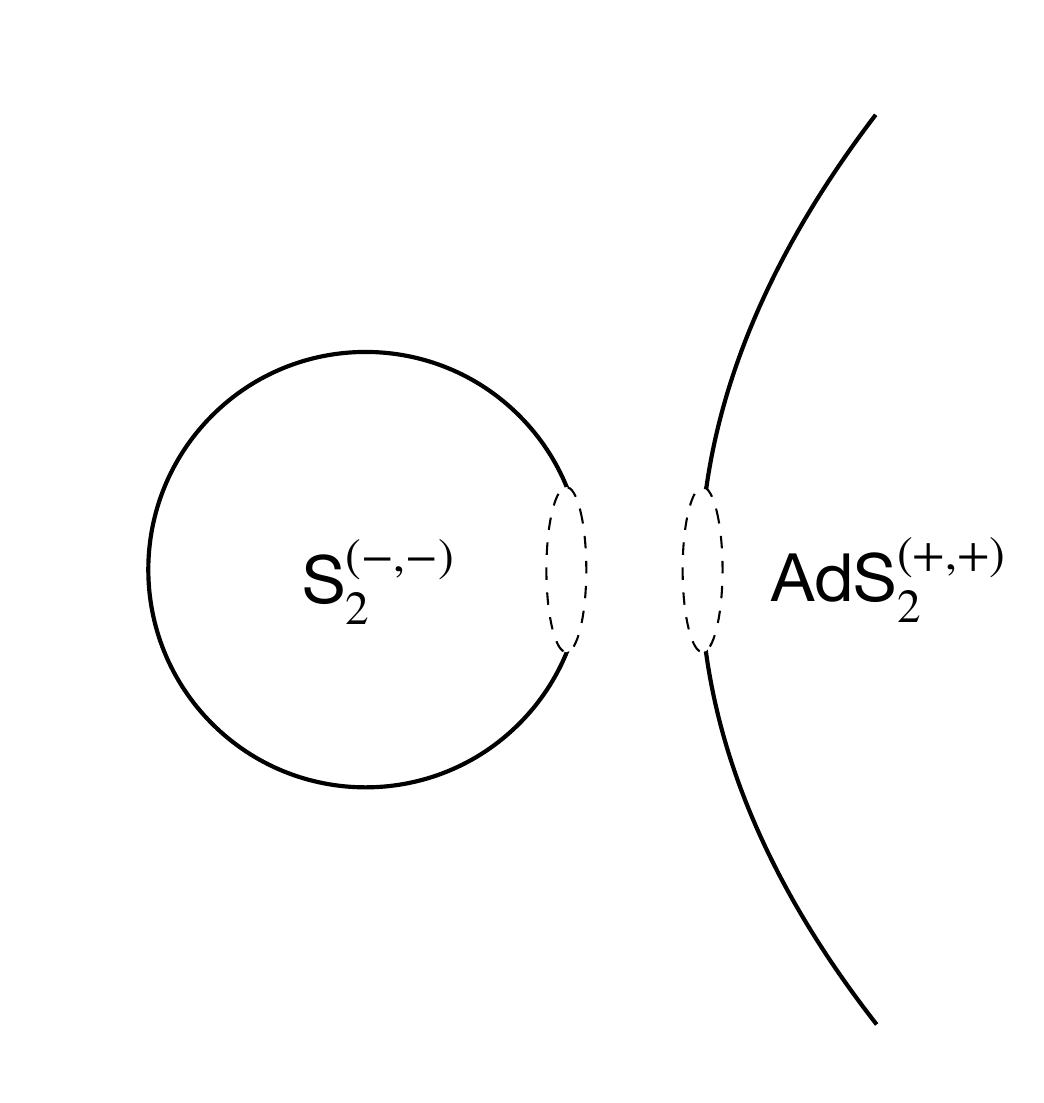}
\caption{
An illustration of the inner horizon saddle. The geometry is complexified near the dotted circles so that it avoids the outer horizon. The superscripts indicate the metric signature.}
\label{fig:innerhorizonschem}
\end{figure}

Hence, while the inner horizon saddle is subleading, its contribution is physically meaningful; 
accompanied by its crucial minus sign, 
it facilitates an appropriate counting of states towards extremality. 
Nevertheless, the bulk gravitational interpretation of this contribution is understudied.
Only recently, in Ref.~\cite{Kruthoff:2024gxc}, has it been proposed that the subleading term in Eq.~\eqref{eq:rho-two-areas} can be understood as the contribution of a semiclassical saddle geometry to the gravitational path integral.
A schematic diagram of this geometry is shown in Fig.~\ref{fig:innerhorizonschem}.
Much like the standard Euclidean cigar geometry computing $e^{A_\text{out}/4G}$,  this geometry is characterized by having a Euclidean time circle which shrinks to zero size in the bulk.
However, instead of capping off at the outer horizon, this geometry caps off at a location associated with the Lorentzian inner horizon.
As shown in Fig.~\ref{fig:innerhorizonschem}, one can visualize this geometry as a standard AdS$_2$ cigar which, before capping off at the outer horizon, complexifies, and transitions to a spherical region with exotic signature which caps off at the inner horizon.

In this work, we follow up on the proposal of Ref.~\cite{Kruthoff:2024gxc}.
First, we demonstrate how microcanonical gravitational boundary conditions, which are appropriate to the computation of $\rho(E)$, naturally give rise to a bulk saddle geometry which caps off at the inner horizon.
We find that this geometry is constrained to have a negative boundary length, interpreted literally as a condition on the line element of the thermal circle.
This unusual fact admits a clean thermodynamic interpretation, for while $E$ is fixed in the microcanonical ensemble, $\beta$ is integrated over, and in gravity this corresponds to an integration over the length of the asymptotic thermal circle.
This integral runs over an inverse Laplace contour in the complex plane, and hence can pick up contributions from saddles with non-positive ``length'' $\beta$.
This is precisely the case for the inner horizon saddle, and points to a more general possibility that negative-$\beta$ saddles may contribute to gravitational path integrals in interesting ways. 

Having understood the classical behavior of the inner horizon saddle, we then turn to questions of stability and quantum corrections.
In particular, we show how quantum corrections give rise to the relative minus sign between the two saddles $e^{A_{\text{out}}/4G}$ and $e^{A_{\text{in}}/4G}$, following the rules of contour integrals and Picard-Lefschetz theory.
Physically, the minus sign arises from instabilities of the inner horizon saddle at one loop. Rotating the integration contours of the wrong-sign Schwarzian modes produces a Jacobian phase $(\mp i)^3=\pm i$.
Combined with a $\mp 1/(2\pi i)$ factor in the inverse-Laplace measure (with the $\pm$ arising from the orientation of the steepest-descent contour through $\beta=-\beta_0$), this yields an unambiguous overall minus sign for the inner-horizon contribution. 
Furthermore, the saddle is robust to perturbative classical sources, unlike geometries with inner horizons in Lorentzian signature~\cite{Poisson:1989zz,Poisson:1990eh}.

Finally, we discuss what it means for the inner horizon saddle to be a complex geometry.
In particular, as noted in \cite{Kruthoff:2024gxc}, the inner horizon saddle violates the  Kontsevich--Segal--Witten criterion~\cite{Kontsevich:2021dmb,Witten:2021nzp} for the allowability of complex metrics in the gravitational path integral.
This criterion has a nuanced status in the literature.
On the one hand, for certain black hole saddles of the superconformal index, KSW violation coincides with the onset of superradiance~\cite{BenettiGenolini:2026raa,BenettiGenolini:2025jwe,Krishna:2026rma}, suggesting the criterion is correlated with genuine physical boundaries in the space of metrics. 
A similar picture arises in the original examples discussed in Ref.~\cite{Witten:2021nzp}.
On the other hand, other works suggest the KSW criterion is overly restrictive: the double cone with matter~\cite{Saad:2018bqo,Chen:2023hra}, the wormhole saddles of~\cite{Bah:2022uyz}, bra-ket wormholes in de Sitter~\cite{Fumagalli:2024msi,Jung:2025fbt}, and the inner horizon saddle studied here all violate bulk KSW despite appearing to yield well-defined physics.  
Motivated by these examples,  
we propose a weakening of the KSW criterion, called the ``spectral KSW'' (sKSW) criterion.
The idea behind the new criterion is to rule out any complex metric whose spectrum of perturbative fluctuations is so spread out in the complex plane that the one-loop determinant cannot be defined, even using contour rotations for wrong-sign modes.
We check that sKSW is satisfied for scalar fields on the inner horizon saddle and in other simple examples, leaving the cases discussed above to future work. 

The outline of the paper is as follows:
\begin{enumerate}
\item In Section \ref{sec:jt}, we argue that the inner horizon geometry provides a sensible contribution to  $\rho(E)$ at the classical level.
We show that it is stable, at least classically, to the addition of perturbative matter sources. 

\item In Section \ref{sec:quantum_corrections}, we use quantum corrections to derive and interpret the overall minus sign on the inner horizon contribution to $\rho(E)$.

\item In Section \ref{sec:ksw}, we discuss KSW violation in the inner horizon geometry. 
We introduce the spectral KSW criterion and consider simple examples.

\item In Section \ref{sec:phantom}, we show that the contour of integration for $\beta$ appears to rule out contributions from sKSW-violating solutions with nontrivial ``winding number'' of the complex geometry. 
\end{enumerate}

\noindent
Meanwhile, in Appendix \ref{app:geodesic}, we interpret fixed-energy two-point functions semiclassically in terms of complexified geodesics associated with the inner horizon geometry.
In Appendix \ref{app:holonomy-rep}, we elaborate upon the structure of two-dimensional gauge theory on the inner horizon saddle. 
In Appendix \ref{app:agmon-details}, we discuss the role of spectral cuts in zeta-regularized determinants.
In Appendix \ref{app:multihorizon}, following Ref.~\cite{Kruthoff:2024gxc}, we take first steps towards a generalization of our results to $2d$ theories with richer horizon structure.

\vspace{10pt}
\noindent
\textbf{A note on units:}
Throughout we shall set $8\pi G=1$, or $1/4G=2\pi$.
We will also set $\ell_{AdS}=1$, which means that temperatures $\beta^{-1}$, energies $E$, radii $r$, etc., are measured in units of $\ell_{AdS}$ or $\ell_{AdS}^{-1}$.
The dilaton $\phi$ and topological expansion parameter $S_0$ are dimensionless.
We eventually set $\phi_b=1/2$. 

\section{The classical inner horizon saddle}
\label{sec:jt}

In this section, we review JT gravity and its thermodynamic ensembles, and we argue that the inner horizon saddle of Ref.~\cite{Kruthoff:2024gxc} is a classical solution of JT gravity with fixed-energy boundary conditions.
We then turn to quantum corrections to this picture in Section \ref{sec:quantum_corrections}.

\subsection{Review: Ensembles and boundary conditions in JT gravity}
\label{sec:ensembles}

We consider JT gravity~\cite{Jackiw:1984je,Teitelboim:1983ux} with action 
\be\label{eq:JT-action}
I_{\text{JT}}=I_{\text{bulk}}+ I_{\text{bdy}}+I_{\text{top}}\qquad
I_{\text{bulk}} = -\frac{1}{2}\int_{{\cal M}} d^2x\,\sqrt{g}\,\bigl(\phi\,R + U(\phi)\bigr)\,,
\ee
where the dilaton potential is $U(\phi) = 2\phi$, and $I_{\text{top}}\equiv -S_0 \chi(\cal M)$ is a topological term proportional to the Euler characteristic.
We take $S_0$ to be very large, which suppresses contributions from non-disk topologies.
This theory has no local degrees of freedom: the dilaton serves as a Lagrange multiplier enforcing $R=-2$, which in turn fixes the local geometry. 
The theory does, however, admit nontrivial boundary dynamics upon introducing a boundary for the spacetime at finite distance, $\partial {\cal M}$.\footnote{A way to think about this is that one is performing a path integral over all manifolds-with-boundary having $R=-2$. 
Each of these geometries can be immersed into a larger asymptotically AdS$_2$ geometry, and from that
perspective they are described by wiggly, possibly self-overlapping, boundaries \cite{Ferrari:2024ndr}.}
The choice of boundary action $I_{\text{bdy}}$ hence plays a crucial role in defining the theory, as we now review. 

When studying black hole thermodynamics with Euclidean path integrals in JT gravity, there are two natural choices for the boundary action (for a recent discussion of these choices, among others, see \cite{Goel:2020yxl}). 
The first choice is a Gibbons-Hawking-York-type prescription,
\be
\label{eq:dd-bcs}
I_{\text{bdy}}^{\text{DD}} =  - \int_{\partial {\cal M}} du\,\sqrt{h}\,\phi (K-1)\,,
\qquad 
\begin{cases}
    \phi|_{\partial {\cal M}} = \phi_b/\epsilon\\
    \int_{\partial {\cal M}} du \sqrt{h}\big|_{\partial {\cal M}}
    = \beta/\epsilon\,,
\end{cases}
\ee
where $u$ parameterizes $\partial {\cal M}$, and $h=g_{uu}$ characterizes the induced metric on $\partial {\cal M}$.
On the right, we have shown the associated ``Dirichlet--Dirichlet'' boundary conditions,
namely fixed renormalized boundary dilaton value $\phi_b$ and fixed renormalized boundary length $\beta$; 
the quantity $1/\epsilon$ represents a large-distance cutoff in the bulk.
The gravitational path integral performed with these ``DD'' boundary conditions, schematically denoted
\begin{equation}
    Z_{\text{DD}}(\beta, \phi_b)=\int_{\substack{\phi_b,\,\beta\\\text{fixed}}} \mathcal{D}g\,\mathcal{D}\phi\; e^{-I_{\text{DD}}},
\end{equation}
is interpreted as the partition function of the theory in the canonical ensemble, denoted $Z(\beta)$.

An alternative choice of boundary term is given by a Brown-York-type prescription \cite{Brown:1992bq,Goel:2020yxl},
\be\label{eq:dn-bcs}
I^{\text{DN}}_{\text{bdy}} = - \int du\,\sqrt{h}\,(\phi K - \partial_n \phi)\,,
\qquad
\begin{cases}
    \phi|_{\partial {\cal M}} = \frac{\phi_b}{\epsilon}
    \\
    \frac{\phi-\partial_n \phi}{\epsilon} \big|_{\partial {\cal M}} = E \,, 
\end{cases}
\ee
where the associated  boundary conditions are ``Dirichlet--Neumann'';  namely, instead of fixing the length of the boundary, one fixes the normal derivative of the dilaton at the boundary.
The quantity $E$ specifies the ADM energy in JT gravity, so this is a fixed-energy boundary condition.
The gravitational path integral performed with DN boundary conditions,
\begin{equation}
    Z_{\text{DN}}(E, \phi_b)=\int_{\substack{\phi_b,\,E\\\text{fixed}}} \mathcal{D}g\,\mathcal{D}\phi\; e^{-I_{\text{DN}}},
\end{equation}
is interpreted as the density of states at energy $E$ in the microcanonical ensemble, denoted $\rho(E)$.

In ordinary statistical mechanics, $\rho(E)$ and $Z(\beta)$ are related by a Laplace transform
\begin{equation}
    Z(\beta) = \int_0^{\infty} dE \rho(E) e^{-\beta E}
    \label{eq:Laplace}
\end{equation}
and inverse-Laplace transform\footnote{In writing this formula, we are treating the density of states as a distribution on the continuous energy axis. To see this, one can write
\begin{equation}
    \rho(E) = \int_0^{\infty} dE' \rho(E') \int_{a-i\infty}^{a+i\infty} \frac{d\beta}{2\pi i} e^{\beta (E-E')}
\end{equation}
and recognize the innermost integral as a representation of $\delta(E-E')$, as opposed to the Kronecker delta.}
\begin{equation}
\label{eq:inverse-laplace-0}
    \rho(E) = \int_{a-i\infty}^{a+i\infty} \frac{d\beta}{2\pi i} e^{\beta E} Z(\beta) \,,
\end{equation}
where the positive parameter $a$ places the integration contour to the right of any non-analyticities. 
Hence, we can reinforce the aforementioned identifications 
\begin{equation}
    Z(\beta) = Z_{\text{DD}}(\beta) \qquad \rho(E) = Z_{\text{DN}}(E)\,,
\end{equation}
by noting that these two quantities are also related by an inverse Laplace transform \cite{Brown:1992bq,Goel:2020yxl}:\footnote{For intuition on this result, write
\be\label{eq:DN-derivation}
\begin{aligned}
Z_{\text{DN}}(E)
&= \int_{\substack{\phi_b,\,E\\\text{fixed}}} \mathcal{D}g\,\mathcal{D}\phi\; e^{-I_{\text{DN}}} 
= \int_{\substack{\phi_b,\,E\\\text{fixed}}} \mathcal{D}g\,\mathcal{D}\phi\; e^{-I_{\text{DD}}}\,e^{\int_{\partial {\cal M}} du\,\sqrt{h}\,(\phi - \partial_n\phi)} \,,
\end{aligned}
\ee
and note that if \textit{both} boundary length $\beta$  and energy $E$ were held fixed, we would have $\int_{\partial\mathcal{M}} du\,\sqrt{h}\,(\phi - \partial_n\phi) = \beta E$, with which eq.~\eqref{eq:DN-derivation} starts to look like an inverse Laplace transform.}
\be\label{eq:DN-derivation2}
\begin{aligned}
Z_{\text{DN}}(E)= \int \frac{d\beta}{2\pi i}\; e^{\beta E}\,\int_{\substack{\phi_b,\,\beta\\\text{fixed}}} \mathcal{D}g\,\mathcal{D}\phi\; e^{-I_{\text{DD}}}
= \int \frac{d\beta}{2\pi i}\,e^{\beta E}\,Z_{\text{DD}}(\beta)\,,
\end{aligned}
\ee
where the integral is taken on an inverse Laplace contour.\footnote{As we explain in Appendix~\ref{app:joukowski}, this formula is also compatible with the representation of $\rho(E)$ that arises from random matrix theory~\cite{Turiaci:2020fjj,Kruthoff:2024gxc}. In that case, the relevant integration parameter is interpreted as the horizon dilaton value $\phi_h$, rather than the inverse-temperature $\beta$.}
Hence, comparing to Eq.~\eqref{eq:inverse-laplace-0}, we see that the DN path integral is equivalent to an inverse Laplace transform of the DD partition function. 
We will make explicit use of this fact in Section \ref{sec:quantum_corrections}.

\subsection{Classical solutions}
\label{sec:jt-setup}

Building on the boundary conditions reviewed in Section~\ref{sec:ensembles}, we now discuss the classical solutions of JT gravity with fixed-energy boundary conditions.

The standard solution of pure JT gravity is the hyperbolic disk or ``Euclidean cigar,'' with a linear dilaton profile:
\be\label{eq:metric-r}
ds^2 = (r^2 - r_+^2)\,d\tau^2 + \frac{dr^2}{r^2 - r_+^2}\,,\qquad \phi(r) = \phi_b r\,.
\ee
The domain of the radial coordinate is $r \in (r_+,\infty)$ or, in view of the fixed-dilaton boundary condition in Eq.~\eqref{eq:dn-bcs}, 
\begin{equation}
    r \in (r_+,1/\epsilon)\,.
    \label{eq:cigar-contour}
\end{equation}
Working henceforth at leading order in the asymptotic cutoff $\epsilon\to0$, the fixed-energy boundary condition in Eq.~\eqref{eq:dn-bcs} gives
\begin{equation}
    E = \frac{1}{2}\phi_b r_+^2\,.
    \label{eq:energy}
\end{equation}
Smoothness at the cap requires introducing the Euclidean time periodicity $\tau\sim \tau+2\pi/r_+$, and hence we can read off the following on-shell relation
\begin{equation}
    \beta_0 \equiv  \sqrt{\frac{2\pi^2 \phi_b}{E}}
     \label{eq:beta-plus}
\end{equation}
between the renormalized boundary length and the energy $E$.

Note that when Wick rotated to Lorentzian signature, Eq.~\eqref{eq:metric-r} describes AdS$_2$ in Rindler coordinates, with a horizon at $r=r_+$ and a nontrivial extension to $r<r_+$.
Moreover, the equal-and-opposite values of the dilaton, namely $\phi_{\pm}=\pm \phi_b r_+$ at $r=r_+$ and $r=-r_+$, further allow one to interpret these loci as the outer horizon and inner horizon of the parent near-extremal black hole, respectively.\footnote{This interpretation is reinforced by the relation between Eq.~\eqref{eq:metric-r} and the BTZ metric in ADM coordinates $(T,R,\varphi)$.
First, one considers the two-dimensional BTZ base metric $-N^2(R) dT^2 + dR^2/N^2(R)$.
Then, one defines $R_0 = (R_\textrm{out}+ R_\textrm{in})/2$ as the midpoint of the two BTZ horizon radii.
With $R = R_0 + \sqrt{2}\epsilon r$, $T=t/(\sqrt{2}\epsilon)$, and $R_\textrm{out/in} = R_0 \pm \sqrt{2}\epsilon r_+$, taking $\epsilon\to 0$ (the near-horizon, near-extremal limit) recovers Eq.~\eqref{eq:metric-r} up to a Wick rotation and rescaling of the AdS radius.
As desired, $r=\pm r_+$ correspond to the horizons of the parent BTZ spacetime.}
Hence, the cigar we have just described, which caps off at $r_+$, is the dimensional reduction of the standard higher-dimensional cigar geometry, which caps off at the charged black hole's outer horizon.
Indeed, for the JT cigar saddle, the on-shell action is a boundary term which can be straightforwardly evaluated, and gives the following classical contribution to $Z_{\text{DN}}(E)=\rho(E)$:
\begin{equation}
\label{eq:first-term}
    \rho(E) \ni \# e^{S_0 + 2\pi\sqrt{2\phi_b E}}\,,
\end{equation}
recovering the first term in Eq.~\eqref{eq:rho-two-areas} with $S_0 + 2\pi\sqrt{2\phi_b E}$ corresponding to the outer horizon area over $4G$ of the parent near-extremal black hole.
We emphasize that while $E$ is the ADM energy in JT gravity, in the higher-dimensional uplift it instead represents the deviation from extremality.

While the cigar is the standard classical solution of interest in JT gravity, Eq.~\eqref{eq:rho-JT} inspires us to seek other classical solutions compatible with fixed-energy boundary conditions.
The equation of motion $R=-2$ completely fixes the local geometry, 
so that the metric can always locally be written in the form \eqref{eq:metric-r}.
However, globally distinct on-shell geometries can be obtained by considering contours in the complex $r$-plane which differ from the real ray in Eq.~\eqref{eq:cigar-contour}.
For example, with $r\in (-r_+, r_+)$, which can be parameterized via $r=r_+ \cos\theta$, Eq.~\eqref{eq:metric-r} describes a two-sphere with minus-minus signature; this exotic signature is what allows the two-sphere to still obey the on-shell condition $R=-2$.
Meanwhile, the inner horizon geometry shown in Fig.~\ref{fig:innerhorizonschem} is obtained using the complex contour shown in Fig.~\ref{fig:r-plane}(b), which runs from $r=-r_+$ to $r=1/\epsilon$, and passes through the complex $r$-plane to avoid capping off at $r_+$.
Note that, following Ref.~\cite{Witten:2021nzp}, we will initially assume that any two homotopic curves in the complex-$r$ plane give rise to physically equivalent complex metrics, so that, for instance, wiggling the contour near some large, positive $r$ is the same as doing nothing.\footnote{See, however, the discussion of contour dependence above Eq.~\eqref{eq:for-p4}.} 

We claim that the inner horizon geometry provides a sensible classical contribution to $\rho(E)$, matching, in particular, the second term in Eq.~\eqref{eq:rho-two-areas}.
To understand this, we need to discuss the inner horizon geometry in more detail.
As with the cigar geometry, the periodicity $\tau\sim \tau+2\pi/r_+$ is sufficient to ensure smoothness at the cap $r=-r_+$,\footnote{We note that the relevant relation $\beta_{in}\approx -\beta_{out}$ is only true in the extremal regime; there are corrections away from extremality, and in this case a conical defect will emerge at the inner horizon if the outer-horizon periodicity is chosen.
The property $\beta_{in}= -\beta_{out}$ is also violated in the generalized dilaton gravity theories that we discuss in Appendix \ref{app:multihorizon}.
} 
and the ADM energy, computed using Eq.~\eqref{eq:dn-bcs}, is again given by Eq.~\eqref{eq:energy}.
Unlike the cigar geometry, however, the radial coordinate $r$ on the inner horizon geometry is complex-valued, and this has several consequences.
Consider exchanging the complex radial coordinate $r$ for a real parameter $\lambda$ parameterizing the curve in Fig.~\ref{fig:r-plane}(b).
Then, the metric has manifestly complex components:
\be\label{eq:metric-r-lambda}
ds^2 = \left(r(\lambda)^2 - r_+^2\right)\,d\tau^2 + \frac{r'(\lambda)^2}{r(\lambda)^2 - r_+^2}\,d\lambda^2\,,
\ee
and manifestly complex area and induced line elements:
\begin{equation}
    \sqrt{g}\,d\tau d\lambda 
    =\sqrt{r'(\lambda)^2}\,d\tau d \lambda
    \qquad
    \sqrt{h}\,d\tau
    =\sqrt{r(\lambda)^2 - r_+^2}\,d\tau\,.
\end{equation}
Here $\sqrt{g}$ and $\sqrt{h}$ are two-sheeted quantities, owing to the fact that the square root function $\sqrt{z}$ has two sheets in the complex-$z$ plane, distinguished by an overall sign. 
We will say that the first branch of $\sqrt{g}$ (or $\sqrt{h}$) is the one which yields a positive answer as $r(\lambda)$ reaches the boundary at $r=+\infty$, while the second branch is the one which yields a negative answer there.
For instance,
\begin{equation}
   \sqrt{h}\overset{r\to+\infty}{\longrightarrow} \begin{cases}
       r \,, \qquad &\text{sheet 1}\\
       - r  \,, \qquad &\text{sheet 2}\,.
   \end{cases}
   \label{eq:first-branch}
\end{equation}

\begin{figure}[t]
\centering
\begin{minipage}[t]{0.48\textwidth}
\centering

\begin{tikzpicture}[scale=0.85]
  \draw[->] (-2.5,0) -- (5,0) node[right] {$\text{Re}\,r$};
  \draw[->] (0,-1.5) -- (0,2) node[above] {$\text{Im}\,r$};

  \draw[thick, red] (1,0) -- (4.5,0);

  \draw[thick, red, -stealth] (1,0) -- (3,0);

  \node[above, red] at (2.8,0) {$\mathcal{S}_1$};
  
  \draw[decorate, decoration={zigzag, segment length=4pt, amplitude=1.5pt}, thick, gray]
    (-1,0) -- (1,0);

  \fill (-1,0) circle (2.5pt);
  \node[below left, font=\small] at (-1,0) {$r=-r_+$};
  \fill (1,0) circle (2.5pt);
  \node[below, font=\small] at (1,-0.15) {$r=r_+$};
\end{tikzpicture}
\vspace{2pt}

(a) Outer horizon contour
\end{minipage}
\hfill
\begin{minipage}[t]{0.48\textwidth}
\centering
\begin{tikzpicture}[scale=0.85]

  \draw[decorate, decoration={zigzag, segment length=4pt, amplitude=1.5pt}, thick, gray]
    (-1,0) -- (1,0);

  \draw[->] (-2.5,0) -- (5,0) node[right] {$\text{Re}\,r$};
  \draw[->] (0,-1.5) -- (0,2) node[above] {$\text{Im}\,r$};

  \def\r{0.4}

  \draw[thick, orange]
    (-1,0) -- ({1-\r},0)
    arc (180:0:\r)
    -- (4.5,0);

  \draw[thick, orange, -stealth] (-1,0) -- (0,0);
  \draw[thick, orange, -stealth]
    ({1-\r},0) arc (180:90:\r);
  \draw[thick, orange, -stealth] ({1+\r},0) -- (3.5,0);

  \node[above, orange] at (3.5,0) {$\mathcal{S}_2$};

  \fill (-1,0) circle (2.5pt);
  \node[below left, font=\small] at (-1,0) {$r=-r_+$};
  \fill (1,0) circle (2.5pt);
  \node[below, font=\small] at (1,-0.15) {$r=r_+$};
\end{tikzpicture}
\vspace{2pt}

(b) Inner horizon contour
\end{minipage}
\caption{
The complex-$r$ plane with the branch cut of $\sqrt{h} = (r^{2}-r_+^2)^{1/2}$ shown as a squiggly line.
\textbf{(a)}~The cigar contour $\mathcal{S}_1$ (red) lives on the first sheet of $\sqrt{h}$. It runs along the real axis from the cap at $r=r_+$ to the boundary at $r\to \infty$, and $\sqrt{h}$ is positive throughout.
\textbf{(b)}~The inner horizon contour $\mathcal{S}_2$ (orange) lives on the second sheet of $\sqrt{h}$. It runs from the cap at $r=-r_+$, towards the outer horizon.
It avoids the outer horizon by going into the complex plane, where $\sqrt{h}$ is complex.
Finally, it ends up at $r\to+\infty$ where $\sqrt{h}$ is negative.
The KSW condition is violated in part of the region where $\sqrt{h}$ is complex (see Section~\ref{sec:ksw-violation-section}).}
\label{fig:r-plane}
\end{figure}

Usefully, the branch structure of disk solutions is highly constrained if we assume\footnote{See  \cite[appendix A]{Witten:2021nzp} and \cite{Louko:1995jw} for comments on applying the Gauss-Bonnet theorem to complex two-dimensional metrics.} the Gauss-Bonnet theorem holds.
Consider applying the Gauss-Bonnet theorem:
\begin{equation}
    2\pi\chi
    =
    \frac{1}{2}\int_{\mathcal{M}}\sqrt{g}\,d^{2}x\, R
    +
    \int_{\partial \mathcal{M}}\sqrt{h}\,du\, k_g \,,
    \label{eq:GB}
\end{equation}
to the on-shell metric \eqref{eq:metric-r}.
Here, $k_g$ is the boundary geodesic curvature.
Remaining agnostic to the location of the smooth cap $r=r_{\text{cap}}$, and remembering that the boundary lies at $r=1/\epsilon$, we have
\begin{equation}
\begin{aligned}
\frac{1}{2}\int_{\mathcal{M}}\sqrt{g}\,d^{2}x\, R
    &=
    \pm \frac{2\pi}{r_+}\left(r_{\text{cap}}-\frac{1}{\epsilon}\right)
    \\
    \int_{\partial \mathcal{M}}\sqrt{h}\,du\, k_g
    &=
    \pm\frac{2\pi}{r_+\epsilon}\,,
\end{aligned}
\label{eq:genius}
\end{equation}
where the $\pm$ signs indicate that we are also allowing ourselves to be agnostic about the branches of $\sqrt{g}$ and $\sqrt{h}$.
However, finiteness of $\chi$ requires the IR-divergent pieces in eq.~\eqref{eq:genius} to cancel, and this happens only when the branch choices for $\sqrt{g}$ and $\sqrt{h}$ are correlated as written.
Further, after the cancellation, we obtain a nontrivial constraint on the location of the cap:
\begin{equation}
    \chi=\pm\frac{r_{\text{cap}}}{r_+}\,.
    \label{eq:chi-cap}
\end{equation}
Since a manifold with disk topology has  $\chi=1$,  we conclude that we must have either
\begin{itemize}
    \item $r_{\text{cap}}=r_+$, and branch choice \#1 ($\sqrt{h}>0$ and $\sqrt{g}>0$ at the boundary), or
    \item $r_{\text{cap}}=-r_+$, and branch choice \#2 ($\sqrt{h}<0$ and $\sqrt{g}<0$ at the boundary) .
\end{itemize}
The first bullet point defines the cigar geometry, while we say the second bullet point defines the inner horizon geometry.

We are now prepared to see how the inner horizon saddle accounts for the second term in Eq.~\eqref{eq:rho-two-areas}.
As mentioned above Eq.~\eqref{eq:first-term}, on-shell actions in pure JT gravity are boundary terms, due to the equation of motion $R=-2$.
Hence, flipping the sign of $\sqrt{h}$ corresponds to flipping the sign of the entire on-shell action (with the exception of the $S_0$ term) relative to the cigar answer \eqref{eq:first-term}, and consequently the inner horizon geometry contributes to $\rho(E)$ as follows: 
\begin{equation}
\label{eq:second-term}
    \rho(E)\ni \# e^{S_0 - 2\pi\sqrt{2\phi_b E}}\,.
\end{equation}
This recovers the second term in Eq.~\eqref{eq:rho-two-areas}, with $S_0 - 2\pi\sqrt{2\phi_b E}$ corresponding to the inner horizon area over $4G$ of the parent near-extremal black hole.
Explicitly, putting together Eqs.~\eqref{eq:first-term} and \eqref{eq:second-term}, we have
\begin{equation}
    \rho(E)=\# e^{S_0 + 2\pi\sqrt{2\phi_b E}}
    + \# e^{S_0 -2\pi\sqrt{2\phi_b E}}\,,
    \label{eq:Zdn}
\end{equation}
in the semiclassical gravity approximation,  as desired.

We emphasize that because the inner horizon lives on the second branch of $\sqrt{h}$, its boundary thermal circle has negative length.
In particular, denoting the boundary length by $\beta/\epsilon$ as in Eq.~\eqref{eq:dd-bcs},
we have
\be
\label{eq:neg-length}
\frac{\beta}{\epsilon}
=\int_0^{2\pi/r_+} d\tau \sqrt{h}\big|_{r_b} 
= - \frac{2\pi}{r_+}\sqrt{r_b^2 - r_+^2} 
\approx 
- \frac{2\pi}{r_+ \epsilon}\,,
\ee
and hence, using Eq.~\eqref{eq:beta-plus}, the renormalized boundary length of the inner horizon saddle reads
\begin{equation}
    \beta= -\beta_0\,,
\end{equation}
and is negative.\footnote{This echoes the well-known phenomenon that inner horizons have negative surface gravity; it would be interesting to make this connection  more precise.}
This fact is important for correctly interpreting the apparent absence of the inner horizon saddle in the canonical ensemble.
Consulting the DD boundary conditions in Eq.~\eqref{eq:dd-bcs}, in light of Eq.~\eqref{eq:neg-length}, we see that the inner horizon saddle provides a classical contribution not only to the density of states $\rho(E)=Z_{\text{DN}}(E)$ with $E>0$, but also to the ordinary partition function $Z(\beta)=Z_{\text{DD}}(\beta)$ when $\beta$ is taken to be negative.
To calculate this contribution, we first compute the on-shell action of the cigar to obtain
\begin{equation}
\label{eq:ZDD-classical}
    Z(\beta) 
    \approx 
    e^{-I_{\text{cigar}}}
    =
e^{S_0+2\pi^2 \phi_b/\beta}\,, \qquad \beta>0\,,
\end{equation}
and then we use that the inner horizon geometry has opposite sign of $\sqrt{h}$, to obtain
\begin{equation}
\label{eq:ZDD-classical-2}
    Z(\beta) 
    \approx 
    e^{-I_{\text{inner}}}
    =
e^{S_0-2\pi^2 \phi_b/|\beta|}\,, \qquad \beta<0\,,
\end{equation}
which is consistent with simply analytically continuing $Z(\beta)$ to negative temperatures. 
From these equations, Eq.~\eqref{eq:Zdn} can be suggestively re-written as
\begin{equation}
    \rho(E) = \# e^{\beta_0 E} Z(\beta_0) 
    +
    \# e^{-\beta_0 E} Z(-\beta_0) \,,
    \label{eq:laplace-hint}
\end{equation}
with $\beta_0$ as in Eq.~\eqref{eq:beta-plus}.
This equation suggests that the cigar and inner horizon saddles appear as saddle points of the inverse Laplace transform \eqref{eq:inverse-laplace-0} which relates $Z_{\text{DN}}$ to $Z_{\text{DD}}$; 
understanding this more precisely will be the topic of Section \ref{sec:two-saddles}.

Before concluding this subsection, let us comment that there are actually two variants of the inner horizon saddle geometry, differing by whether the $r$-contour passes above or below $r_+$ in the complex plane, and it is natural to ask which of these contours is actually contributing.
The two contours are not continuously deformable into one another, because such a deformation would necessarily involve crossing $r_+$, and the geometry pinches off there.\footnote{One cannot continuously deform the contour ``through infinity'' either, since this would introduce extra asymptotic boundaries.}
To distinguish the two contours, 
imagine continuously dragging the cap location in the cigar geometry from $r_+$ to $r_-$ along the circle $r(t)=r_+ e^{i t}$.
Depending on whether we travel counterclockwise or clockwise, the final cap location reads either
$r_-=e^{i\pi}r_+$ (with the inner horizon contour running above $r_+$) or $r_-=e^{-i\pi}r_+$ (with the inner horizon contour running below $r_+$).
Using the smoothness relation $\beta=2\pi/r_{\textrm{cap}}$,
we can equivalently refer to the original contour in Fig.~\ref{fig:r-plane} as the $\beta=|\beta_0|e^{-i\pi}$ contour, and the latter contour as the $\beta=|\beta_0|e^{i\pi}$ contour.
The meaning behind this nomenclature will become more apparent in Section \ref{sec:quantum_corrections}.
Of course, one can also imagine contours which wind several times around the branch cut $(-r_+,r_+)$ of $\sqrt{h}$, as shown in Fig.~\ref{fig:innerhorizonbubble} below.
We will discuss the role of all these geometries in Section \ref{sec:phantom}.

\subsection{Stability under classical perturbations}
\label{sec:stability}

We now briefly step away from pure JT gravity and turn to a stability analysis for the inner horizon saddle in the presence of matter fields.\footnote{We thank Adam Levine for discussions on the material in this section.}
It is natural to be concerned about the stability of the inner horizon saddle, because in Lorentzian signature, matter perturbations typically blueshift as they approach an inner horizon, and this effect is expected to destroy its structure through backreaction~\cite{Matzner:1979zz,Penrose:1979azm,Poisson:1989zz,Poisson:1990eh}.
Despite this Lorentzian instability, we will show that the Euclidean inner horizon saddle is stable to perturbations, 
due to the nature of the boundary conditions required at the cap. 

Let us discuss these boundary conditions. 
We must be careful to distinguish between smoothness conditions on the metric, which simply follow from the equations of motion (in particular, $R=-2$ is violated by a conical singularity at the cap), and regularity conditions on matter fields.
For matter fields, regularity at the cap is not a consequence of the equations of motion; instead, it follows from the behavior of the on-shell action.
As we will see below, solutions of the equations of motion that blow up at the cap have divergent on-shell action and hence are suppressed, both when one is interested in the cigar geometry and when one is interested in the inner horizon geometry.
Consequently, the dominant contributions are always those which are regular at the cap. 
On the cigar geometry, this regularity condition is equivalent, under analytic continuation to Lorentzian signature, to the Hartle-Hawking prescription for regularity on both the past and future horizons.
Meanwhile, on the inner horizon geometry, the relevant boundary condition is regularity at $-r_+$. 

This boundary condition automatically makes the Euclidean inner horizon saddle stable under classical perturbations, as we now illustrate with the example of a static field profile. 
\begin{figure}[t]
    \centering
    \includegraphics[width=0.6\textwidth]{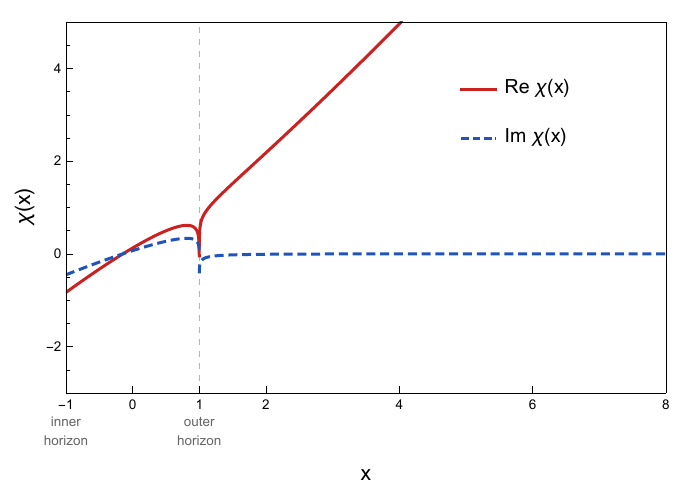}
    \caption{The real and imaginary parts of $\chi(x)$ for $m^{2}\ell^{2}=5/2$ and $J=1$. 
    The solution is regular at the inner horizon $x=-1$ (the cap), and has a singularity at the outer horizon $x=1$.
    The singularity is harmless because the inner horizon contour passes above $x=1$  in the complex $x$-plane.}
    \label{fig:chiprofile}
\end{figure}
Consider JT gravity coupled to a massive scalar field:
\be\label{eq:action-matter}
I_{\text{bulk}} = -\frac{1}{2}\int d^2x\,\sqrt{g}\,\bigl(\phi\,R + 2\phi\bigr)+\frac{1}{2}\int d^2x\,\sqrt{g}\,\bigl((\partial\chi)^2 + m^2\chi^2\bigr)\,
\ee
with $m^2\geq 0$.
We assume that the scalar field is perturbative; that is, we impose the boundary  condition 
\begin{equation}
    \chi(x)\overset{x\to\infty}{\to} Jx^{|\Delta_-|}\,,\quad
    \Delta_{\pm} = \frac{1}{2}(1 \pm \sqrt{1+4m^2})\,,
\end{equation}
with $J$ being a perturbatively small constant.\footnote{Integrating out the dilaton enforces $R=-2$, so $\chi$ does not backreact on the local geometry.
However, we must keep $J$ perturbative in order to control the order-$J^2$ dilaton backreaction.
In particular, the amount of backreaction grows with $x$, so we impose an IR cutoff at a large but finite value of $x=x_b$, and we take $J^2$ to be sufficiently small with respect to $x_b$.
}
Here, we are working in the rescaled coordinates $x = r/r_+$, $\theta = r_+\tau$, in which the metric and leading-order dilaton read
\be\label{eq:metric-x}
ds^2 = (x^2 - 1)\,d\theta^2 + \frac{dx^2}{x^2 - 1}\,,\qquad \phi = \phi_h\,x\,,
\ee
with $\theta\sim \theta+2\pi$, the outer horizon at $x = 1$, and the inner horizon at $x = -1$. 
For this metric, the scalar field equation reduces to
\be
\bigl((x^2-1)\,\chi'\bigr)' - m^2\,\chi = 0\,,
\ee
which is the Legendre equation
\be\label{eq:legendre}
\bigl((1-x^2)\,\chi'\bigr)' + |\Delta_-|(|\Delta_-|+1)\,\chi = 0\,.
\ee
The general solution to the Legendre equation is
\be\label{eq:chi-solution}
\chi(x) = c_1\,P_{|\Delta_-|}(x) + c_2\,Q_{|\Delta_-|}(x)\,,
\ee
where $P_{\ell}(x)$ and $Q_{\ell}(x)$ are Legendre functions of the first and second kind, with degree $\ell$.
Two boundary conditions are needed to fix $c_1$ and $c_2$: regularity at the cap, and the asymptotic condition $\chi(x)\sim Jx^{|\Delta_-|}$ as $x\to\infty$.

For the cigar geometry, the cap lies at $x=1$, and regularity at $x=1$ can be achieved by setting $c_2=0$:
\begin{equation}
    \chi^{\text{outer}}(x)\propto P_{|\Delta_-|}(x)\,.
\end{equation}
Meanwhile, for the inner horizon geometry, the cap lies at $x=-1$, and regularity at $x=-1$ can be achieved by analytically continuing $P_{|\Delta_-|}(-x)$ to the $x>1$ region:
\begin{equation}
\label{eq:scalar-solution}
    \chi^{\text{inner}}(x)=
  c_{0}P_{|\Delta_-|}(-x-i0^+)
    =
  c_{0}\left (e^{-i\pi |\Delta_-|} P_{|\Delta_-|}(x) - \frac{2}{\pi}\sin(\pi |\Delta_-|)Q_{|\Delta_-|}(x) \right )
    \, .
\end{equation} 
The overall normalization, $c_0$, is fixed by the boundary condition $\chi(x) \sim J\,x^{|\Delta_-|}$ as $x \to \infty$, giving
\be\label{eq:c0-solution}
c_0 = J\,\frac{2^{|\Delta_-|}\,\Gamma(|\Delta_-|+1)^2}{\Gamma(2|\Delta_-|+1)}e^{i \pi|\Delta_-|}\,.
\ee 
We comment that the regular combination $\chi^{\text{inner}}$ is typically singular at the outer horizon, but this is harmless since the inner horizon contour avoids $x=1$ in the complex $x$-plane.

Explicitly, on the inner horizon saddle, since $\chi$ satisfies the equation of motion~\eqref{eq:legendre}, integrating by parts reduces the scalar contribution to the on-shell action to a boundary term: 
\be\label{eq:on-shell-action-matter}
I_b = -
\pi
\,\bigl[(x^2-1)\,\chi'\,\chi\bigr]_{x = -1}^{x \to \infty}\,,
\ee
where the overall minus sign comes from the branch choice for $\sqrt{g}$ derived below eq.~\eqref{eq:GB}.
For generic noninteger $|\Delta_-|$, if we had incorrectly imposed regularity at the outer horizon via $\chi = P_{|\Delta_-|}(x)$, then $\chi$ would diverge logarithmically at $x=-1$ and $I_b$ would diverge. 
On the other hand, with the appropriate inner horizon boundary conditions, $\chi$ and $\chi'$ are finite as $x\to-1$, so the contribution to $I_b$ from $x=-1$ vanishes due to the factor of $x^2-1$.
The contribution at $x \to \infty$ is handled by standard holographic renormalization. 
This provides evidence that the inner horizon is physically meaningful rather than an artifact of the exactly solvable nature of JT gravity. 

To close, let us comment that we can also frame this result in the language of the AdS/CFT correspondence.
In the AdS/CFT correspondence, studying the behavior of a massive field on a gravitational saddle amounts to turning on an appropriate deformation in the boundary theory.
Specifically, one can deform the boundary theory by coupling a source $J$ to a single-trace scalar operator $\mathcal{O}$ of conformal dimension~$\Delta$,
\be\label{eq:deformed-action}
S_{\text{def}} = S_{\text{CFT}} + J \int d^d x\, \mathcal{O}(x)\,.
\ee
If the deformation is irrelevant ($\Delta > d$), this translates to a boundary condition on a bulk scalar field $\chi$ with mass $m$ obeying
\begin{equation}
    m^2\ell_{\text{AdS}}^2 = \Delta(\Delta - d)>0\,.
    \label{eq:ads-cft-mass}
\end{equation}
Explicitly,  near the conformal boundary, $\chi$ admits the expansion
$\chi \sim J\, z^{d-\Delta}  + \langle\mathcal{O}\rangle\, z^{\Delta} + \cdots$ in the Fefferman--Graham coordinate $z$, where here $\Delta$ denotes the largest positive root of Eq.~\eqref{eq:ads-cft-mass}, and dimensional analysis identifies the source $J$ with the coefficient of the leading asymptotic $z^{d-\Delta}$, which is a non-normalizable mode of $\chi$~\cite{Witten:2001ua}.\footnote{Technically, extracting
finite one-point functions $\langle\mathcal{O}\rangle$ from the bulk solution requires holographic renormalization: one evaluates the on-shell bulk action on a radial cutoff surface and adds covariant boundary counterterms to cancel the divergences order by order in the Fefferman--Graham expansion.  See Refs.~\cite{Henningson:1998gx,Balasubramanian:1999re,deHaro:2000vlm} for the original development, and Ref.~\cite{Skenderis:2002wp} for a review.}
We see that this agrees with how we defined $J$ above;
hence, the above result suggests that the inner horizon is stable to perturbative irrelevant deformations in the boundary theory.

Curiously, the bulk field $\chi$ in Eq.~\eqref{eq:scalar-solution} is complex-valued, even for real $J$, and 
as a result, the one-point function $\langle\mathcal{O}\rangle$ extracted from the subleading asymptotic on the inner horizon saddle can be complex. 
For the pure-JT partition function, the corresponding cigar and inner-horizon contributions combine to give a real microcanonical answer.
It would be interesting to determine 
whether an analogous reality condition holds for source-dependent matter observables.

\section{Quantum corrections and the minus sign}
\label{sec:quantum_corrections}

We now return to pure JT gravity.
In the previous section, we argued that the inner horizon saddle can provide a sensible classical contribution to the DN path integral, recovering Eq.~\eqref{eq:Zdn}.
However, it is a basic fact about saddle point analysis that not all saddle points need to contribute to the integral of interest, so the $\#$ coefficients in Eq.~\eqref{eq:Zdn} could very well be zero.

In this section, we show that the inner horizon saddle, together with its accompanying minus sign, does get picked up as a saddle point in the DN path integral, or, equivalently, the inverse Laplace transform \eqref{eq:DN-derivation2}.
Specifically, we carry out a steepest descent analysis for the inverse Laplace transform of the quantum-corrected partition function $Z_{\text{DD}}$.
Then, we discuss how the minus sign arises from negative mode instabilities on the inner horizon geometry at one loop.

\subsection{Review: The Schwarzian theory at one loop}
\label{sec:beta-3-2}

As we have just mentioned, our discussion of the inner horizon saddle contribution requires understanding quantum corrections to $Z_{\text{DD}}$.

In the previous section, we derived $Z(\beta)=Z_{\text{DD}}(\beta)$ in the classical approximation:
\begin{equation}
    Z(\beta)\approx e^{S_0+2\pi^2 \phi_b/\beta}\,.
\end{equation}
The one-loop correction to this answer is well-known
\cite{Maldacena:2016upp,Engelsoy:2016xyb,Cotler:2016fpe,Bagrets:2016cdf,Stanford:2017thb}:
\be\label{eq:Z-beta}
Z(\beta) = C \beta^{-3/2}\,e^{S_0 + 2\pi^2\phi_b/\beta} \qquad C = \frac{\phi_b^{3/2}}{\sqrt{2\pi}},
\ee
and moreover, $Z(\beta)$ is known to be one-loop exact, so this answer captures the full gravity path integral with $\text{DD}$ boundary conditions at disk level. 

In this section, we review a derivation of this result.
The first step is to reduce JT gravity with DD boundary conditions to the Schwarzian theory~\cite{Maldacena:2016upp,Stanford:2017thb} which has action
\be\label{eq:schwarzian-action}
I_{\text{Sch}} = -\phi_b\int_0^\beta du\;\{f(u),u\}\,,\qquad \{f,u\} \equiv \frac{f'''}{f'} - \frac{3}{2}\left(\frac{f''}{f'}\right)^2,
\ee
where $u$ is the proper time along the boundary, and $f(u)$ is the boundary reparametrization mode, subject to $\text{SL}(2,\mathbb{R})$ gauge equivalence.

The classical saddle $f(u) = \tan(\pi u/\beta)$ reproduces the on-shell action $I_{\text{Sch}}^{\text{cl}} = -2\pi^2\phi_b/\beta$, as expected.
Meanwhile, we can expand around the classical solution as
\begin{equation}
    f(u) = \tan\left(\frac{\pi u}{\beta}+ \frac{\epsilon(u)}{2}\right)\,.
\end{equation}
One finds
\begin{equation}
    I_{\text{quad}}  = \frac{\phi_b}{2} \left(\frac{\beta}{2\pi}\right)^2 \int_{0}^{\beta} du\,  \epsilon\,\left(\partial_u^4 + \frac{4\pi^2}{\beta^2} \partial_u^2\right)\,\epsilon\,.
\end{equation}
To improve the symmetry of this expression, it is useful to define a rescaled integration variable $u\to2\pi u/\beta$:
\begin{equation}
    I_{\text{quad}}  = \frac{1}{2}\int_{0}^{2\pi} du\,\epsilon\,\left[\frac{2\pi\phi_b}{\beta} \left(\partial_u^4 + \partial_u^2\right)\right]\,\epsilon\,.
\end{equation}
Now, we expand in a basis of real, orthogonal eigenmodes of the quadratic fluctuation operator,
\begin{equation}
    \epsilon(u) = \frac{\epsilon_0}{2} +\sum_{n=1}^{\infty} a_n \cos(n u)+b_n \sin(n u)\,,
    \label{eq:cos-sin-basis}
\end{equation}
with $\epsilon_0,a_n,b_n\in \mathbb{R}$, since $\epsilon$ is real-valued.
This gives
\begin{equation}
\begin{aligned}
    I_{\text{quad}} 
    &= \sum_{n=2}^{\infty} \left(\frac{\pi^2\phi_b}{\beta}n^2(n^2 -1 )\right)(a_n^2+b_n^2)
    \,,
\end{aligned}
\end{equation}
where we have applied 
$\int du \cos(nu)\cos(mu)=\pi \delta_{m,n}$, and similarly for $\sin$, and we have noticed that  $\epsilon_0$, $a_1$, and $b_1$ are zero modes.
Defining the quadratic coefficient for each non-zero mode by
\begin{equation}
    \lambda_n = \frac{\pi^2\phi_b}{\beta}n^2(n^2 -1 )\,,
    \label{eq:stable-modes}
\end{equation}
each integer $n\ge 2$ contributes two copies of $\sqrt{\pi/\lambda_n}$, namely $\pi/\lambda_n$, to the one-loop partition function.  
Meanwhile, the three zero modes  do not contribute to the determinant; they are gauge directions, divided out by $\text{Vol}(\text{SL}(2,\mathbb{R}))$.
Hence, ignoring the measure factors for now, we have
\begin{equation}
    Z_{\text{1-loop}}
    =
    \prod_{n=2}^{\infty}\frac{\pi}{\lambda_n}
    =
    e^{\sum_{n=2}^{\infty}\log(\frac{\pi}{\lambda_n})}\,,
    \label{eq:z-1-loop-schw}
\end{equation}
which can be dealt with by zeta regularization;
namely by writing
\begin{equation}
\label{eq:original-zeta-sum}
\sum_{n=2}^{\infty}\log(\frac{\pi}{\lambda_n})
=
-\sum_{n=2}^{\infty}\log(n^2(n^2-1))
+ \log\left(\frac{\beta}{\pi \phi_b}\right)\sum_{n=2}^{\infty}1\,,
\end{equation}
and, recognizing $\sum_{n=2}^{\infty}1 = \zeta(0)-1=-3/2$, we have, up to an overall numerical constant,
\be
Z(\beta) \sim  \left(\beta/\phi_b\right)^{-3/2}\,e^{S_0 + 2\pi^2\phi_b/\beta}\,,
\ee
as desired.
Computing the overall numerical constant requires deriving the path integral measure --- and in particular, its $n$-dependence --- from first principles (see, e.g., \cite{Stanford:2017thb,Cotler:2024xzz}), and we will not undertake this here.

Note that if the measure of the full Schwarzian field, including the $\text{SL}(2,\mathbb{R})$ zero-mode sector, is rescaled uniformly by an $n$-independent factor $\#$, then a consistent zeta regularization gives
\begin{equation}
    \prod_{n=-\infty}^{\infty} \#
    \,=\,
    e^{\sum_{n=-\infty}^{\infty} \log \#}
    \,=\,
    e^0\, = \, 1\,.
\end{equation}
Paradoxically, this fact (which we derived using zeta regularization) can be used to avoid using zeta regularization in the previous computation.
That is, if one adopts a measure which pairs each mode with a factor of $\beta^{-1/2}$, then the  contribution of these measure factors for $n\ge 2$ cancels with the infinite product of $\beta$'s in Eq.~\eqref{eq:z-1-loop-schw}, 
while the contribution of the measure factors for $n=0,1$ immediately gives the desired three copies of $\beta^{-1/2}$ in the one-loop contribution to $Z(\beta)$.

\subsection{Deriving the minus sign}
\label{sec:two-saddles}

We now show how the inner horizon saddle, together with its accompanying minus sign, gets picked up as a saddle point in the DN path integral.
For simplicity, we will use Eq.~\eqref{eq:DN-derivation2}, which states that the full DN gravitational path integral reduces to the inverse Laplace transform of the DD partition function.

From Eq.~\eqref{eq:DN-derivation2}, $\rho(E)=Z_{\mathrm{DN}}(E)$ is given by
\begin{equation}
\rho(E) = C e^{S_0} \int \frac{d\beta}{2\pi i} \beta^{-3/2} e^{\beta E} e^{2\pi^2 \phi_b/\beta}\,.
\label{eq:zdn-for-saddle-analysis}
\end{equation}
In the semiclassical regime
\begin{equation}
2\pi\sqrt{2\phi_b E}\gg1\,,
\end{equation}
the above integral is well-approximated by saddle point methods. 
The phase function $f(\beta) = \beta E + 2\pi^2\phi_b/\beta$ has two\footnote{See Section \ref{sec:phantom} for comments on the saddles which live on the second sheet of $\beta^{-3/2}e^{\beta E+2\pi^2\phi_b/\beta}$.} critical points at $\beta=\pm \beta_0$, with $\beta_0$ as in Eq.~\eqref{eq:beta-plus}, and at these points, we have $f(\beta_\pm) = \pm 2\pi\sqrt{2\phi_b E}$, consistently with the classical result in Eq.~\eqref{eq:laplace-hint}.
To derive the steepest descent curves associated with the $\beta_\pm$ saddle points, we first recall that the curve of steepest descent ${\cal J}_P$ attached to a given point $P$ must satisfy $\text{Im}(f({\cal J}_P))=\text{Im}(f(P))$.
In the present context,
\begin{equation}
    \text{Im}(f(\beta_+)) = 0 = \text{Im}(f(\beta_-))\,,
\end{equation}
and one finds that the locus $\text{Im}f(\beta)=0$ must either lie on the real $\beta$ axis, or along the circle
\begin{equation}
    \beta = \beta_0 e^{i \theta}\,,
    \label{eq:steepest-circle}
\end{equation}
with $\theta \in [0,2\pi)$.
To identify where the steepest descent lines actually lie within this locus, we write
\begin{equation}
    f''(\beta_\pm)= \pm 4\pi^2\phi_b/\beta_0^3\,.
\end{equation}
This equation tells us that $f(\beta)$ has a local minimum along the real axis at $\beta_+$.
To obtain a Gaussian profile with a local maximum, the contour must therefore run through $\beta_+$ parallel to the imaginary axis.
Hence, the steepest descent direction at $\beta_+$ is along the imaginary direction (parallel to the original inverse Laplace contour), and moves along the circle \eqref{eq:steepest-circle}. Meanwhile, the steepest descent direction at $\beta_-$ is along the real axis.

\begin{figure}[t]
\centering
\begin{minipage}[t]{0.48\textwidth}
\centering
\begin{tikzpicture}[x=0.75pt, y=0.75pt, yscale=-1, line width=0.75pt, scale=0.75, every node/.style={scale=0.75}]
\draw[<->] (360,24) -- (360,278);
\draw[<->] (222,150) -- (498,151);
\draw[->, color=bromwichgreen, dashed] (480,280.5) -- (480,27.5);
\draw[decorate, decoration={zigzag, segment length=4pt, amplitude=1.5pt}]
  (229,150) -- (360,150);
\draw[color=contourred] (295.42,125.39)
  .. controls (305.53,99.81) and (330.63,81.71) .. (360,81.71)
  .. controls (398.3,81.71) and (429.35,112.51) .. (429.35,150.5)
  .. controls (429.35,188.49) and (398.3,219.29) .. (360,219.29)
  .. controls (324.55,219.29) and (295.31,192.91) .. (291.16,158.86);
\draw[->, color=contourred] (295.42,125.39) -- (358,145);
\draw[->, color=contourred] (358,145) -- (231,145);
\draw[->, color=contourred] (237,159) -- (291.16,158.86);
\draw[color=saddleblue] (292,145) circle (4.25pt);
\draw[color=saddleblue] (289,142.7) -- (295,147.3);
\draw[color=saddleblue] (295,142.7) -- (289,147.3);
\draw[color=saddleblue] (429.5,150.75) circle (4.25pt);
\draw[color=saddleblue] (426.5,148.45) -- (432.5,153.05);
\draw[color=saddleblue] (432.5,148.45) -- (426.5,153.05);
\node[anchor=north west, color=saddleblue] at (432,124) {$\beta_{+}$};
\node[anchor=north west, color=saddleblue] at (271,119) {$\beta_{-}$};
\node[anchor=north west, color=contourred] at (367,58) {$\mathcal{C}_{2}$};
\node[anchor=north west, color=bromwichgreen] at (482,47) {$\mathcal{C}_{1}$};
\node[anchor=north west] at (497,138) {$\text{Re}\,\beta$};
\node[anchor=north west] at (343,-1) {$\text{Im}\,\beta$};
\end{tikzpicture}
\vspace{2pt}

(a) Contour deformation
\end{minipage}
\hfill
\begin{minipage}[t]{0.48\textwidth}
\centering
\begin{tikzpicture}[x=0.75pt, y=0.75pt, yscale=-1, line width=0.75pt, scale=0.75, every node/.style={scale=0.75}]
\draw[<->] (360,24) -- (360,278);
\draw[<->] (222,150) -- (498,151);
\draw[->, color=bromwichgreen, dashed] (480,280.5) -- (480,27.5);
\draw[decorate, decoration={zigzag, segment length=4pt, amplitude=1.5pt}]
  (229,150) -- (360,150);
\draw[color=saddleblue] (292,145) circle (4.25pt);
\draw[color=saddleblue] (289,142.7) -- (295,147.3);
\draw[color=saddleblue] (295,142.7) -- (289,147.3);
\draw[color=saddleblue] (429.5,150.75) circle (4.25pt);
\draw[color=saddleblue] (426.5,148.45) -- (432.5,153.05);
\draw[color=saddleblue] (432.5,148.45) -- (426.5,153.05);
\node[anchor=north west, color=saddleblue] at (432,154) {$\beta_{+}$};
\node[anchor=north west, color=saddleblue] at (271,119) {$\beta_{-}$};
\node[anchor=north west, color=bromwichgreen] at (482,47) {$\mathcal{C}_{1}$};
\node[anchor=north west] at (497,138) {$\text{Re}\,\beta$};
\node[anchor=north west] at (343,-1) {$\text{Im}\,\beta$};
\draw[->, color=saddleblue, line width=1.5pt] (429.5,150.75) -- (490,150.75);
\draw[->, color=saddleblue, line width=1.5pt] (292,145)
  .. controls (293,81) and (315,45) .. (342,45)
  .. controls (357,45) and (371,63) .. (381,87)
  .. controls (401,137) and (417,131) .. (490,130);
\end{tikzpicture}
\vspace{2pt}

(b) Anti-thimbles through $\beta_\pm$
\end{minipage}
\caption{\textbf{(a)}~The contour deformation from the inverse Laplace contour $\mathcal{C}_1$ (green, dashed) to the steepest descent contour $\mathcal{C}_2$ (red, solid).
The wiggly line along the negative real axis is the branch cut of $\beta^{-3/2}$.
\textbf{(b)}~The anti-thimbles (blue) through the saddle points $\beta_+$ and $\beta_-$ both intersect the original contour $\mathcal{C}_1$ (green, dashed), as is necessary for both saddles to contribute to the integral.  }
\label{fig:contour-deformation}
\end{figure}

Clearly, it is impossible to deform the original contour so that it crosses both saddle points in the correct direction, and also lives in the locus $\text{Im}(f(\beta)) = 0$.
However, this issue is easily resolved using an $i\epsilon$ prescription in which $f(\beta)$ is deformed to
\begin{equation}
    f(\beta) = \beta E e^{i \epsilon} + 2\pi^2\phi_b/\beta\,.
\end{equation}
With $\epsilon$ infinitesimal but nonzero, the relevant constant-phase loci are deformed so that the original contour $\mathcal C_1$ can be deformed to a contour $\mathcal C_2$ crossing both saddle points in the appropriate directions, with the segment through each saddle $\beta_\pm$ lying on $\operatorname{Im}f(\beta)=\operatorname{Im}f(\beta_\pm)$.
See Fig.~\ref{fig:contour-deformation}(a) for the resulting contour ``thimbles'' with infinitesimal negative $\epsilon$. 
In Fig.~\ref{fig:contour-deformation}(b) we have shown the associated anti-thimbles --- these represent directions of steepest ascent, along which we drag our original contour so that it intersects the saddle points of interest.

Note that if we had chosen positive $\epsilon$ instead of negative $\epsilon$, we would have a similar \ref{fig:contour-deformation}(a) plot, except with (1) a reflection about the horizontal axis and (2) the orientation of the arrows reversed.
The combination of (1) and (2) has no effect on the orientation of the contour near $\beta_+$, but it reverses the orientation of the contour near $\beta_-$. 
Importantly, we will find that the final result is independent of the choice of $i\epsilon$ prescription due to a competing effect.

In particular, the $\beta^{-3/2}$ prefactor has a branch cut along the negative real $\beta$ axis. 
The inner horizon saddle point lies on this branch cut, so the $i\epsilon$ prescription is essential for meaningfully evaluating its contribution. 
For the prescription of Fig.~\ref{fig:contour-deformation}(a) ($\epsilon<0$), the steepest-descent contour
through the inner horizon saddle runs from $\beta=0$ to $\beta=-\infty$ along the upper lip of the branch cut,
i.e.\ through the copy of the saddle reached by continuation from the upper half-plane,
$\beta_-=|\beta_0|e^{i\pi}$, for which
\begin{equation}
    \beta_-^{-3/2} = |\beta_0|^{-3/2}e^{-(3/2)i\pi} =  i|\beta_0|^{-3/2}\,.
\end{equation}
For $\epsilon>0$, the contour instead runs from $\beta=-\infty$ to $\beta=0$ along the lower lip of the branch
cut, through $\beta_-=|\beta_0|e^{-i\pi}$, for which
\begin{equation}
    \beta_-^{-3/2} = |\beta_0|^{-3/2}e^{(3/2)i\pi} =  -i|\beta_0|^{-3/2}\,.
\end{equation}
This relative minus sign is canceled by the orientation flip mentioned above, so there is ultimately an unambiguous overall minus sign for the inner horizon saddle. 
(The $i$ is eliminated by the division by $2\pi i$ in the inverse Laplace transform.)

Hence, when we apply the saddle approximation to Eq.~\eqref{eq:DN-derivation2}, we obtain
\begin{equation}
    \rho(E) \approx
    \sqrt{\frac{2\pi}{f''(\beta_0)}}
    \left(
    \frac{1}{2\pi} e^{\beta_0 E} Z(\beta_0)
    -
    \frac{1}{2\pi} e^{-\beta_0 E} |Z(-\beta_0)|\right)
    \,,
\end{equation}
yielding the desired minus sign on the inner horizon saddle.
Note that, upon simplification, this saddle approximation actually matches the exact answer\footnote{It is easy to compute the exact answer \eqref{eq:exact} from the identity
\begin{equation}
\label{eq:bessel}
   I_{\nu}(z)\left(\frac{z}{2}\right)^{-\nu} 
   =  \frac{1}{2\pi i} \int_{a-i\infty}^{a+i\infty}ds\, s^{-\nu-1} e^{s+z^2/(4s)}\,,
\end{equation}
with $s=\beta E$, $\nu=1/2$, and $z=\sqrt{8\pi^2 \phi_b E}$.} for $\rho(E)$ in Eq.~\eqref{eq:rho-JT}, namely
\begin{equation}
\rho(E) = \frac{e^{S_0}\,\phi_b}{2\,\pi^2}\,\sinh\!\sqrt{8\pi^2\phi_b E}\,,
\label{eq:exact}
\end{equation}
and in this sense, the inverse Laplace integral is one-loop exact.

We emphasize that the one-loop prefactor $\beta^{-3/2}$ makes the relative sign of the two saddle contributions particularly transparent: the change of prefactor phase compensates the orientation change of the steepest-descent contour. On the other hand, for a more general fractional prefactor $\beta^{-p}$, the decomposition into dominant and subdominant saddle contributions exhibits a Stokes phenomenon whereby the coefficient of the subdominant exponential can differ between the two $i\epsilon$ prescriptions.
See Section \ref{sec:gauge} and Appendix \ref{sec:saddle-comparison} for a further discussion of  the compensating-sign mechanism  for inner horizon saddles with matter.

\subsection{The meaning of the minus sign}
\label{sec:minus-meaning}

What is the physical meaning of the minus sign derived in the previous subsection?
A natural guess is that the minus sign is related to instabilities.
While we showed in Section \ref{sec:stability} that classical matter content does not destabilize the Euclidean inner horizon, at the quantum level the sign flip on $\beta$
converts the stable quadratic fluctuations of Eq.~\eqref{eq:stable-modes} into unstable negative modes, and stability again becomes a concern.

In this section, we demonstrate that performing a Polchinski contour rotation~
\cite{Gibbons:1978ac,Polchinski:1988ua,Ivo:2025yek,Shi:2025amq,Maldacena:2024spf} for each aforementioned wrong-sign mode indeed accounts for the overall minus sign of the inner horizon
contribution to $\rho(E)$.\footnote{We thank Douglas Stanford for extensive discussion on this point.} 
In this sense, the negative sign accompanying the inner horizon saddle is not an
accident of the inverse Laplace contour, but instead captures a physical instability.
We will discuss this instability from the perspective of the canonical ensemble, so as to take advantage of established machinery in the Schwarzian theory, but it would be interesting to analyze it from the microcanonical perspective in future work.

Keeping track of phases, Eq.~\eqref{eq:laplace-hint} becomes
\begin{equation}
    \rho(E) \sim \frac{i}{2\pi i}e^{\beta_0 E} Z(\beta_0) 
    +
    \frac{1}{2\pi i} e^{-\beta_0 E} Z(-\beta_0) \,,
\end{equation}
where the first $i$ comes from the fact that the contour of integration runs along the \textit{imaginary} $\beta$ direction near the outer horizon saddle. 
Hence, the overall sign of the second term (the inner horizon term) is determined by the phases of $Z(-\beta_0)$, which can be nontrivial due to loop-level instabilities of the inner horizon geometry.

To compute this phase, 
first note that a more precise formula for the inner horizon contribution reads
\begin{equation}
    \frac{\mp 1}{2\pi i} e^{-\beta_0 E} Z(\beta_0 e^{\pm i \pi})\,.
    \label{eq:unambiguous-again}
\end{equation}
The overall $\mp1$ comes about because, as discussed in the previous subsection, the orientation of the contour near the inner horizon saddle may be reversed depending on the $i\epsilon$ prescription used.
We have already seen that this sign ambiguity is canceled by a sign ambiguity in $Z(\beta_0 e^{\pm i \pi})$, namely $Z(\beta_0 e^{\pm i \pi})\sim\pm i$. In this subsection, we will see how it arises from an entirely different perspective.

To derive the phase of the partition function on the inner horizon saddle, we revisit our derivation of $Z(\beta)$ in Section \ref{sec:beta-3-2}.
When the boundary's induced metric has negative $\sqrt{h}$, the overall sign of the Schwarzian action flips. 
If we keep the same real basis of fluctuations as in Eq.~\eqref{eq:cos-sin-basis}, the quadratic action reads
\begin{equation}
\begin{aligned}
    I_{\text{quad}} 
    &= - \sum_{n=2}^{\infty} \left(\frac{\pi^2\phi_b}{\beta_0}n^2(n^2 -1 )\right)(a_n^2+b_n^2)
    \,,
\end{aligned}
\end{equation}
and hence, introducing an $i\epsilon$ prescription, we say that all the nonzero mode coefficients have become negative: 
\begin{equation}
    \lambda_n = \frac{\pi^2\phi_b}{\beta_0 e^{\pm i\pi}}n^2(n^2 -1 )\,.
\end{equation}
Following the Polchinski rotation prescription~\cite{Gibbons:1978ac,Polchinski:1988ua,Ivo:2025yek,Shi:2025amq,Maldacena:2024spf}, for each mode, say $a_n$, with a wrong-sign Gaussian
$\exp(-e^{\mp i\pi}|\lambda_n|a_n^2)$,
the steepest descent rotation $a_n \to \pm ia_n$ contributes a Jacobian factor of $\pm i$ to the final result, while the three $\text{SL}(2,\mathbb{R})$ zero modes are not rotated.
The total Jacobian from the nonzero modes is
\be\label{eq:J-product}
J_{\text{total}} = 
\prod_{n=2}^{\infty}(\pm i)^2
=
\prod_{n=2}^{\infty}e^{\pm i \pi} = e^{\pm i\pi\sum_{n=2}^{\infty}1}= e^{\mp 3i\pi/2}\,,
\ee
where, as above, we have regulated the divergent sum over 1 to give $-3/2$.\footnote{The value $\zeta_H(0,2) = -3/2$ is the JT analogue of the integer $n_- = -3$ that Chen--Stanford--Tang--Yang~\cite{Chen:2025jqm} compute for lukewarm de Sitter black holes, giving phase $(-i)^{-3} = -i$.}
Hence, the phase of $Z(e^{\pm i \pi} \beta_0)$ is indeed $e^{\mp3  i\pi/2}=\pm i$, giving an overall unambiguous\footnote{Note that in order to get an unambiguous answer, we have prescribed that the $a_n\to\pm i a_n$ rotations be correlated with the $i\epsilon$ prescription on $\beta$.
While this is a sensible assumption, it is a nontrivial one; see Section \ref{sec:non-topological} for further comments on sign ambiguities associated with contour rotations for wrong-sign modes.
} minus sign in Eq.~\eqref{eq:unambiguous-again}.

We note that this result could equivalently have been obtained by naively setting $\beta= \beta_0 e^{\pm i \pi}$ in the original zeta sum \eqref{eq:original-zeta-sum}, without worrying about contours for wrong-sign modes.
The observation that zeta regularization can automatically account for contour rotations underlies our spectral KSW proposal in Section \ref{sec:ksw}.

\section{A spectral KSW criterion}
\label{sec:ksw}

In this section, we analyze the Kontsevich--Segal--Witten (KSW) criterion~\cite{Kontsevich:2021dmb,Witten:2021nzp} for the inner horizon geometry. 
We argue that while the KSW criterion is violated, the violation is harmless: we propose a weakening of the KSW
criterion, called the spectral KSW criterion, which permits the inner horizon saddle while continuing to exclude some  unphysical saddles.

\subsection{The KSW criterion and its violation}
\label{sec:ksw-violation-section}

We begin by introducing the KSW criterion.
The KSW criterion is a conjectured necessary criterion for a complex metric to provide a valid saddle point of the gravitational path integral.
Roughly, it says that the path integral of a free $p$-form gauge field on such a background must converge.
More precisely, it asks that the complex metric be diagonalizable at any point in a real basis as $g_{ij} = \alpha_i\,\delta_{ij}$ with $\alpha_i \in \mathbb{C}$, and that it obey the pointwise condition
\be\label{eq:ksw}
\sum_{i=1}^{d}|\text{Arg}\,\alpha_i| < \pi\,.
\ee
Clearly, Euclidean metrics ($\text{Arg}\,\alpha_i = 0$) are allowable.
Meanwhile, Lorentzian metrics saturate the bound --- because one has $\alpha_0\propto -1$ for the time component --- but become allowable upon introducing an $i\epsilon$ regulator, $\alpha_0\to -1+i\epsilon$. This can be interpreted as a geometric version of the statement that Feynman $i\epsilon$ prescriptions are required to make sense of many standard integrals in quantum field theory.

To see why Eq.~\eqref{eq:ksw} is connected to  convergence of the path integral, 
note that Eq.~\eqref{eq:ksw} implies the two properties $\text{Re}\sqrt{g}>0$ and $\text{Re}(\sqrt{g}g^{ij})>0$. 
For a free, real scalar field, the action reads
\begin{equation}
\label{eq:free-scalar-action}
    I = \frac{1}{2}\int_{\mathcal{M}} d^d x \sqrt{g} \left(
    g^{ij}\partial_i \phi \partial_{j}\phi
    +
    m^2 \phi^2\right)
\end{equation}
and we see that the above two properties imply that the action has positive real part.
In particular, after integrating by parts and dropping boundary terms, we have positivity of the real part of
\begin{equation}
\label{eq:free-scalar-action-2}
    I = \frac{1}{2}\int_{\mathcal{M}} d^d x \sqrt{g}\,
    \phi\left(-\nabla^2+m^2 \right)\phi\,.
\end{equation}
This is enough to guarantee that the path integral converges at finite field cutoff, because convergence of Gaussian integrals is controlled by the sign of the real part of the exponent.
Additionally, Eq.~\eqref{eq:ksw} implies the analogous generalization to $p$-form gauge theories. 

We now show that the inner horizon geometry violates Eq.~\eqref{eq:ksw}.
It is convenient to pass to the proper radial coordinate $\rho$, defined by $r = r_+\cosh\rho$, in which the metric and dilaton read
\be\label{eq:JT-metric}
ds^2 = \sinh^2\!\rho\;d\theta^2 + d\rho^2\,,\qquad
\phi = \phi_h\,\cosh\rho\,.
\ee
The outer horizon sits at $\rho = 0$, the inner horizon at $\rho = i\pi$, and the boundary at $\rho_b = \log(|\beta|/(\pi\epsilon))$.
The inner horizon contour therefore runs in the complex $\rho$ plane from the cap at $\rho = i\pi$ to the boundary at $\rho \to +\infty$, as shown in Fig.~\ref{fig:rho-plane}(b).
We will show that whatever path the $\rho$-contour takes between these two points, it must traverse a point at which the KSW condition is violated.

\begin{figure}[t]
\centering
\begin{minipage}[t]{0.48\textwidth}
\centering
\begin{tikzpicture}[scale=0.85]
  \draw[->] (-0.5,0) -- (5,0) node[right] {$\text{Re}\,\rho$};
  \draw[->] (0,-0.5) -- (0,4) node[above] {$\text{Im}\,\rho$};
  \draw[decorate, decoration={zigzag, segment length=4pt, amplitude=1.5pt}, thick, gray]
    (0,0) -- (0,3);
  \node[left, font=\small] at (0,1.5) {$\frac{i\pi}{2}$};
  \draw[dashed, gray] (0,3) -- (4.8,3);
  \node[left, font=\small] at (0,3) {$i\pi$};
  \draw[thick, red, ->] (0,0) -- (2.5,0);
  \draw[thick, red, ->] (2.5,0) -- (4.5,0);
  \node[red, above right, font=\small] at (3.5,0.1) {$\mathcal{S}_1$};
  \fill (0,0) circle (2pt);
  \node[below left, font=\small] at (0,0) {$\rho=0$};
\end{tikzpicture}
\vspace{2pt}

(a) Outer horizon contour
\end{minipage}
\hfill
\begin{minipage}[t]{0.48\textwidth}
\centering
\begin{tikzpicture}[scale=0.85]
  \draw[->] (-0.5,0) -- (5,0) node[right] {$\text{Re}\,\rho$};
  \draw[->] (0,-0.5) -- (0,4) node[above] {$\text{Im}\,\rho$};
  \draw[decorate, decoration={zigzag, segment length=4pt, amplitude=1.5pt}, thick, gray]
    (0,0) -- (0,3);
  \node[left, font=\small] at (0,1.5) {$\frac{i\pi}{2}$};
  \draw[dashed, gray] (0,3) -- (4.8,3);
  \draw[thick, orange,
    postaction={decorate, decoration={markings,
      mark=at position 0.25 with {\arrow{>}},
      mark=at position 0.6 with {\arrow{>}},
      mark=at position 0.9 with {\arrow{>}}}}]
    (0,3) .. controls (1.2,3) and (1.5,2.3) .. (1.5,1.5)
    .. controls (1.5,0.7) and (2,0) .. (3,0)
    -- (4.5,0);
  \node[orange, above right, font=\small] at (3.5,0.1) {$\mathcal{S}_2$};
  \fill (0,3) circle (2pt);
  \node[left, font=\small] at (0,3) {$\rho=i\pi$};
\end{tikzpicture}
\vspace{2pt}

(b) Inner horizon contour
\end{minipage}
\caption{The saddle contours $\mathcal{S}_1$ (cigar) and $\mathcal{S}_2$ (inner horizon) in the complex-$\rho$ plane. 
The two choices of sign of $\sqrt{h} = \pm \sinh\!\rho$ are inherited from the two branches of $(r^2-r_+^2)^{1/2}$ in the $r$-plane.
The squiggly line connecting $\rho = 0$ to $\rho = i\pi$ is the image of the $r$-plane branch cut under the map from $r$ to $\rho$. \textbf{(a)}~$\mathcal{S}_1$ lives on the first sheet of $\sqrt{h}$ and runs along the real axis from $\rho = 0$ to the boundary at $\rho_b$.  \textbf{(b)}~$\mathcal{S}_2$ lives on the second sheet; it caps off at $\rho = i\pi$ and runs to the real axis, passing through $\text{Im}\,\rho = \pi/2$, where the KSW bound is violated.}
\label{fig:rho-plane}
\end{figure}

We parametrize the $\rho$-contour by a real coordinate $\lambda$,  in terms of which the metric \eqref{eq:JT-metric} becomes
\be\label{eq:metric-lambda}
ds^2 = \underbrace{\sinh^2\!\rho(\lambda)}_{\alpha_1}\;d\theta^2 + \underbrace{\rho'(\lambda)^2}_{\alpha_2}\;d\lambda^2\,.
\ee
To show KSW violation, note that any continuous path connecting the cap at $\rho = i\pi$ ($\text{Im}\,\rho = \pi$) to the boundary at $\rho \to +\infty$ ($\text{Im}\,\rho = 0$) must cross the line $\text{Im}\,\rho = \pi/2$. 
At any such crossing, one has $\sinh^2\rho = -\cosh^2(\text{Re}\,\rho)$, which is real and negative, so $|\text{Arg}(\alpha_1)| = \pi$ saturates, and hence violates, the KSW bound, which is a strict bound.

\begin{figure}[t]
\centering
\includegraphics[width=0.48\linewidth]{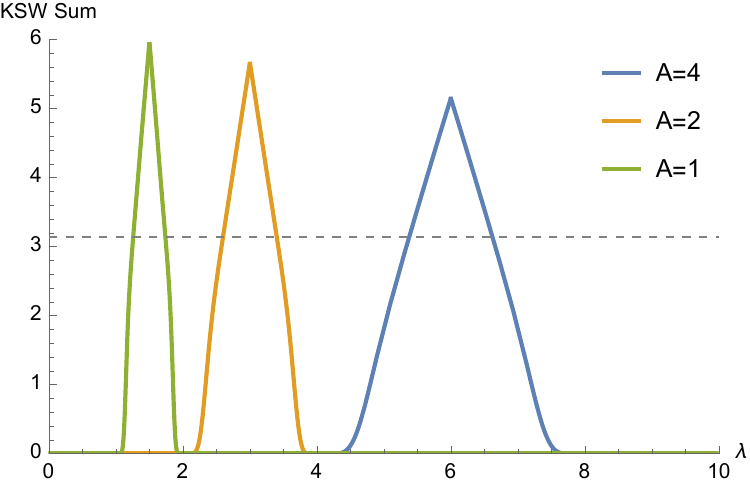}%
\hfill
\includegraphics[width=0.48\linewidth]{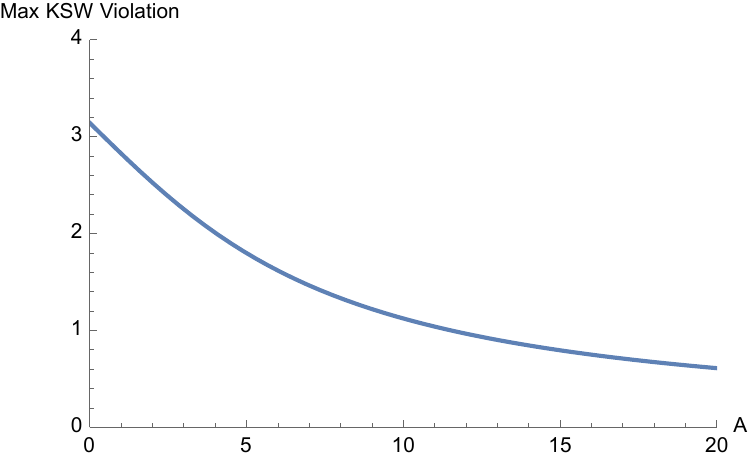}
\caption{\textbf{Left:} KSW sum along the inner horizon contour for various stretching parameters $A$.  The bound $\pi$ (dashed) is violated near the midpoint of the transition region.  \textbf{Right:} Maximum KSW violation vs.\ $A$.  The violation decreases as $O(1/A)$ but never vanishes.}
\label{fig:ksw-violation}
\end{figure}

To see the violation in an example, we construct an explicit family of contours in the $\rho$ plane, parameterized by a positive stretching parameter $A$.
We define this curve as $\rho(\lambda) = \lambda + i\pi$ for $\lambda < A$, a smooth $C^\infty$ interpolation for $A \leq \lambda \leq 2A$, and $\rho(\lambda) = \lambda$ for $\lambda > 2A$. The interpolation function is given by 
\begin{equation}\label{foot:stretch}
    \rho(\lambda) = \lambda + \frac{i\pi}{1+e^{-A((\lambda-A)^{-1} + (\lambda-2A)^{-1})}}\,, \quad A<\lambda<2A \ ,
\end{equation}
and the maximum KSW violation occurs when the contour crosses $\text{Im}\rho=\pi/2$, namely when $\lambda = 3A/2$.
At this point, as we have already mentioned, we have $|\text{Arg}(\alpha_1)| = \pi$, already saturating the KSW bound.
The second eigenvalue adds $|\text{Arg}(\rho'^2)| > 0$ for any finite $A$ (since $\rho'$ has a nonzero imaginary part in the transition region), giving a total KSW sum
\be\label{eq:ksw-violation}
|\text{Arg}\,\alpha_1| + |\text{Arg}\,\alpha_2| > \pi
\ee
exceeding the KSW bound --- see Fig.~\ref{fig:ksw-violation}.
The KSW violation coming from the $|\text{Arg}(\rho'^2)|$ term achieves its maximum at $\lambda = 3A/2$ and reads
\begin{equation}
    |\text{Arg}(\rho'^2)|=2\arctan \frac{2\pi}{A}
    \overset{A\to\infty}{\to}\frac{4\pi}{A}\,.
\end{equation}
We see that the excess violation above $\pi$ scales as $O(1/A)$ for large $A$ (Fig.~\ref{fig:ksw-violation}), so the violation can be made arbitrarily small by stretching the contour, but it never vanishes.

\subsection{Gauge fields on the inner horizon saddle}
\label{sec:gauge}

In this subsection, we consider JT gravity coupled to a gauge theory.
We reinterpret known results in this context~\cite{Kapec:2019ecr,Iliesiu:2019lfc} in terms of inner
horizon saddles.
By doing so, we will find
mechanisms by which a KSW-violating geometry can, nevertheless, lead
to well-defined physical answers.

In particular, the original KSW
papers~\cite{Witten:2021nzp,Kontsevich:2021dmb} identified two key obstructions to studying quantum field theories on KSW-violating backgrounds:
\begin{itemize}
    \item \textit{Local obstruction:} The path integral over perturbative fluctuations may diverge, when using the original contour of integration for the fields.
    \item 
    \textit{Global obstruction:}
    In the specific context of $p$-form gauge theory, 
    the partition function involves a sum over quantized integer fluxes, which may also diverge. 
\end{itemize}
In subsection \ref{sec:non-topological} below, we discuss how the local obstruction can be remedied using contour rotations at the one-loop level; 
this generalizes, for instance, the example considered in Section
\ref{sec:minus-meaning}.
Meanwhile, the second point cannot be addressed with contour rotations, since the integers labeling flux sectors cannot be meaningfully rotated into the complex plane.
Nevertheless, as we now argue, inner-horizon saddle contributions remain well-defined in the microcanonical density of states once the inverse Laplace transform is performed sector by sector.

In two spacetime dimensions, Yang--Mills theory has no local propagating degrees of freedom.
Any two-form on a two-dimensional manifold is proportional to the
``volume'' (really, area) form, so we may write
\begin{equation}
\label{eq:f-vol}
    F = f\,\text{vol}\,,
\end{equation}
where $f$ is Lie-algebra-valued for a nonabelian gauge group.
The Yang--Mills equations imply that $f$ is covariantly constant.
The remaining degrees of freedom are therefore global rather than local.
In the quantum theory, it is convenient to organize the partition function as a sum over representation sectors.
In each representation sector, the on-shell action is proportional to the area of the disk, denoted $\mathcal{A}$, and this already gives us insight into gauge theory on the inner horizon saddle.
As discussed in Section \ref{sec:jt-setup}, the inner horizon has opposite sign for the volume form relative to the cigar geometry, and hence we have $\mathcal{A}>0$ on the cigar geometry, and  
$\mathcal{A}<0$ on the inner horizon geometry.
In practice, we will prefer to express quantities in terms of the renormalized boundary length, $\beta$, which is related to the area as $\mathcal{A}=\beta/\epsilon$, at leading order in $\epsilon$ --- see
Appendix~\ref{app:holonomy-rep} for details.
The upshot is that the cigar and inner horizon on-shell actions are related by a sign flip on $\beta$, as in the pure JT analysis of Section \ref{sec:jt}.

We now address the sum over discrete gauge sectors.
As alluded to above, it will be convenient to organize this sum in terms of unitary irreducible representations of the gauge group, using the methods reviewed in Appendix \ref{app:holonomy-rep}.
For JT gravity coupled to a gauge field with compact gauge group
$G$~\cite{Iliesiu:2019lfc}, the partition function can be expressed as a sum over
representations as follows:
\be\label{eq:JTYM-partition}
Z(\beta) = Z_{\text{JT}}(\beta)\sum_R(\dim R)^2\,
        e^{-\beta E_0(R)}\,,\qquad
        E_0(R) \equiv e^2 C_2(R)/2
\ee
where $e$ is the gauge coupling, $C_2(R)$ is the quadratic Casimir in
the representation $R$, and
$Z_{\text{JT}}(\beta)$ is the pure JT partition function \eqref{eq:Z-beta}.\footnote{A factor of the Dynkin index
has been suppressed, following the conventions of
\cite{Iliesiu:2019lfc}.}
Note that we specialize to the standard disk-level JT--Yang--Mills setup of Ref.~\cite{Iliesiu:2019lfc}, with boundary conditions for which the Schwarzian and gauge sectors factorize; see Appendix~\ref{app:holonomy-rep} for the coupling and cutoff conventions.
Note also that we are imposing standard thermal boundary conditions, and in particular, vanishing  chemical potential for gauge charge.
In general, a chemical potential may be introduced by imposing Dirichlet boundary conditions
that fix the gauge holonomy around the thermal circle, e.g.~to $U = e^{i\theta}$ in the $U(1)$ case.
Here, we are setting $\theta=0$ and hence $U = \mathbf{1}$.\footnote{For nonzero $\theta$, each Boltzmann weight in
\eqref{eq:JTYM-partition} would get multiplied by the character $\chi_R(e^{i\theta})/\dim R$.
We give a self-contained derivation of
Eq.~\eqref{eq:JTYM-partition} for $G = U(1)$, and general chemical
potential, in Appendix~\ref{app:holonomy-rep}.}
Finally, the degeneracy factor $(\dim R)^2$ in eq.~\eqref{eq:JTYM-partition} arises because the heat kernel on
$G$ assigns weight $(\dim R)\,\chi_R(U)$ to each sector, and
$\chi_R(\mathbf{1}) = \dim R$.

We now consider the inverse Laplace transform which converts $Z(\beta)$ to the microcanonical density of states. 
This integral is well-defined with the original inverse Laplace contour; however, there is an obstruction to deforming the contour to reach a putative inner horizon saddle with $\beta<0$.
At positive $\beta$, the Boltzmann weight in
Eq.~\eqref{eq:JTYM-partition} suppresses large-Casimir representations
so that the sum converges.
Meanwhile, at negative $\beta$, the weight becomes
$\exp(+|\beta|E_0(R))$, growing exponentially with $C_2(R)$, and
the sum diverges.
Hence, on this KSW-violating metric, the representation sum appears to diverge,
as anticipated by Ref.~\cite{Witten:2021nzp}.
To make matters worse, one cannot even define the negative-$\beta$ answer by analytically
continuing the positive-$\beta$ answer across the imaginary axis.
For example, consider again $G=U(1)$.
In this case, representations are labeled by an integer $m$ with
$\dim R_m=1$ and quadratic Casimir $m^2$; hence, the representation
sum reduces to the Jacobi theta function
\begin{equation}\label{eq:thetareppart}
    \sum_{m\in\mathbb{Z}}e^{-e^2\beta\,m^2/2}
    = \theta_3(0,\,e^{-e^2\beta/2})\,,\qquad
    q = e^{-e^2\beta/2}\,.
\end{equation}
The convergence boundary $|q|=1$ corresponds to $\text{Re}(\beta)=0$,
and the theta function has singularities at $q = e^{2\pi i\,p/r}$ for
all $p/r\in\mathbb{Q}$. Since these rational points are dense on the
unit circle, the imaginary axis is a wall of singularities past which
no naive analytic continuation exists.
Hence, one naively expects that no saddles with $\beta<0$ can
contribute to the inverse Laplace transform of
Eq.~\eqref{eq:thetareppart}.

This issue may be resolved by considering each individual
term
\be\label{eq:Z-R}
Z_R(\beta) \propto \beta^{-3/2}\,
        e^{2\pi^2\phi_b/\beta\,-\,\beta\,E_0(R)}
\ee
in Eq.~\eqref{eq:JTYM-partition} separately.
Each term admits an analytic continuation to the left half-plane of  $\beta$, where $\text{Re}\beta<0$.
Therefore, in each sector, the density of states may be computed by
inverse Laplace transform:
\be\label{eq:rho-sectors}
\rho_{R}(E) = 
        \int_{a-i\infty}^{a+i\infty}\frac{d\beta}{2\pi i}\,
        e^{\beta E}\,Z_R(\beta)\,.
\ee
For $E>E_0(R)$, the contour may be deformed to include a saddle at $\beta<0$, 
yielding a $\sinh$ with a shifted energy argument:
\be\label{eq:rho-R}
\begin{split}
\rho_R(E) &\;=\; \frac{e^{S_0}\,\phi_b}{2\,\pi^2}\,
        \Theta\!\left(E - E_0(R)\right)\,
        \sinh\!\Bigl(\,2\pi\sqrt{2\phi_b\bigl(E - E_0(R)\bigr)}\,\Bigr)\,,  
\end{split}
\ee
where $E_0(R)$ was defined in Eq.~\eqref{eq:JTYM-partition}, and where we exclude the degeneracy factor $(\dim R)^2$ in the definition of $\rho_R(E)$.
This is just the JT density of states in
Eq.~\eqref{eq:rho-JT}, evaluated at a Casimir-shifted energy
$E\to E - E_0(R)$.
We see that each representation sector has both an outer- and an inner-horizon saddle contribution, in the same sense as
Eq.~\eqref{eq:rho-JT}. 
Summing over representations, the total density of states is finite because, at fixed energy, only finitely many representation sectors are allowed.
Similarly, splitting the $\sinh$ into two exponentials, the total sum over outer horizon saddles, or inner horizon saddles, is also finite.

Hence, by reinterpreting a known calculation, we have shown that $\rho(E)$ admits an inner horizon saddle contribution even when JT gravity is coupled to a gauge field. 
To achieve this, our prescription is to evaluate the inverse Laplace transform sector by sector before summing over sectors. 
This gives a density of states expressed as a sum of outer- and inner-horizon saddle contributions, with the expected positivity, finiteness, and continuity properties.\footnote{See Appendix~\ref{sec:saddle-comparison} for a discussion of this mechanism in the flux-lift representation.}
Curiously, when $G=U(1)$, the gauge-field configuration on the inner-horizon saddle appears to be complex; see Appendix~\ref{sec:saddle-comparison}. 
This complicates the usual interpretation of the gauge field as a real connection for a compact gauge group, and we discuss the physical interpretation of this in Section~\ref{sec:discussion}.

We speculate that the analysis in this section extends to
higher-dimensional inner horizon saddles.
Of course, in higher dimensions, an essential new feature is that the
gauge field is no longer topological, and so obtaining a finite
contribution from the path integral on a KSW-violating metric may require an
analytic continuation of the gauge field.
Conceptually, it is useful to decompose discrete gauge-sector data from the continuous fluctuations within each sector.
Depending on the problem, the discrete labels may arise from topology, flux data, or representation sectors (see Appendix~\ref{sec:gaugeexh} for an explicit example in the context of $\textrm{AdS}_{2}\times S^{2}$).
The global KSW obstruction arises in the sum over such discrete sectors, while the local KSW obstruction concerns the functional integral over continuous fluctuations within a fixed sector. 
Our approach would be to address these two obstructions separately, by choosing an appropriate order of integration and then by using contour deformations for the continuous field variables.

\subsection{The spectral KSW criterion}
\label{sec:non-topological}

In this section, we propose a weakening of the KSW criterion, called the spectral KSW criterion, and we show that the scalar field version of the criterion permits the inner horizon saddle in JT gravity.

To motivate the spectral KSW criterion, we begin with a general  discussion of matter fields on a manifold $\mathcal{M}$ with complex metric $g$.
We will focus on the case of scalar field theory and defer a detailed discussion of other matter content to future work.
The  path integral for a free, real, minimally coupled scalar  reads
\begin{equation}
    Z[g_{\mu \nu}]= \int_{\mathcal{C}_0}\mathcal{D}\phi \,  e^{-I}\,,
    \qquad
    I=\frac{1}{2}
    \int_{\mathcal{M}}d^dx \sqrt{g}\, \phi \,  \mathcal{O}\,   \phi
    \,, \qquad  \mathcal{O}\equiv -\nabla^2 + m^2\,,
    \label{zm}
\end{equation}
where the contour of integration, $\mathcal{C}_0$, runs over all real field profiles $\phi(x)$ on $\mathcal{M}$.
To make this equation more precise, we must specify both the measure $\mathcal{D}\phi$ and the contour of integration.
To do so, we  introduce a bilinear pairing on field space: 
\begin{equation}
   \langle \chi|\eta \rangle_g =  \int_\mathcal{M} d^d x \sqrt{g}\; \chi(x) \eta(x) \,.
   \label{inner-prod} 
\end{equation}
For a real Riemannian metric, this pairing is the usual positive-definite inner product on real field profiles. 
For a complex metric, it is generally complex-valued; 
however, we can still use it to define the measure $\mathcal{D}\phi$.
In particular, we can introduce coordinates $c_n$ on field space by expanding $\phi(x)$ in an arbitrary real basis:
\begin{equation}
    \phi(x) =\sum_n c_n \chi_n(x)\,,
    \qquad \chi_n(x)\in \mathbb{R}\,, \qquad c_n\in \mathbb{R}\,,
\end{equation}
and we can form the Gram matrix
\begin{equation}
    G_{nm}=\langle \chi_n|\chi_m \rangle_g\,.
\end{equation}
Then, the volume element on field space formally induced from Eq.~\eqref{inner-prod} is
\begin{equation}
    \mathcal{D}\phi =\sqrt{\textrm{det}G}\prod_n d c_n\,,
    \label{eq:inducedvol}
\end{equation}
where note that computing $\sqrt{\textrm{det}G}$ may require imposing branch choices and a choice of regularization, for instance a truncation to the first $N$ basis modes.

Altogether, we arrive at the following formula:
\begin{equation}
    Z_N[g_{\mu \nu}]= \sqrt{\det G}\int_{\mathbb{R}^N} \prod_n dc_n \,   e^{-\frac{1}{2}c^T K c}\,,
    \qquad 
     K_{nm} \equiv \int_{\mathcal{M}}d^dx \sqrt{g} \,  \chi_n \,  \mathcal{O}\,  \chi_m\,,
    \label{shock}
\end{equation}
which defines the scalar field theory with mode cutoff $N$ on the complex background $g$. 
We can suggestively rewrite Eq.~\eqref{shock} by  choosing our real basis $\chi_n$ to be orthonormal with respect to an auxiliary positive-definite inner product:
\begin{equation}
   \langle \chi|\eta \rangle_{g_0} =  \int_\mathcal{M} d^d x \sqrt{g_0}\; \chi(x)\,   \eta(x) \,,
   \label{inner-prod-2} 
\end{equation}
where here $g_0$ is an auxiliary real Riemannian reference metric on $\mathcal{M}$. 
We also rewrite the action $I$ via
\begin{equation}
    I = \frac{1}{2}
    \int_{\mathcal{M}}d^dx \sqrt{g_0} \,  \phi \,  \tilde{\mathcal{O}} \,  \phi\,,
    \qquad 
    \tilde{\mathcal{O}}\equiv s(x)\mathcal{O}\,,\qquad
    s(x)\equiv \frac{\sqrt{g}}{\sqrt{g_0}}\,.
    \label{otilde}
\end{equation}
We see that we can view $I$ as
defining a scalar field theory on the real background $g_0$, albeit with an unusual quadratic fluctuation operator $\tilde{\mathcal{O}}=s(x) \mathcal{O}$.
In this modified theory, the natural field space inner product is given by Eq.~\eqref{inner-prod-2}.
So, up to the $\sqrt{\det G}$ prefactor  in Eq.~\eqref{shock}, the original partition function $Z[g]$ is equivalent to the partition function of the modified theory on $g_0$:
\begin{equation}
    Z[g]= \sqrt{\det G}\;
\tilde{Z}[g_0]\,.
\end{equation}
Since $\sqrt{\det G}$ cancels out when computing normalized observables,
we say the matter theory defined by
$Z[g]$ is well-defined if and only if the  matter theory defined by
$\tilde{Z}[g_0]$ is well-defined.

We now turn our attention to the well-definedness of $\tilde{Z}[g_0]$, which is given by the $N\to\infty$ limit of $\tilde{Z}_N[g_0]$:
\begin{equation}
   \tilde{Z}_N[g_0]= \int_{\mathbb{R}^N} \prod_n dc_n e^{-\frac{1}{2}c^T K c}\,, \qquad
   K_{nm} = \int_{\mathcal{M}}d^dx \sqrt{g_0} \,\chi_n \,\tilde{\mathcal{O}}\,\chi_m\,.
   \label{firsttime}
\end{equation}
It is nontrivial for this $\tilde{Z}_N[g_0]$ integral to converge when the original metric $g$ appearing in the definition of $K$ is complex.
The integral converges absolutely if and only if we have $\textrm{Re}K>0$; conversely, if $\textrm{Re}K$ has negative eigenvalues then the integral will diverge.
(The condition $\textrm{Re}K\ge 0$ provides a delicate intermediate case.)
A natural, albeit schematic, proposal is then the following:
\begin{quote}
\textbf{Proposal 1:}
An admissible complex metric $g$ is one for which  $\tilde{Z}[g_0]$ converges as written.
\end{quote}
This proposal is closely connected to the KSW criterion, as we explain in detail further below.

To relax Proposal 1, 
let us rewrite $\tilde{Z}[g_0]$ in a different form. 
Note that $K$ is symmetric ($K^T=K$); let us additionally assume $K$ is invertible.
Then, we can change variables as follows:
\begin{equation}
    c= K^{-1/2}x\,,
\end{equation}
and up to factors of $2\pi$, we obtain
\begin{equation}
   \tilde{Z}_N[g_0]= (\det K)^{-1/2}
   \int_{K^{1/2}\mathbb{R}^N} \prod_n \frac{dx_n}{\sqrt{2\pi}}   e^{-\frac{1}{2}x^T x}\,.
   \label{discussrot}
\end{equation}
The integral in the above expression only converges absolutely for $\textrm{Re}K>0$, and it diverges if $\textrm{Re}K$ has negative eigenvalues.
However, suppose we allow ourselves to rotate the $K^{1/2}\mathbb{R}^N$  contour to  $\mathbb{R}^N$ by hand, in order to force the integral to converge. 
In this case, we simply obtain 
\begin{equation}
    \tilde{Z}_N[g_0]=(\det K)^{-1/2}\overset{N\to\infty}{\longrightarrow}  (\det \tilde{\mathcal{O}})^{-1/2}
    \,.
\end{equation} 
However,
even after removing the possibly divergent integral,
it remains nontrivial  to show that the remaining object $\det \tilde{\mathcal{O}}$ is  well-defined.
$\tilde{\mathcal{O}}$ is a differential operator on an infinite-dimensional Hilbert space, and moreover $\tilde{\mathcal{O}}$ need not be self-adjoint.
Hence, defining its determinant requires careful branch choices and regularization, and in some cases it may not even be possible.

This leads us to propose weakening the original proposal to:
\begin{quote}
\textbf{Proposal 2:}
An admissible complex metric $g$ is one for which $\det\tilde{\mathcal{O}}$ converges.
\end{quote}
This schematic proposal is a precursor of the spectral KSW criterion, as we explain below.

One may first wonder whether Proposals 1 and 2  depend on a choice of regularization scheme for $\tilde{Z}[g_0]$ and $\det \tilde{\mathcal{O}}$.
Indeed, it is nontrivial that either of these quantities should remain well defined in the continuum   $N\to\infty$ limit.
In this context, our preferred regularization scheme is zeta function regularization, because it automatically implements the mode rotation for wrong-sign modes discussed below Eq.~\eqref{discussrot} in the context of Proposal 2. 
To see this, consider the case where $\tilde{\mathcal{O}}$ has a discrete spectrum $\lambda_n$, and write
\begin{equation}
    \tilde{Z}[g_0]=(\det \tilde{\mathcal{O}})^{-1/2} = e^{\frac{1}{2}\zeta_{\tilde{\mathcal{O}}}'(0)}\,
    \qquad
    \zeta_{\tilde{\mathcal{O}}}(s) = \sum_n \lambda_n^{-s}\,,
    \qquad 
    \lambda_n^{-s}=|\lambda_n|^{-s}\,e^{-is\arg_\theta\lambda_n}\,,
    \label{dett}
\end{equation}
where $\arg_\theta z\in(\theta,\theta+2\pi)$ indicates that we have chosen the $\arg$ function with branch cut along a radial ray $L_\theta=(0,\infty)e^{i\theta}$.
Note that $\det \tilde{\mathcal{O}}$ is independent of the position of the spectral cut $L_\theta$ with respect to isolated eigenvalues, since increasing $\textrm{arg}_\theta \lambda_n$ by an integer multiple of $2\pi$ sends $e^{-\zeta_{\tilde{\mathcal{O}}}'(0)}$ to itself.
By contrast, $(\det\tilde{\mathcal{O}})^{-1/2}=e^{\frac{1}{2}\zeta_{\tilde{\mathcal{O}}}'(0)}$ changes sign each time $L_\theta$ is moved across a simple isolated eigenvalue. More generally, crossing an eigenvalue of odd multiplicity flips this sign, while crossing one of even multiplicity leaves it unchanged. This sign is the zeta-function counterpart of the choice between the two convergent rotations of the corresponding wrong-sign mode. In this regularization scheme, 
adding a single wrong-sign mode formally contributes a finite quantity to the partition function:
\begin{equation}
    (\det \tilde{\mathcal{O}})^{-1/2}
    \to 
    (\det \tilde{\mathcal{O}})^{-1/2}|\lambda_n|^{-1/2}e^{-\frac{i}{2}\arg_\theta  \lambda_n}\,,
\end{equation}
which  precisely matches the contribution of the initially divergent Gaussian integral
\begin{equation}
    \int_{\mathbb{R}} \frac{dc_n}{\sqrt{2\pi}} e^{-\frac{1}{2}\lambda_n c_n^2}
\end{equation}
after it has been rotated to the convergent contour $\lambda_n^{-1/2}\mathbb{R}$, as specified below Eq.~\eqref{discussrot}.
Hence, as claimed, zeta regularization automatically encodes the contour rotations described in the context of Proposal 2.

We now explain how to associate Proposals 1 and 2 with pointwise constraints on the complex metric $g$.
Let us begin with Proposal 1.
We treat the zero-mode sector separately and work with the reduced
Gaussian path integral.\footnote{In the context of zeta regularization, this means that we define a reduced determinant
$\det\,\!'\tilde{\mathcal{O}}$ by restricting $\tilde{\mathcal{O}}$
to an appropriate subspace where it is invertible.
Under assumptions A1 and A2, this subspace is simply
$(\textrm{ker}\tilde{\mathcal{O}})^\perp$.
}
The following two conditions are then sufficient for the reduced Gaussian path integral to have a well-defined zeta-regularized value:
\begin{itemize}
    \item[A1.] $\tilde{\mathcal{O}}$ is elliptic as a differential operator on a compact manifold with suitable boundary conditions,
    \item[A2.] $\textrm{Re}\tilde{\mathcal{O}}\ge0$.
\end{itemize}
At finite mode cutoff,  Condition A2, namely $\textrm{Re}K\geq 0$,  ensures that the real-contour integral converges as written (or with a suitable $i\epsilon$ prescription);
Condition A1 serves to address extra subtleties arising in the continuum limit.
Note also that in practice, compactness of the manifold can always be achieved using an IR cutoff.

Crucially, for appropriate boundary conditions, the KSW criterion \eqref{eq:ksw}  implies Conditions A1 and A2.
In particular, if we write  $\phi=\sum_n c_n \chi_n$, then we have
\begin{equation}
   c^T \textrm{Re}( K )c= \int_{\mathcal{M}}d^dx \sqrt{g_0}\,\left[\textrm{Re}(sg^{\mu \nu}) \,\partial_\mu\phi \,\partial_\nu \phi +\textrm{Re}(s) m^2\phi^2\right]
    \,.
\end{equation}
The scalar KSW constraints  $\textrm{Re}(\sqrt{g})>0$ and $\textrm{Re}(\sqrt{g}g^{\mu\nu})>0$ guarantee that this integral is nonnegative,\footnote{In fact, for $m^2>0$, the scalar KSW constraints guarantee the strict inequality $\textrm{Re}K>0$, which also guarantees the absence of zero modes.}
which gives $\textrm{Re}\tilde{\mathcal{O}}\ge 0$.
The fact that KSW also implies ellipticity  will become clear below.
In this sense, Proposal 1 is tightly associated with the pointwise KSW constraints on the metric.

We now turn to Proposal 2.
The following weaker set of criteria is sufficient for  $\tilde{\mathcal{O}}$
to admit a well-defined zeta-regularized reduced determinant \cite{Seeley:1967ea,
Seeley1969ANALYTICEO,
Braverman2005RefinedAT,Kontsevich:1994xe}:
assumption A1 above, together with
\begin{itemize}
    \item[A2$'$.] $\tilde{\mathcal{O}}$ admits a principal cut: a radial ray $L_\theta$ which avoids all attainable values of the principal symbol of $\tilde{\mathcal{O}}$.
\end{itemize}
The ellipticity requirement in A1 and the principal-cut condition A2$'$ combine into a single pointwise constraint on the metric.
We begin with A2$'$.
The principal symbol  of a general second-order differential operator
\begin{equation}
    -a^{\mu\nu}(x) \nabla_\mu \nabla_\nu + b^\mu(x) \nabla_\mu + c(x)
    \label{belowthm}
\end{equation}
is defined as 
\begin{equation}
    p_2(x,\xi) = a^{\mu\nu}(x) \xi_\mu \xi_\nu\,,
    \label{p2-first-time}
\end{equation}
where $x$ is any point in $\mathcal{M}$ and $\xi$ is any  real covector.
Intuitively speaking, the principal symbol controls the high-energy part of the spectrum, where the highest-derivative term dominates over all lower derivative terms.
In our scalar field case, the principal symbol of $\tilde{\mathcal{O}}$ reads
\begin{equation}
    p_2(x,\xi) = s(x)g^{\mu \nu}\xi_\mu \xi_\nu\,.
\end{equation}
Hence, in order to obey Condition A2$'$, it is sufficient to obey the pointwise constraint that $\sqrt{g}g^{\mu \nu}\xi_\mu \xi_\nu$ always avoids a common fixed radial cut $L_\theta$.

The ellipticity requirement in A1 imposes a separate requirement, but also a local one.
Ellipticity is the property that the principal symbol \eqref{p2-first-time} is everywhere nonvanishing for $\xi\ne 0$.
Intuitively, it is the requirement that high-frequency modes always have a nontrivial leading-order cost.\footnote{For instance, the operator $-\partial_x^2+m^2$ on $\mathbb{R}^2 \ni(x,y)$ fails to be elliptic because it annihilates (at order $N^2$) functions of $y$ with a large frequency proportional to $N$.}
In the scalar field case, this is simply the constraint that  $\sqrt{g}g^{\mu \nu}\xi_\mu \xi_\nu$ fails to attain zero for $\xi\ne 0$. 


This leads us to propose relaxing the original KSW criterion:
\begin{quote}
\textbf{KSW criterion:}
An admissible complex metric $g$ is one which satisfies
the local conditions \eqref{eq:ksw}.
\end{quote}
to the following:
\begin{quote}\label{def:sKSW}
\textbf{Spectral KSW criterion:}
An admissible complex metric $g$ is one for which $\tilde{\mathcal{O}}$ admits  a principal cut and is elliptic,
for any free scalar or $p$-form gauge theory.
\end{quote}
For a scalar field,
this criterion imposes 
that for nonzero $\xi$,
$\sqrt{g(x)}g^{\mu \nu}(x)\xi_\mu \xi_\nu$ nowhere intersects $L_\theta\cup \{0\}$, for some common radial ray $L_\theta$. 
For a gauge field, $p_2$ becomes matrix-valued, and it is then the spectrum of this $p_2$ matrix, with appropriate gauge-fixings understood, which must avoid $L_\theta\cup \{0\}$. Although real Lorentzian spacetimes violate this criterion, they lie on the boundary of the allowed domain, with their amplitudes defined as boundary values through a specified $i\epsilon$ limiting prescription, just as for the original KSW criterion. 

Three features of our spectral KSW criterion (sKSW) now deserve comment.

First, we emphasize that sKSW, like the original KSW criterion, is proposed as a \textit{necessary} criterion, not a sufficient one.
Satisfying sKSW guarantees nothing about whether a saddle $g$ actually contributes to the gravitational path integral. 
In particular, even when the matter partition function on $g$ is well-defined, the saddle may still be ruled out by global features of the integration contour that local spectral analysis cannot see.

Secondly, we emphasize that for scalar fields, the spectral KSW criterion is strictly weaker than the original KSW criterion.
KSW requires that $\sqrt{g}g^{\mu\nu}$ has positive real part, and this automatically gives the desired ellipticity property and principal cut for spectral KSW (for instance, the negative real axis).
The implication is strict, at least at the scalar field level, because there are nontrivial cases where KSW is violated but sKSW is satisfied, 
as exhibited by the examples below. 

Thirdly, we comment on the importance of having a principal cut $L_\theta$ for $\sqrt{g}g^{\mu \nu}\xi_\mu \xi_\nu$.
While we have presented it as part of the sufficient criteria for the convergence of $\det \tilde{\mathcal{O}}$, the existence of a principal cut is, in many cases,  necessary for $\det \tilde{\mathcal{O}}$ to define a continuous functional on the \textit{space} of backgrounds.
In particular, $\sqrt{g}g^{\mu \nu}\xi_\mu \xi_\nu$ controls the high-energy part of the spectrum of $\tilde{\mathcal{O}}$,  
and so the absence of a principal cut is often associated with the presence of asymptotic accumulation regions in the spectrum which fill all directions in the complex plane.
In this case, a small variation of the background metric can cause an infinite number of eigenvalues to cross the spectral cut $L_\theta$ used to define the complex exponential $\lambda_n^{-s}$ in Eq.~\eqref{dett}, and this 
then leads to a discontinuous jump in the value of $\det \tilde{\mathcal{O}}$ across two continuously related backgrounds. 
We give examples of this pathological effect below and in Appendix \ref{app:agmon-details}.

\subsubsection{Examples}

We  now apply the spectral KSW criterion in examples.
We focus on the scalar field case throughout;
additional analysis is required to determine whether the $p$-form part of sKSW holds in these examples.

\begin{figure}[t]
\centering
\begin{tikzpicture}
\begin{axis}[
    name=left,
    width=6.5cm, height=5.5cm,
    scale only axis,
    title={\textbf{(s)KSW satisfied}: $\Omega = 1/2 + i/2$},
    title style={font=\small, yshift=-2pt},
    xlabel={$\mathrm{Re}\,\lambda$},
    ylabel={$\mathrm{Im}\,\lambda$},
    xlabel style={font=\small},
    ylabel style={font=\small},
    tick label style={font=\footnotesize},
    xmin=-40, xmax=460,
    ymin=-110, ymax=240,
    axis x line=bottom,
    axis y line=left,
    axis line style={-},
  ]
  \addplot[
    only marks,
    mark=*,
    mark size=0.8pt,
    darkblue,
    opacity=0.6,
  ] table[x=re, y=im] {ksw_safe.dat};
\end{axis}
\begin{axis}[
    name=right,
    at={(left.east)},
    anchor=west,
    xshift=1.4cm,
    width=6.5cm, height=5.5cm,
    scale only axis,
    title={\textbf{(s)KSW violated}: $\Omega = 1/2 + 3i/2$},
    title style={font=\small, yshift=-2pt},
    xlabel={$\mathrm{Re}\,\lambda$},
    ylabel={$\mathrm{Im}\,\lambda$},
    xlabel style={font=\small},
    ylabel style={font=\small},
    tick label style={font=\footnotesize},
    xmin=-200, xmax=200,
    ymin=-240, ymax=680,
    axis x line=bottom,
    axis y line=left,
    axis line style={-},
  ]
  \addplot[
    only marks,
    mark=*,
    mark size=0.8pt,
    darkblue,
    opacity=0.6,
  ] table[x=re, y=im] {ksw_viol.dat};
\end{axis}
\end{tikzpicture}
\caption{Scatter plot of the scalar Laplacian spectrum
$\lambda_{n,k}=\bigl(2\pi n/\beta-k\Omega\bigr)^{2}+k^{2}$
on the torus with rotating identifications, with $\beta=2\pi$ and $|n|,|k|\le 12$.
\textbf{Left:} for $|\mathrm{Im}\,\Omega|<1$ (here $\Omega=1/2+i/2$) the
eigenvalues and principal symbols lie in a wedge of the right half-plane,
so a principal
cut exists outside the wedge. The metric satisfies both sKSW and KSW.
\textbf{Right:} for $|\mathrm{Im}\,\Omega|>1$ (here $\Omega=1/2+3i/2$)
the wedge opens past the imaginary axis and the asymptotic spectrum
accumulates in every direction.}
\label{fig:spectrum}
\end{figure}

\paragraph{Complex torus:} First consider the flat, real torus
\begin{equation}
ds^2 = d\tau^2 + d\phi^2\,,
\end{equation}
with $\tau\sim \tau+\beta$ and $\phi\sim \phi+2\pi$. The scalar Laplacian  is
\begin{equation}
\nabla^2=\partial_\tau^2+\partial_\phi^2 ,
\end{equation}
and we consider the operator $\mathcal{O}=-\nabla^2+m^2$ with $m^2>0$. 
This operator is invertible.
Its principal symbol reads
\begin{equation}
p_2(\xi)=\xi_\tau^2 + \xi_\phi^2\,,
\end{equation}
and is constrained to lie in $\mathbb{R}_{>0}$ for nonzero $\xi$.
We see that $\mathcal{O}$ is invertible,
elliptic,
and admits a principal cut (e.g. the negative real axis $L_\pi$),
and so  this geometry satisfies spectral KSW.

Now consider the same torus but with the thermal identification replaced by the following complex identification:
\begin{equation}
(\tau,\phi)\sim(\tau+\beta,\phi+\beta\Omega).
\label{eq:complex-identification}
\end{equation}
Here, $\Omega$ is a complex ``angular potential'' giving rise to a complexification of the angular coordinate $\phi$.
To revert to a real angular coordinate, we introduce
\begin{equation}
\tilde{\phi}=\phi-\Omega\tau ,
\end{equation}
so that we simply have the real identification $\tilde{\phi}\sim\tilde{\phi}+2\pi$. The price to be paid is that the metric now has complex coefficients:
\begin{equation}
ds^2=d\tau^2+\left(d\tilde{\phi}+\Omega d\tau\right)^2 .
\label{eq:equatorial}
\end{equation}
Conveniently, however, the determinant remains real: $\det g=1$.
Hence, there is no need to introduce the auxiliary metric $g_0$, and we can set  $s(x)=1$ in Eq.~\eqref{otilde}.
The relevant scalar Laplacian is
\begin{equation}
\nabla^2=\left(\partial_\tau-\Omega\partial_{\tilde{\phi}}\right)^2+\partial_{\tilde{\phi}}^2 ,
\end{equation}
and we consider the operator  $\tilde{\mathcal{O}}=-\nabla^2+m^2$, with $m^2>0$.
$\tilde{\mathcal{O}}$ is invertible for $|\textrm{Im}\Omega|<1$.
Its principal symbol reads
\begin{equation}
p_2(\xi)=\left(\xi_\tau-\Omega\xi_{\tilde{\phi}}\right)^2+\xi_{\tilde{\phi}}^2 ,
\end{equation}
and is nonzero for every nonzero real covector (hence $\tilde{\mathcal{O}}$ is elliptic) 
away from the special cases with $|\textrm{Im}\Omega| = 1$.
Finally, for $|\textrm{Im}\Omega| < 1$,
the principal symbol values $p_2(\xi)$ are constrained to lie in an angular wedge in the right half-plane:
\begin{equation}
    p_2(\xi) \in \{z\in \mathbb{C}:|\arg z|\le \arcsin|\textrm{Im}\Omega|<\frac{\pi}{2}\}\,,
    \label{wedge}
\end{equation}
so that  the negative real axis furnishes a principal cut.
We see that  $\tilde{\mathcal{O}}$ is invertible, elliptic, and admits a principal cut for $|\textrm{Im}\Omega| < 1$,
and so spectral KSW is satisfied in this case.

This example also illustrates the idea that the principal symbol controls the high-energy part of the spectrum.
Here, the spectrum of $-\nabla^2$ is discrete:
\begin{equation}
\lambda_{n,k}=\left(\frac{2\pi n}{\beta}-k\Omega\right)^2+k^2, \qquad n,k\in\mathbb{Z}.
\label{eq:scalar-spectrum-torus}
\end{equation}
For $|\operatorname{Im}\Omega|<1$, the entire spectrum is restricted to the wedge \eqref{wedge}, 
but accumulates towards infinity towards each radial ray in the wedge --- see Fig.~\ref{fig:spectrum}. 
In particular, the spectrum accumulates towards infinity along all of the directions accessible to the principal symbol.

From the spectrum \eqref{eq:scalar-spectrum-torus}, we can directly compute $\det\tilde{\mathcal{O}}$ in zeta regularization.
In the massless case, 
 it reads
\begin{equation}
   \det\,\!'(-\nabla^2) 
   =
   \beta^2\,\eta\left(\frac{\beta}{2\pi}(\Omega+i)\right)^2
   \,
   \eta \left(\frac{\beta}{2\pi}(-\Omega+i)\right)^2
   \,,
   \label{eq:dettorus}
\end{equation} 
where the prime indicates removal of the zero mode.
This expression is well-defined for $|\textrm{Im}\Omega|<1$.
In the massive case ($m^2>0$), the formula \eqref{eq:dettorus} becomes more complicated, but it is also well-defined for $|\textrm{Im}\Omega|<1$, consistently with  spectral KSW.

On the other hand, for $|\textrm{Im}\Omega| > 1$,
the principal symbol values $p_2(\xi)$  explore the full complex plane as $\xi$ varies.
Similarly,  the spectrum has asymptotic accumulation directions for all angles in the complex plane (again, see Fig.~\ref{fig:spectrum}).
Hence, spectral KSW is violated for $|\operatorname{Im}\Omega|>1$. 
Indeed, 
the determinant \eqref{eq:dettorus} is ill-defined for $|\operatorname{Im}\Omega|>1$.

Two comments are now in order.

First, note that the $|\textrm{Im}\Omega| =1$ threshold matches the KSW condition in this example.
Necessity of $|\textrm{Im}\Omega|<1$ for KSW allowability is immediate: the submanifold $\tilde{\phi}=-\tau\, \text{Re}(\Omega)$ of Eq.~\eqref{eq:equatorial} has purely real line element $ds^2 = (1 -\text{Im}(\Omega)^2)d\tau^2$, and $|\textrm{Im}\Omega|<1$ is the condition that the coefficient of $d\tau^2$ is positive.
Sufficiency follows by explicitly diagonalizing Eq.~\eqref{eq:equatorial} with the real coordinates
\begin{equation}
    \tilde{\phi} = r_1+r_2    \qquad
    \tau = \frac{r_1}{x}+\frac{r_2}{y}\qquad
    \begin{cases}
        x\equiv \sqrt{1-\text{Im}(\Omega)^2}-\text{Re}\Omega\\
        y\equiv \frac{|\Omega|^2-1}{x}\,,
    \end{cases}
\end{equation}
in terms of which Eq.~\eqref{eq:equatorial} becomes
\begin{equation}
    ds^2 =
    \frac{2}{x^2}\left[1-\text{Im}(\Omega)\left(\text{Im}(\Omega)-i\sqrt{1-\text{Im}(\Omega)^2}\right)\right] dr_1^2
    +
    \left[\frac{1+(y+\Omega)^2}{y^2}\right]dr_2^2\,.
    \label{eq:torus-diagonalized}
\end{equation}
From this, it follows that the KSW condition \eqref{eq:ksw} reduces to $|\textrm{Im}\Omega|<1$. 

Secondly, we comment that the $|\text{Im}\Omega|=1$ threshold has physical significance for rotating AdS black holes.
In particular, the torus \eqref{eq:complex-identification} 
with $\Omega_H\equiv i\Omega\in \mathbb{R}$
is the conformal boundary of the BTZ geometry with horizon velocity $\Omega_H$.
Hence, the $|\textrm{Im}\Omega|=1$ threshold coincides with the $|\Omega_H|=1$ extremality threshold of the BTZ geometry, below which its quasi-Euclidean metric is KSW-allowable \cite{Witten:2021nzp}.
This suggestive result is discussed in more detail in Section \ref{sec:discussion}.
Moreover, preliminary calculations suggest the conclusion that both KSW and sKSW violation occur at the $|\textrm{Im}\Omega|=1$ threshold generalizes to
\begin{equation}
ds^2 = d\tau^2 + d\theta^2 + \sin^2 \theta (d\tilde{\phi}+ \Omega d\tau)^2\,,
\end{equation}
which can be thought of as a higher-dimensional version of the torus \eqref{eq:equatorial}, and the conformal boundary of a spinning geometry in AdS$_4$ gravity.

\paragraph{Inner horizon saddle:}
We now demonstrate that the inner horizon saddle satisfies scalar sKSW. 
Consider again the complex metric \eqref{eq:metric-lambda}.
In this section, we adopt a new radial coordinate 
$w=\coth(\rho/2)$, under which the metric reads
\begin{equation}
    ds^2 = \frac{4}{(w(\lambda)^2-1)^2}(w'(\lambda)^2 d\lambda^2+w(\lambda)^2 d\theta^2)\,.
\end{equation}
The contour defining the geometry runs from $\rho=i\pi$, or equivalently $w=0$, to a large, real value of $\rho$, call it $\rho=\rho_b\gg1$.
In the $w$ coordinate, the cutoff radius lies at $w_b=\coth(\rho_b/2)>1$, where $w_b\to1^+$ as $\rho_b\to\infty$.

Unlike in the torus example above, here the volume form is complex:
\begin{equation}
    \sqrt{g}
    =
    \frac{4ww'}{(w^2-1)^2}\,.
\end{equation}
Following Eq.~\eqref{otilde}, we introduce a real Riemannian reference metric
\begin{equation}
ds_0^2=d\lambda^2 + \lambda^2 d\theta^2\,,
\qquad
\sqrt{g_0}
    = \lambda \,,
\end{equation}
together with the quantity
\begin{equation}
    s(x) = \frac{\sqrt{g}}{\sqrt{g_0}}=\frac{4ww'}{\lambda(w^2-1)^2}.
\end{equation}

The sKSW criterion then concerns the operator $\tilde{\mathcal{O}}=s(x)(-\nabla^2+m^2)$.
Its
principal symbol reads
\begin{equation}
    p_2 = \frac{1}{\lambda}\left(\frac{1}{a}\xi_\lambda^2 +a\xi_\theta^2\right)\qquad 
    a=\frac{w'}{w}\,,
    \label{p2-inner}
\end{equation}
and its real part reads
\begin{equation}
   \textrm{Re} p_2 =\frac{ (\log|w|)'}{\lambda}\left(\frac{1}{|a|^2}\xi_\lambda^2 +\xi_\theta^2\right)\,.
   \label{realp2inner}
\end{equation}
The coefficient $(\log|w|)'$ is positive as long as $|w|$ is strictly increasing along the contour.
To realize this choice, we can take the contour to have the form $w(\lambda)=\lambda e^{i\varphi(\lambda)}$, with $\lambda=|w(\lambda)|\in[0,w_b]$.
We can also choose $\varphi(\lambda)$ to be a small, smooth bump which vanishes near $\lambda=0,w_b$, and is negative at $\lambda=1$, so as to avoid the singularities at $w=\pm 1$.\footnote{In the complex $\rho$ plane, this contour reaches $\rho_b$ from the right (i.e.~from larger real $\rho$) instead of from the left.
Despite this counterintuitive behavior, 
the calculations in Section \ref{sec:jt} all carry over, as long as one continuously transports the square-root branches and boundary normal orientation from the original inner-horizon contour of Fig.~\ref{fig:rho-plane}.}
With these choices,  
$\tilde{\mathcal{O}}$ is elliptic; $p_2$ vanishes only when $\xi$ vanishes.
Moreover, $p_2$ will always lie in the right half-plane, allowing for a principal cut along the negative real axis. 
Invertibility of $\tilde{\mathcal{O}}$ also holds for a generic choice of $m^2>0$ and cutoff value $\rho_b$. 
Hence, spectral KSW is satisfied for the inner horizon saddle.

While this result depends on the particular choice of contour  $w(\lambda)$ taken in the complex plane,
we emphasize that this contour-dependence is not new --- it exists also in the original KSW criterion.
In particular, the cigar saddle satisfies KSW with $\rho\in(0,\infty)$, but violates KSW if  we deform the real-axis contour to a contour which crosses the line $\textrm{Im}\rho=\pi/2$.
Hence, both KSW and spectral KSW should be understood as criteria applied to a preferred representative of the complexified geometry.

Before concluding this example, we recall that the simplest manifestation of KSW violation on the inner horizon saddle is the presence of infinitely many negative modes localized to the boundary of the spacetime (Section \ref{sec:minus-meaning}), where the induced volume element is negative.
While the Schwarzian modes do not obey free scalar field dynamics, it is still illustrative to revisit this boundary theory from the perspective of spectral KSW. 
The quadratic fluctuation operator  for the Schwarzian modes
\begin{equation}
    \tilde{\mathcal{O}}
    =
    \frac{2\pi \phi_b}{\beta}(\partial_u^4 + \partial_u^2)
    \label{eq:for-p4}
\end{equation}
with $\beta<0$ is elliptic, invertible after removing zero modes, and admits a principal cut (the principal symbol $p_4(\xi) = \frac{2\pi \phi_b}{\beta}\xi^4$ is strictly negative for nonzero $\xi$).
Hence, the Schwarzian theory on the boundary of the inner horizon saddle has a zeta-regularizable one-loop determinant, 
consistent with our explicit calculations in Section \ref{sec:minus-meaning}.

\paragraph{Complex torus with a chiral fermion:}
Above, we argued that the existence of a principal cut is important for defining a zeta-regularized determinant continuously across a family of backgrounds.
We now illustrate what can go wrong  when no principal cut exists, using a single chiral fermion (see also supplementary examples in Appendix~\ref{app:agmon-details}).
We emphasize that this example is intended only to illustrate the importance of a principal cut --- our spectral KSW criterion itself is formulated in terms of bosonic matter and hence is agnostic to the behavior of the fermion partition function studied here.

Consider again the torus \eqref{eq:complex-identification}.
For simplicity, we take $\Omega$ to be real.
We rewrite the identifications in terms of
$\tau_1 =\beta\Omega/2\pi$ and $\tau_2=\beta/2\pi$
via
\begin{equation}
    \phi\sim \phi+2\pi 
    \qquad 
    (\tau_E,\phi) \sim (\tau_E + 2\pi \tau_2, \phi+2\pi \tau_1)\,,
\end{equation}
so that, with
\begin{equation}
    \begin{cases}
        z\equiv \frac{1}{2\pi}(\phi+i \tau_E)
        \\
        \bar{z}\equiv \frac{1}{2\pi}(\phi-i \tau_E)
    \end{cases}
    \qquad
     \begin{cases}
        \tau\equiv \tau_1 + i \tau_2
        \\
        \bar{\tau}\equiv \tau_1- i \tau_2\,,
    \end{cases}
\end{equation}
we have simply
\begin{equation}
    z\sim z+1  \qquad z\sim z+\tau\,,
\end{equation}
and similarly with bars, which describes the flat complex torus
$\mathbb{C}/(\mathbb{Z}+\tau \mathbb{Z})$,
with metric
\begin{equation}
ds^{2}\propto dz\,d\bar z \,.
\end{equation}
We consider a complex chiral fermion with the action
\begin{equation}
S_{\psi}
=
\frac{1}{2\pi}
\int d^2 z\,
\bar\psi\,\partial_{\bar z}\psi .
\end{equation}
The quadratic fluctuation operator is simply $\partial_{\bar z}$,
and its principal symbol reads
\begin{equation}
    p_1(\xi) = i\pi(\xi_\phi + i\xi_{\tau_E})\,.
\end{equation}
As $\xi$ varies, the principal symbol covers all of $\mathbb{C}$, and there is no principal cut. 

We will show that the absence of a principal cut detects a subtle pathology in the zeta-regularized determinant $\det \partial_{\bar{z}}$:
while it is well-defined for a fixed choice of $\beta$ and $\Omega$, it does not provide a well-defined functional on the space of backgrounds when the spectral cut is held fixed.\footnote{Note that a similar issue was analyzed in~\cite{Alvarez-Gaume:1983ihn,Witten:1985xe,Losev:1989fe}, from a different perspective.
In particular, the chiral fermion exhibits a gravitational anomaly, and it would be interesting to connect this more precisely to the present calculation.}

For concreteness, we choose periodic boundary conditions around both cycles; then, the eigenfunctions of $\partial_{\bar{z}}$ are 
\begin{equation}
f_{k,n}(z,\bar z)=e^{a z+b \bar z},
\end{equation}
with
\begin{equation}
a+b=2\pi i k,
\qquad
a\tau+b\bar\tau=2\pi i n,
\qquad
k,n\in \mathbb{Z}.
\end{equation}
Thus, identifying $\lambda_{k,n}$ with $b$ and
solving for $b$, we have
\begin{equation}
\partial_{\bar z} f_{k,n}=\lambda_{k,n} f_{k,n}, \quad \textrm{where} \quad \lambda_{k,n}
=
\frac{2\pi i(k\tau-n)}{\tau-\bar{\tau}}\,,
\end{equation}
or, in terms of $\beta$ and $\Omega$,
\begin{equation}
    \lambda_{k,n}=- \pi \left[ \left(\frac{2\pi n}{\beta}-k \Omega\right) - ik\right]\,,
\end{equation}
which can be thought of as the ``square root'' of Eq.~\eqref{eq:scalar-spectrum-torus}.
This spectrum is a rank-two lattice in the complex plane, and 
its asymptotic accumulation directions fill all directions in the complex plane,
consistently with the absence of a principal cut for $\partial_{\bar{z}}$.

With $D\equiv \partial_{\bar{z}}$, the zeta-regulated determinant is defined by
\begin{equation}
    \det  D = e^{-\zeta_{D}'(0)}\,
    \qquad
    \zeta_{D}(s) = \sum_{k,n} \lambda_{k,n}^{-s}\,,
    \qquad 
        \lambda^{-s}=|\lambda|^{-s}\,e^{-is\arg_\theta\lambda}\,,
\end{equation}
where we shall choose the traditional spectral cut $\theta=-\pi$, and we remove 
 zero modes later where necessary.
We will also employ a prescription in which we  zeta-regularize each sum over $n$ before handling the sum over $k$.
We begin with $k=0$.
In this case,  all eigenvalues $\lambda_{k=0,n}$ with $n>0$ lie on the negative real axis, so we additionally assign the upper lip prescription $(-x)^{s} = e^{i\pi s}x^{s}$ for positive $x$.
Performing the sum over $n\ne 0$ gives
\begin{equation}
    \zeta_{D^{(k=0)}}(s) = 
    \left(\frac{2\pi^2}{\beta}\right)^{-s}(1+e^{-i\pi s})\zeta_R(s)\,,
    \qquad
    \det\,\!' D^{(k=0)} = -\frac{i\beta}{\pi}\,,
\end{equation}
where the prime indicates our removal of the $k=n=0$ zero mode.
For $k> 0$, 
all the eigenvalues live in the upper half-plane, and the sum over $n$ gives
\begin{equation}
    \zeta_{D^{(k)}}(s) = 
    \left(\frac{2\pi^2}{\beta}\right)^{-s}
    \left[\zeta_H(s, A k) + e^{-i\pi s}\zeta_H(s,1-A k)\right]\,,
    \label{eq:epsilon-0-answer}
\end{equation}
where we define $A=\frac{\beta }{2\pi}(\Omega+i)$.
For $k<0$, all the eigenvalues instead live in the lower half-plane, and the only change is that the $e^{-i\pi s}$ factor above becomes $e^{+i\pi s}$.
After differentiating, we obtain 
\begin{equation}
\det D^{(k)} = 
    \begin{cases}
        1-e^{2\pi i k A}\,, \qquad &k>0
        \\
        1-e^{-2\pi i k A}\,, \qquad &k<0 \,.
    \end{cases}
    \label{fermiondet0}
\end{equation}
Hence, with our iterated zeta-regularization prescription, the full determinant  reads
\begin{equation}
    \det\,\!' D = 
    -\frac{i\beta}{\pi}
    \prod_{k=1}^{\infty} 
   ( 1-e^{2\pi i k A})^2
   =
   -\frac{i\beta}{\pi}
   e^{-\frac{i\pi}{6}A}\eta(A)^2\,.
   \label{fermiondet1}
\end{equation}
We see that $\det'\partial_{\bar{z}}$ appears well-defined for any fixed choice of $\beta>0$ and $\Omega\in \mathbb{R}$.
However, as we now show, it does not define a continuous functional on the space of backgrounds when the $\arg$ cut is held fixed.

In particular, when we give $\beta$ a small imaginary shift, 
\begin{equation}
    \beta\to \beta + i\epsilon\,,
\end{equation}
and take the limit $\epsilon\to 0^+$, we do not recover Eq.~\eqref{fermiondet1}.
To see this, note that the $i\epsilon$ prescription leaves the $k=0$ calculation unchanged; it pushes the eigenvalues on the negative axis upwards, 
where the relation $(-x)^{s} = e^{i\pi s}x^{s}$ that we prescribed before now 
holds automatically.
Similarly, for $k>0$,
any real-axis crossing induced by the $i\epsilon $ prescription occurs at $\textrm{Re}\lambda_{k,n}>0$ for sufficiently small $\epsilon$, and so  the  $i\epsilon$ prescription does not push any $k>0$ eigenvalues across the negative-real-axis branch cut.
On the other hand, for $k<0$, we have  
\begin{equation}
    \textrm{Im}\lambda_{-|k|,n}=- \pi|k| +\frac{2\pi^2 n\epsilon}{\beta^2+\epsilon^2} \,.
\end{equation}
As $n$ increases, these eigenvalues cross the real axis at
\begin{equation}
    n_*\approx \frac{|k|(\beta^2+\epsilon^2)}{2\pi \epsilon}\,,
\end{equation}
where  $\textrm{Re}\lambda_{-|k|,n}$ is negative (again, assuming a sufficiently small $\epsilon$).
Hence, for every $\epsilon>0$, we have infinitely many modes with $n>n_*$ lying above the negative real axis; 
by contrast, with  $\epsilon= 0$, all of the modes lie below the real axis.
A careful calculation with $\epsilon\to 0^+$ gives
\begin{equation}
    \det D^{(k)} = 
    1-e^{2\pi i k A}\,, \qquad k< 0\,,
\end{equation}
which clearly differs from the $\epsilon=0$ answer in Eq.~\eqref{fermiondet0}. 
Indeed, the full determinant with $\epsilon=0^+$ (and after an additional zeta regularization step) gives
\begin{equation}
    \det\,\!'D =-\frac{i\beta}{\pi}\prod_{k=1}^{\infty}(1-e^{2\pi ikA})(1-e^{-2\pi ikA})
    =
    -\frac{\beta}{\pi}\eta(A)^2\,,
\end{equation}
which has a different functional form altogether compared with Eq.~\eqref{fermiondet1}.

We conclude that the zeta-regularized quantity $\det' D$ fails to define a continuous functional  on the space of backgrounds when the spectral cut is held fixed.
Due to the absence of a principal cut, the accumulation regions in the spectrum are permitted to occupy all directions in the complex plane.
A tiny change in the background geometry then causes an infinite number of eigenvalues to cross the $\arg$ branch cut, propagating changes to $\det' D$ and rendering it discontinuous on the space of backgrounds.

\section{Ruling out extra saddles}
\label{sec:phantom}

In this last section, we briefly return to pure JT gravity, 
to comment on a possible gravity interpretation of certain saddle points in $\beta$ excluded by the integration contour for $\rho(E)$.

To discuss these saddle points, we must first develop a bulk geometric way to think about complex values of $\beta$.
In Section \ref{sec:quantum_corrections},  we computed $\rho(E)$ using an inverse-Laplace transform of $Z(\beta)$.
This inverse Laplace integral had two saddle points in the complex-$\beta$ plane, at $\beta=\pm \beta_0$, corresponding to the fact that there are only two choices of boundary length  such that the corresponding on-shell solutions have the desired value $E$ of the ADM energy.
More generally, other values of $\beta$ along the integration contour, which are generally complex, describe renormalized boundary lengths for the following family of geometries:
\begin{equation}
    ds^2_\beta = \left[r^2-r_c(\beta)^2\right] d\tau^2 + \frac{dr^2}{r^2-r_c(\beta)^2}\,,
    \qquad 
    r_c(\beta) = \frac{2\pi}{\beta}\,,
    \qquad
    \tau\sim \tau+2\pi/r_c\,,
\end{equation}
which smoothly cap off at $r_c(\beta)$ and run up to the IR cutoff value $r_b=1/\epsilon$, as required by the fixed-dilaton boundary condition.
These geometries are generically off-shell of the fixed-energy boundary condition; the energy reads $2\pi^2\phi_b/\beta^2$ and equals $E$ only for the particular choices $\beta = \pm \beta_0$.
Their action reads $-S_0-2\pi^2 \phi_b/\beta$, consistently with the integrand of the inverse-Laplace contour.

What happens as we vary $\beta$?
Let us start from $\beta=\beta_0$, hence $r_c=r_+=2\pi/\beta_0$,
and move to $\beta=e^{i\pi}\beta_0$, hence
$r_c=r_-=e^{-i\pi}r_+$.
We can visualize this as continuously dragging the cap position of the contour clockwise from $r_+$ to $r_-$ in the complex plane,
providing a family of geometries which interpolate  between the cigar and inner horizon saddles.
Moreover, we can formally associate the result of a full-circle rotation $\beta=e^{2\pi i}\beta_0$ with a contour starting at $r_+$ and winding clockwise once around $r_-$ before running towards the cutoff value $r_b$.

\begin{figure}[t]
\centering
\begin{minipage}[b]{0.48\textwidth}
\centering
\begin{tikzpicture}[scale=0.85]
  \draw[decorate, decoration={zigzag, segment length=4pt, amplitude=1.5pt}, thick, gray]
    (-1,0) -- (1,0);

  \draw[->] (-2.5,0) -- (5,0) node[right] {$\text{Re}\,x$};
  \draw[->] (0,-1.5) -- (0,2) node[above] {$\text{Im}\,x$};

  \draw[thick, orange,
    postaction={decorate, decoration={markings,
      mark=at position 0.07 with {\arrow{stealth}},
      mark=at position 0.21 with {\arrow{stealth}},
      mark=at position 0.34 with {\arrow{stealth}},
      mark=at position 0.47 with {\arrow{stealth}},
      mark=at position 0.58 with {\arrow{stealth}},
      mark=at position 0.70 with {\arrow{stealth}},
      mark=at position 0.88 with {\arrow{stealth}}}}]
    (-1,0) -- (0.7,0.08)
    arc (165.07:-165.07:0.3105)
    -- (-0.7,-0.08)
    arc (338.20:21.80:0.3231)
    -- (0.5,0.16)
    arc (180:0:0.5)
    -- (4.5,0.16);

  \fill (-1,0) circle (2.5pt);
  \node[above left, font=\small] at (-1,0.05) {$x=-1$};
  \fill (1,0) circle (2.5pt);
  \node[below, font=\small] at (1,-0.15) {$x=1$};
\end{tikzpicture}

\vspace{2pt}
(a) Contour in the $x$-plane
\end{minipage}
\hfill
\begin{minipage}[b]{0.48\textwidth}
\centering
\includegraphics[width=\linewidth]{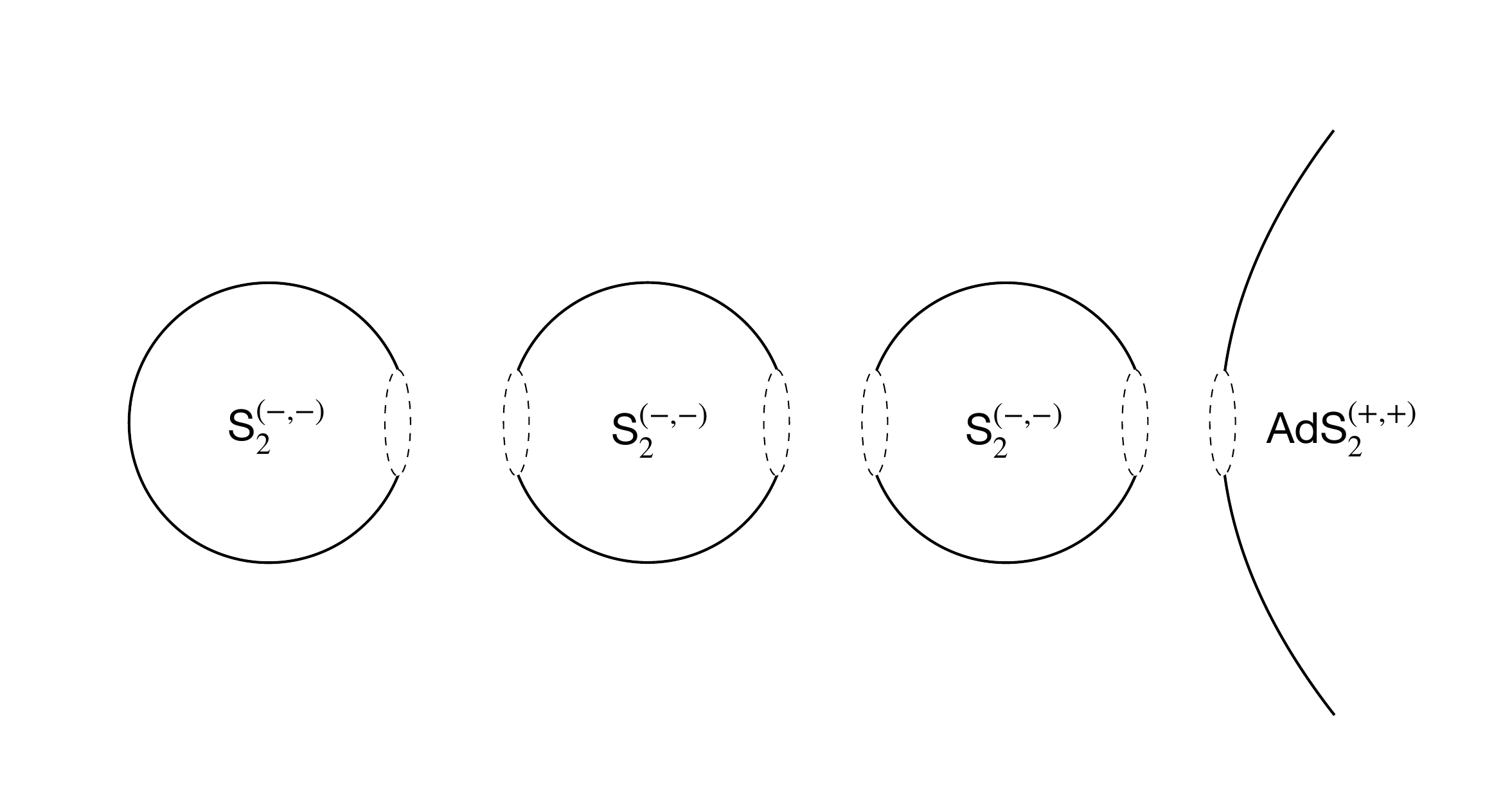}

\vspace{2pt}
(b) Geometric picture
\end{minipage}
\caption{\textbf{(a)}~A winding contour in the complex $x=r/r_+$ plane, with the branch cut of $\sqrt{h}=\sqrt{x^{2}-1}$ shown as a squiggly line. Starting at the cap $x=-1$, the contour runs to $x=1$, turning once around it, then runs to $x=-1$, turning once around it, and finally escapes to $x\to\infty$ through the upper half-plane. \textbf{(b)}~A winding geometry which caps off at the inner horizon, with three antisphere regions; the dotted circles correspond to the intermediate complex region.}
\label{fig:innerhorizonbubble}
\end{figure}

Clearly, one can imagine solutions with generic winding numbers, so that this construction produces candidate winding geometries labeled by
\begin{equation}
    \beta=\beta_0 e^{2\pi i n}\,,
    \label{eq:even-winders}
\end{equation}
which correspond to solutions which cap off at the outer horizon, and
whose $r$-plane contour winds $n$ times around the branch cut
$(r_-,r_+)$,
giving rise to $2|n|$ anti-spheres.
Likewise, every choice of
\begin{equation}
    \beta=\beta_0 e^{i \pi (2n+1)}\,,
    \label{eq:odd-winders}
\end{equation}
corresponds to a solution which caps off at the inner horizon and
whose $r$-plane contour winds $n+\tfrac12$ times around the branch cut.
These
solutions give rise to $|2n+1|$ anti-spheres.  See Fig.~\ref{fig:innerhorizonbubble} for an example, together with a visualization of the winding geometry in terms of spheres with minus-minus signature, glued together by regions where the metric is complex.

In this system, the two natural $i\epsilon$ bookkeeping choices $\beta=\beta_0 e^{\pm i\pi}$ pair with the two canonical choices for the inner-horizon contour, differing by whether the contour runs above or below $r_+$ in the $r$-plane. These $i\epsilon$ prescriptions played a key role in the steepest descent and contour-rotation analysis of Section \ref{sec:quantum_corrections}.
In particular, the value of the partition function $Z(e^{i\pi}\beta_0)$ differs from $Z(e^{-i\pi}\beta_0)$, and we now see that we can associate these values with the distinct choices of inner horizon contours just described.
This suggests that the bulk geometry accounting for the inner-horizon contribution to $\rho(E)$ differs depending on the $i\epsilon$ prescription at play, even though, as shown in Section~\ref{sec:two-saddles}, the physical density of states is the same in either case.

More interestingly, the above discussion suggests a mechanism by which higher winding solutions are excluded from contributing to $\rho(E)$.
At the level of the inverse Laplace integral, we postulate that the winding saddle geometries \eqref{eq:even-winders}-\eqref{eq:odd-winders} correspond to the fact that the one-loop term $-(3/2)\log\beta$ is infinite-sheeted, and hence there are infinitely many images of the original $\beta_+$ and $\beta_-$ saddle points on higher sheets of the Riemann surface of $\log\beta$. 
These ``higher winding'' saddle points do not contribute to the density of states computed with an inverse Laplace integral through the Picard-Lefschetz method.\footnote{Concretely,
to remove the branch structure in the inverse Laplace integral,
one can use a uniformization map such as $\beta = z^{-2}$.
We have verified that the saddle intersection numbers are nonvanishing only for the inner and outer horizon saddles, with an appropriate $i\epsilon$ prescription understood.}

Moreover, a simple calculation suggests that the winding saddles violate spectral KSW. 
Consider again  Eq.~\eqref{realp2inner}:
\begin{equation}
   \textrm{Re} p_2 = \frac{(\log|w|)'}{\lambda}\,\left(\frac{1}{|a|^2}\xi_\lambda^2 +\xi_\theta^2\right)\,,
   \qquad 
   a=\frac{w'}{w}\,.
\end{equation}
On the inner horizon contour, we can take $|w|$ to be strictly increasing, which guarantees $\textrm{Re}p_2>0$.
On the other hand, we now consider a winding saddle which starts at the $w=0$ cap; 
note that a similar argument will work for winding contours which start at the outer horizon cap.
In this case, $|w|$ initially increases from zero, giving $(\log|w|)'>0$ there.
Eventually, however, the contour is forced  to go from a point with $|w|>1$ (where the contour crosses $\textrm{Im}\rho=0$) to a point with $|w|<1$ (where the contour crosses $\textrm{Im}\rho=\pi$), so there must exist a point where $|w|$ is decreasing, giving $(\log|w|)'<0$ there.
Since $(\log|w|)'$ must cross zero, the operator is no longer elliptic; moreover, 
similar arguments reveal that $p_2$ attains every possible phase in the complex plane.
Taken together, these facts suggest that the winding saddles violate spectral KSW.

Hence, disk-level JT gravity appears to provide a simple example in which infinitely many candidate saddle
points are excluded from the gravitational path integral by global features of the integration contour, 
and moreover these exclusions are consistent with our 
spectral KSW criterion. 

\section{Discussion}
\label{sec:discussion}

We have shown that the inner horizon saddle in two-dimensional dilaton gravity has a clean origin in the microcanonical density of states: 
it is one of two saddle points of the gravitational path integral with fixed-energy boundary conditions,  corresponding to a geometry with negative boundary length. 
Its crucial minus sign is fixed by the one-loop Schwarzian determinant and its instabilities.
Meanwhile, its violation of the bulk KSW criterion appears harmless: the scalar sKSW condition (Section~\ref{sec:non-topological}) is satisfied, suggesting that a mode-by-mode field space rotation renders the relevant matter path integrals well-defined at one loop. 
We now mention three key highlights of our results.

First, our classical stability analysis (Section~\ref{sec:stability}) has practical significance: real-world actions contain irrelevant operators that cannot all be tuned away, and our perturbative scalar analysis indicates that the inner horizon saddle survives them.
The argument relies only on regularity at the location where the geometry caps off, so it should extend beyond scalar matter. 

Secondly, we emphasize that following our discussion in this paper, the density of states depends directly on geometry behind the horizon, via the inner-horizon contribution to Eq.~\eqref{eq:rho-two-areas}.
Other boundary probes of the interior exist---two-sided observables in the thermofield double~\cite{Maldacena:2001kr}, analytically continued Green's functions and quasinormal spectra~\cite{Festuccia:2005pi,AliAhmad:2026wem}, infalling-observer reconstructions~\cite{Jafferis:2020ora,Gao:2021tzr}, the large-$N$ operator-algebraic framework~\cite{Leutheusser:2021frk}, the island formula~\cite{Almheiri:2020cfm}, and $T\bar T$ deformations~\cite{AliAhmad:2025kki}---but the microcanonical density of states is distinctive in its simplicity: it is a single-boundary, time-independent, equilibrium quantity that nevertheless probes behind the outer horizon.

Thirdly, our analysis illustrates a broader lesson about saddle inclusion and exclusion in the gravitational path integral.
On the inclusion side, new contributing saddles can be found by allowing the boundary length to become negative or complex --- or put differently, saddles of the inverse Laplace transform simply need not lie at positive $\beta$.
On the exclusion side, it has long been argued that the gravitational path integral admits more candidate saddles than actually contribute~\cite{Halliwell:1988ik,Halliwell:1989vu,Halliwell:1989dy};
and indeed, in the JT context, an infinite tower of homotopically inequivalent winding geometries exists, yet Picard--Lefschetz selects only two (Section~\ref{sec:phantom}),
consistently with our spectral KSW criterion. 

We now turn to subjects for future work.

\paragraph{Higher-dimensional generalizations:}
Since JT gravity arises in the near-extremal limit of higher-dimensional black holes ~\cite{Maldacena:1998uz,Almheiri:2014cka,Ghosh:2019rcj}, it is natural to think that the inner horizon saddle studied here can be meaningfully lifted to higher-dimensional theories of gravity.\footnote{Indeed, see the discussion section of Ref.~\cite{Bagchi:2012xr} for a promising appearance of inner horizon saddles in boundary duals to 3$d$ gravity.}
Indeed, under fairly general assumptions, the on-shell action of a higher-dimensional cigar (up to a $\beta E$ term) gives the outer horizon area over $4G$~\cite{Gibbons:1976ue,Brown:1992bq},
and we expect that similar arguments guarantee that the on-shell action of a higher-dimensional inner horizon geometry will give the inner horizon area over $4G$.
However, subtleties of interpretation arise due to the structure of the quantum-corrected partition function
away from extremality.
For instance, in three bulk spacetime dimensions, the canonical partition function $Z(\beta)$ typically develops a boundary in the complex $\beta$-plane, given by a dense accumulation of singularities~\cite{Maloney:2007ud}, obstructing the analytic continuation to $\beta<0$ that in JT gravity leads to the inner horizon saddle. Nevertheless, this problem is reminiscent of the one encountered in Section~\ref{sec:gauge}, where the partition function of JT gravity coupled to a $U(1)$ gauge field also develops a wall of singularities. We speculate that the resolution may be similar; that is, exchanging the order of integration so that the $\beta$ integral takes place earlier on may render the final answer convergent.
Note that even if the inner horizon saddle sensibly contributes, another subtlety in the higher-dimensional setting is that there will be far more contributions to $\rho(E)$ than just the cigar and inner horizon saddles (see, e.g.~\cite{Dijkgraaf:2000fq,Manschot:2007ha}).
A natural non-perturbative test would be the BFSS matrix model~\cite{Banks:1996vh}, but accessing an inner-horizon branch requires nonzero $\mathrm{SO}(9)$ chemical potential, which destabilizes the matrix moduli space.\footnote{
A nonzero angular potential produces a negative centrifugal term
along commuting flat directions of the BFSS moduli space,
schematically
\begin{equation}
V_{\rm cent}\propto-\Omega^2\,\mathrm{Tr}(X^iX^i),
\end{equation}
so the rotating ensemble is unstable for nonzero $\Omega$.
The BMN matrix model~\cite{Berenstein:2002jq} adds harmonic mass terms which can stabilize
the relevant rotating directions for sufficiently small $|\Omega|$,
although it is unclear whether its bulk dual admits an inner-horizon
branch in this regime.
}

\paragraph{sKSW for gauge fields:} 
In Appendix~\ref{sec:saddle-comparison}, we found that the gauge field on the inner horizon saddle is generically complex.  This is a subtle feature. In an ordinary gauge theory, the gauge field is a connection for a compact gauge group, and complexifying its integration contour obscures this interpretation \cite{Witten:2010cx}. Nevertheless, contour deformations involving gauge-field directions are not unprecedented. A familiar example is the semiclassical expansion around a sphaleron. Sphaleron backgrounds~\cite{Klinkhamer:1984di}, which play an important role in electroweak baryogenesis and in related baryon- and lepton-number violating processes, possess a negative mode. Treating this unstable fluctuation requires deforming the integration contour for the corresponding gauge-field fluctuation into the complexified configuration space~\cite{Arnold:1987mh}. Thus, the need to complexify gauge-field directions is (seemingly) a familiar feature of semiclassical expansions around saddles with gauge-field negative modes. A slight difference to our case is that the background gauge connection, not the fluctuations, seems to require analytic continuation.

\paragraph{sKSW at higher loops:}
The sKSW criterion of Section~\ref{sec:non-topological} is tailored to one-loop physics; it is a condition on the quadratic fluctuation operator.
It would be interesting to explore whether this condition implies convergence at higher loops as well.
The mode-by-mode field contour rotation that cures the one-loop computation might propagate to higher-loop computations and cure them too; if it does, sKSW would extend naturally from a one-loop diagnostic into a condition for all-order convergence.
Whether this actually happens, we do not know.

\paragraph{sKSW for Kerr-AdS:}
An important task for the future is to analyze the spectral KSW criterion on the various backgrounds mentioned in the introduction section.
In particular, it would be interesting to analyze the quasi-Euclidean BTZ and Kerr--AdS$_4$ geometries. 
Here, one could check whether sKSW fails at $|\Omega_H|=1$, as suggested by the toy model of Section~\ref{sec:non-topological}.
This is where bulk KSW for the relevant saddles is known to be saturated~\cite{Witten:2021nzp} (see also \cite{BenettiGenolini:2025jwe,BenettiGenolini:2026raa,Krishna:2026rma}). 
Agreement between these thresholds would clarify whether the sKSW criterion shares the original KSW criterion's sensitivity to the onset of physical, superradiant instabilities.

\section*{Acknowledgments}

We thank Ahmed Almheiri, Andreas Blommaert, Luca Iliesiu, Jonah Kudler-Flam, Henry Lin, Rob Myers, Juan Maldacena, and Sabrina Pasterski for valuable discussions. We are particularly grateful to Victor Ivo for extensive discussions concerning negative modes and their history in the gravitational path integral. 
We thank Adam Levine and Douglas Stanford for valuable discussions and sharing unpublished notes.
We also thank Federico Capeccia for detailed comments on the draft.
We used Claude (Anthropic) and Gemini Deep Think (Google DeepMind) as research aides for identifying the spectral Kontsevich-Segal-Witten criterion.

Research at Perimeter Institute is supported in part by the Government of Canada through the Department of Innovation, Science and Economic Development Canada and by the Province of Ontario through the Ministry of Colleges and Universities. JC acknowledges the support of the Natural Sciences and Engineering Research Council of Canada (NSERC) through a Vanier Canada Graduate Scholarship [funding reference number CGV--192707]. AH is grateful to the Simons Foundation as well as the Edward and Kiyomi Baird Founders' Circle Member Recognition for their support.

\appendix

\section{Two-point function cross-check}
\label{app:geodesic}

In this appendix, we demonstrate that the microcanonical two-point function at energy $E$ in JT gravity may be derived, in the semiclassical approximation, by a sum over both the cigar saddle (with $\beta=\pi/\sqrt{E}$) and the inner horizon geometry (with $\beta=-\pi/\sqrt{E}$), with the appropriate inclusion of matter propagators.

Let us briefly review the standard probe approximation in JT gravity.
In the semiclassical Schwarzian regime, for a light probe operator and $0<\tau<\beta$, the disk two-point function is~\cite{Ghosh:2019rcj,Turiaci:2024cad}
\be\label{eq:2pt-disk}
\langle\mathcal{O}_{\tau}\,\mathcal{O}_0\rangle_{\beta}
\equiv
\frac{\text{tr}(e^{-\beta H}\,\mathcal{O}_\tau\,\mathcal{O}_0)}{\text{tr}\,e^{-\beta H}}
= 
e^{-\Delta\ell_{\text{disk}}}
=\left(\frac{\beta}{\pi}\sin\frac{\pi\tau}{\beta}\right)^{-2\Delta}\,,
\ee
where $\ell_{\text{disk}}$ denotes the length of a bulk geodesic stretching between the operator insertions on the boundary.
This geodesic length can be derived as follows.
On the hyperbolic disk \eqref{eq:JT-metric}, geodesics have a conserved angular momentum $L = \sinh^2\!\rho\;\theta'$, where prime denotes differentiation with respect to an affine parameter.
The geodesic equation then reduces to
\be\label{eq:geodesic-eom}
\rho'^2 = 1 - \frac{L^2}{\sinh^2\!\rho}\,,
\ee
with turning point at $L = \sinh\rho_0$.
From these expressions for $\rho'$ and $\theta'$, one can obtain $\rho$ as a function of $\theta$:
\begin{equation}
\frac{\tanh(\rho(\theta))}{\tanh(\rho_0)}=
\frac{1}{\cos\theta}\,,
\end{equation}
which, by setting $\tanh(\rho)=1$, also tells us the opening angle as a function of $\rho_0$:
\begin{equation}
    \Delta \theta = 2\arccos(\tanh\rho_0) \,.
    \label{eq:geod-turning-point}
\end{equation}
Some computations then give  the regularized geodesic length as
\be\label{eq:geodesic-length}
\ell = 2\log\!\left(\frac{\beta}{\pi\epsilon}\sin\frac{\Delta\theta}{2}\right)
= 2\log\!\left(\frac{\beta}{\pi\epsilon}\sech\rho_0\right),
\ee
at cutoff $\rho_b = \log(\beta/(\pi\epsilon))$.
Setting $\Delta\theta/(2\pi) = \tau/\beta$, the probe two-point function $e^{-\Delta\ell}$ reproduces \eqref{eq:2pt-disk} up to cutoff-dependent normalization.

Before turning to the inner horizon case, let us first comment on the analytic structure of this result in the separation $\tau$. 
In particular, the disk correlator~\eqref{eq:2pt-disk} can be thought of as a Euclidean time-ordered two-point function.
For $0<\tau<\beta$, the correlator is purely real.
Meanwhile, under the reversal $\tau<0$, the analytically continued correlator becomes complex, and an $i\epsilon$ prescription is required to specify its phase.
If $\tau$ is taken to be negative along
$\text{Im}\tau =\pm \epsilon$, with $\epsilon$ an infinitesimal positive quantity --- equivalently $\tau=e^{\pm i \pi}|\tau|$
--- then the phase reads $e^{\mp 2\pi i \Delta}$. 

In fact, there is a bulk interpretation for this phase.
In AdS/CFT, it is known that analytic continuations of two-point functions in the separation 
can be computed by studying geodesics in complexified bulk geometries \cite{Kraus:2002iv,Fidkowski:2003nf,Festuccia:2005pi,AliAhmad:2026wem}.
In the present case, the required bulk complexification associated with sending $\tau \to e^{\pm i\pi}\tau$ on the boundary, or equivalently $\Delta\theta \to e^{\pm i\pi}\Delta\theta$, can be derived using a simple trick.
Notice that continuously sending $\Delta\theta \to e^{\pm i\pi}\Delta\theta$ will send $\rho_0 \to \rho_0 \mp i\pi$ in Eq.~\eqref{eq:geod-turning-point}.
Moreover, if $\rho_0$ solves the turning-point equation $L^2 = \sinh^2\rho_0$, then so does $\rho_0 \mp i\pi$.
This suggests that the geodesic associated with a negative $\tau$ separation on the boundary is a complexified geodesic with turning point at $\rho_0 \mp i\pi$. 

Indeed, the length associated with this geodesic recovers the correct two-point function.
From Eq.~\eqref{eq:geodesic-length},
taking $\rho_0 \to \rho_0 \mp i\pi$ shifts the geodesic length by $\pm 2\pi i$:\footnote{Explicitly, sending $\rho_0$ to $\rho_0 \mp i\pi$ sends $\text{sech}\,\rho_0 \to e^{\pm i\pi}\text{sech}\,\rho_0$, contributing $2\log(e^{\pm i\pi}) = \pm 2\pi i$ to the length.}
\be\label{eq:length-shift}
\ell = \ell_{\text{disk}} \pm 2\pi i\,.
\ee
Hence, the negative-$\tau$ two-point function is
\be\label{eq:2pt-disk-neg-tau}
\langle\mathcal{O}_{-\tau}\,\mathcal{O}_0\rangle_{\beta}
=
e^{-\Delta(\ell_{\text{disk}} \pm 2\pi i)} = e^{\mp 2\pi i\Delta}\,e^{-\Delta\ell_{\text{disk}}} =  \left(\frac{\beta}{\pi}\sin\frac{\pi\tau e^{\pm i\pi}}{\beta}\right)^{-2\Delta}\,,
\ee
which,  as desired, reproduces the analytic continuation of the disk correlator to negative $\tau$.

So far, we have focused on the physics of the cigar geometry.
Let us now turn to the inner horizon geometry.
In the main text, we saw that the inner horizon geometry is a classical (albeit complex) solution to JT gravity with negative-length boundary conditions, and hence one expects that correlators computed from bulk geodesics in the inner horizon geometry 
are characterized by $\beta<0$ and $\tau<0$.\footnote{
To see the connection between negative $\tau$ and negative signed
boundary length, note that under $\tau\to e^{i\theta}\tau$ the
boundary metric transforms as
$d\tau^2\to e^{2i\theta}d\tau^2$.
Transporting a branch of its square root continuously gives
$\sqrt{h}\,d\tau\to e^{i\theta}\sqrt{h}\,d\tau$, which becomes a sign flip for $\theta=\pm \pi$.}
By adapting the computations above, we find that the geodesic has a turning point at $\text{Im}(\rho_0)=\pi$, and this yields
\be\label{eq:2pt-inner-geodesic}
\langle\mathcal{O}_{-\tau}\,\mathcal{O}_0\rangle_{-\beta}
=  \left(\frac{\beta}{\pi}\sin\frac{\pi\tau e^{\pm i\pi}}{\beta}\right)^{-2\Delta}\,,
\ee
consistently with analytically continuing \eqref{eq:2pt-disk} to negative $\tau$ and negative $\beta$ simultaneously.
We emphasize that the geodesic with turning point at $\rho_0+i\pi$ plays two very different roles depending on the context.
From the perspective of the cigar geometry, it lives in an analytic continuation of the original geometry, while from the perspective of the inner horizon geometry, it is possible to take it as a bona fide geodesic in the original geometry. 

Following this logic to its natural conclusion, we have that the geodesic with turning point at $\rho_0\in \mathbb{R}$ \textit{also} plays two very different roles:
From the perspective of the disk solution, it is a bona fide geodesic in the geometry.
Meanwhile, from the perspective of the inner horizon solution, it lives in an analytic continuation of the original geometry, which allows one to understand the continuation of Eq.~\eqref{eq:2pt-inner-geodesic} back to positive $\tau$.

Having understood the gravity interpretations of the thermal two-point function analytically continued to negative $\beta$ and/or $\tau$, we now turn to the microcanonical two-point function in JT gravity.
The microcanonical two-point function can be obtained by inverse Laplace transforming the canonical correlator $\text{tr}(e^{-\beta H}\,\mathcal{O}_\tau\,\mathcal{O}_0)$ at fixed
$\tau$.
For the standard ordering with $\tau>0$, we choose the initial Bromwich contour within the thermal analyticity domain, for example with $\operatorname{Re}\beta>\tau$.
Then, applying Eq.~\eqref{eq:2pt-disk}, the connected microcanonical two-point function at energy $E$ and Euclidean time separation $\tau$ is
\be\label{eq:2pt-micro}
\langle\mathcal{O}_{\tau}\,\mathcal{O}_0\rangle_{E}
= \frac{1}{\rho(E)}\int\frac{d\beta}{2\pi i}\,e^{\beta E}\,Z(\beta)\left(\frac{\beta}{\pi}\sin\frac{\pi\tau}{\beta}\right)^{-2\Delta}.
\ee
In the semiclassical limit, the factor of $(\frac{\beta}{\pi}\sin\frac{\pi\tau}{\beta})^{-2\Delta}$ does not appreciably shift the location of the saddle points, and hence we can take the saddle points to lie at the locations $\beta_\pm$ as given in Eq.~\eqref{eq:beta-plus}: 
\begin{equation}
\beta_+ = -\beta_- = \pi/\sqrt{E}\,.
\end{equation}
While $(\frac{\beta}{\pi}\sin\frac{\pi\tau}{\beta})^{-2\Delta}$ has various singularities in the region $|\beta|\le \tau$ (an essential singularity at $\beta=0$, and singularities at $\beta=\tau/n$ for $n\in \mathbb{Z}\backslash\{0\}$), this does not affect our analysis of the sign of the subleading saddle point in Section \ref{sec:two-saddles} if we take $\tau<\pi/\sqrt{E}$, because the saddles themselves then obey $|\beta_\pm|>\tau$.
In particular, there will be a relative minus sign between the two saddles, but it arises from the classical action and the Schwarzian zero mode measure, not the matter propagators.

The final result in the saddle point approximation reads
\be\label{eq:2pt-micro-saddle}
\rho(E)\langle\mathcal{O}_{\tau}\,\mathcal{O}_0\rangle_{E} \sim e^{2\pi\sqrt{E}}\left(\frac{\beta_+}{\pi}\sin\frac{\pi\tau}{\beta_+}\right)^{-2\Delta}
\left[1+\ldots \right]
- e^{-2\pi\sqrt{E}}\left(\frac{|\beta_-|}{\pi}\sin\frac{\pi\tau}{|\beta_-|}\right)^{-2\Delta}\left[1+\ldots \right]\,,
\ee
consistently with having to sum over contributions from both the positive- and negative-$\beta$ bulk geometries in the fixed-energy ensemble.
Here, the $\ldots$ account for additional non-Gaussian corrections and small shifts of the saddle point locations away from $\beta_\pm$.

At leading semiclassical order, the normalized correlator is independent of the exponentially suppressed inner-horizon contribution and of the omitted subleading saddle corrections, and it reduces to
\be\label{eq:2pt-micro-simple}
\langle\mathcal{O}_{\tau}\,\mathcal{O}_0\rangle_{E} \simeq \left(\frac{1}{\sqrt{E}}\sin(\sqrt{E}\,\tau)\right)^{-2\Delta},
\ee
which is the expected leading equivalence-of-ensembles result.

\section{\texorpdfstring{Review: Maxwell theory in AdS${}_{2}$}{Review: Maxwell theory in AdS2}}
\label{app:holonomy-rep}

In this appendix, we derive the representation sum \eqref{eq:JTYM-partition} for $G=U(1)$ by evaluating the gauge theory path integral on the disk from first principles. We begin with pure $U(1)$ Yang--Mills theory on a disk $D^2$. The action is
\be\la{eq:SYM-disk}
S_{\text{YM}}
=
\frac{1}{2e^2}\int_{D^2}F\wedge \star F\,,
\ee
where $F=dA$. The corresponding gauge theory path integral is
\be\la{eq:gauge-PI}
Z(\beta,\theta)
=
\int \frac{\mathcal{D}A}{\text{Vol}(\mathcal{G})}\,
e^{-S_{\text{YM}}[A]}\,,
\ee
where $\theta$ stands for a fixed-holonomy boundary condition $U=e^{i\theta}$, while $\beta$ stands for the renormalized circumference of the disk (see Section~\ref{sec:ensembles}).

We compute $Z(\beta,\theta)$ in two equivalent ways: as a sum over flux sectors in subsection \ref{sec:flux-sum} and as a sum over representations in subsection \ref{sec:rep-sum}. We then show in subsection \ref{sec:saddle-comparison} that the saddle-point structures of the two representations differ in an interesting way despite their exact equivalence.

\subsection{The partition function from a sum over fluxes}
\label{sec:flux-sum}

On the disk, the gauge field strength is proportional to the ``volume'' (really, area) form:
\begin{equation}
    F=f \text{vol}\qquad F\wedge \star F = f^2 \text{vol}\,,
\end{equation}
where $f$ has the on-shell interpretation of a constant magnetic flux density on the disk.
The total flux can be parameterized by an arbitrary real variable $\tilde{\theta}$:
\begin{equation}
\tilde\theta=\int_{D^2}F ={\cal A}f,
\end{equation}
where we have defined ${\cal A}$ as the disk area --- related to $\beta$ via\footnote{Explicitly, for the cigar, we have $\mathcal{A}_{\text{cigar}}= 2\pi(\cosh\rho_b-1)$ and
$\beta_{\text{cigar}}/\epsilon=2\pi \sinh\rho_b$, so that these quantities match at leading order in the cutoff.
For the inner horizon saddle,
we have $\mathcal{A}_{\text{inner}}= -2\pi(\cosh\rho_b+1)$, 
and $\beta_{\text{inner}}/\epsilon=- 2\pi \sinh\rho_b$, so these quantities again match at leading order in the cutoff.
The overall signs in the inner horizon case come from the synchronized sign flip on $\sqrt{h}$ and $\sqrt{g}$ discussed in Section \ref{sec:jt-setup}.} $\beta/\epsilon=\mathcal{A}$ ---
and used that $f$ is constant on shell of the equations of motion.
By contrast, the boundary holonomy of the gauge field is compact; 
applying Stokes' theorem 
\begin{equation}
    \text{Hol}(\partial D^2)
    =
    e^{i \oint_{\partial D^2}A}
    =
    e^{i \int_{D^2}F}
    =
    e^{i\tilde{\theta}}
    \,,
\end{equation}
the holonomy only detects the total flux modulo integer multiples of
$2\pi$. 
Thus, fixing a value $e^{i\theta}$ for the boundary holonomy does not fix the total flux. 
Instead, the partition function receives contributions from every lift of $\theta$ to the real
line,
\begin{equation}
\tilde\theta_n=\theta+2\pi n,
\qquad
n\in\mathbb Z .
\end{equation}
The disk partition function at fixed holonomy $e^{i\theta}$ is therefore obtained
by first computing the contribution from each flux-lift sector,
and then summing over $n$:
\begin{equation}
\label{eq:Z-as-sum}
Z(\beta,\theta)
=
\sum_{n\in\mathbb Z}
Z_{\text{flux}}(\theta+2\pi n)\,,
\end{equation}
where we suppress the $\beta$ dependence in $Z_{\text{flux}}$ for simplicity.

For the first step, we note that 
for a fixed value $\tilde\theta$ of the flux, the on-shell action reads
\begin{equation}
\label{eq:classical-action-lifted}
    S_{\text{YM}}
=
\frac{1}{2e^2}\int_{D^2}F\wedge \star F
=
\frac{1}{2e^2}f^2 \mathcal{A}
=
\frac{1}{2e^2}
\frac{\tilde{\theta}^2}{\mathcal{A}}\,,
\end{equation}
and, since the quadratic fluctuation integral around a solution is independent of the flux, we obtain 
\begin{equation}
    Z_{\text{flux}}(\tilde{\theta}) \propto e^{-\frac{\tilde{\theta}^2}{2t}}\qquad t\equiv e^2 \mathcal{A}=e^2\beta/\epsilon\,.
    \label{eq:Z-sector}
\end{equation}
To fix the normalization, we interpret the path integral quantum mechanically.
We introduce a Hilbert space of states $|\tilde{\theta}\rangle$ with fixed flux, characterized by
\begin{equation}
\label{eq:theta-state-normalization}
\langle \tilde\theta_f|\tilde\theta_i\rangle
=
\delta(\tilde\theta_f-\tilde\theta_i),
\qquad
\mathbf{1}
=
\int_{\mathbb R}d\tilde\theta\,
|\tilde\theta\rangle\langle\tilde\theta| .
\end{equation}
With $q$ defined as the operator conjugate to $\tilde{\theta}$, free-particle evolution in $\tilde{\theta}$ is generated by
\begin{equation}
    H=q^2/2\,,
\end{equation}
and in particular we find that the heat kernel
\begin{equation}
\label{eq:two-boundary-kernel}
K_t(\tilde\theta_f,\tilde\theta_i)=
    \langle \tilde{\theta}_f|e^{-t H}|\tilde{\theta}_i\rangle=
    \frac{1}{\sqrt{2\pi t}}e^{-(\tilde{\theta}_f-\tilde{\theta}_i)^2/(2t)}
\end{equation}
has precisely the form of Eq.~\eqref{eq:Z-sector}.
This object obeys the following gluing property
\begin{equation}
    K_{t+t'}(\tilde\theta_f,\tilde\theta_i)
    =
    \int_{\mathbb{R}} d\theta K_t(\tilde\theta_f,\theta)
    K_{t'}(\theta,\tilde\theta_i)\,,
\end{equation}
and, following e.g.~\cite{SimmonsDuffinNotes}, we prescribe that the path integral must also obey this property.
This fixes the normalization for us; 
we can identify $K_t(\tilde\theta_f,\tilde\theta_i)$ with the output of the Yang-Mills path integral with appropriate flux boundary conditions:
\begin{equation}
K_t(\tilde\theta_f,\tilde\theta_i)
=
\int_{\tilde{\theta}_i}^{\tilde{\theta}_f}\frac{\mathcal{D}A}{\text{Vol}(\mathcal{G})}\,
e^{-S_{\text{YM}}[A]}\,,
\end{equation}
and in particular,
\begin{equation}
K_t(\tilde{\theta})\equiv K_t(\tilde\theta,0)
=
Z_{\text{flux}}(\tilde{\theta})\,.
\end{equation}
Then, using Eq.~\eqref{eq:Z-as-sum}, we obtain 
\begin{equation}
\label{eq:Z-theta-winding}
Z(\beta,\theta)
=
\frac{1}{\sqrt{2\pi t}}
\sum_{n\in\mathbb Z}
\exp\left[-\frac{(\theta+2\pi n)^2}{2t}\right],
\end{equation}
where $n$ runs over all flux-lift sectors compatible with the holonomy $\theta$.
In the purely thermal case --- that is, setting the boundary holonomy to the identity, $\theta=0$ ---
this gives
\begin{equation}
\label{eq:Z-theta-winding-thermal}
Z(\beta)
=
\frac{1}{\sqrt{2\pi e^2 \beta}}
\sum_{n\in\mathbb Z}
\exp\left[-\frac{(2\pi n/e)^2}{2\beta}\right]\,.
\end{equation}
Note that here and below, we take the Yang--Mills coupling appearing in the bulk action to scale with the cutoff as $e^2(\epsilon)=\epsilon\,e_{\mathrm{eff}}^2$
while holding $e_{\mathrm{eff}}^2$ and the renormalized boundary length $\beta$ fixed as $\epsilon\to0$. Equivalently,
\begin{equation}
t\equiv e^2(\epsilon)\mathcal A \to e_{\mathrm{eff}}^2\beta.
\end{equation}
We henceforth write $e^2\equiv e_{\mathrm{eff}}^2$.

\subsection{Converting the partition function to a sum over representations}
\label{sec:rep-sum}

In the previous subsection, we wrote Eq.~\eqref{eq:Z-theta-winding} as a sum over flux-lift sectors labeled by $n$.
The total flux $\int_{D^2}F=\theta + 2\pi n$ can be called magnetic rather than electric, since the Euclidean disk has no timelike direction.

We now use Poisson resummation to rewrite Eq.~\eqref{eq:Z-theta-winding} in terms of the conjugate variable $m \in \mathbb{Z}$,
the quantized electric flux, where this notion requires an appropriate Wick rotation to Lorentzian signature.\footnote{
In particular, if we Wick rotate the outward direction on the disk in radial quantization,  the operator $q$ conjugate to $\tilde{\theta}$ in the previous subsection is precisely the Lorentzian electric flux.
}
Define $g(x) = \exp(-x^2/2t)$; its Fourier transform is
\be\la{eq:fourier-gaussian}
\hat{g}(m) = \int_{-\infty}^{\infty} dx\, e^{-imx}\,g(x) = \sqrt{2\pi t}\; e^{-t m^2/2}\,.
\ee
The Poisson summation formula,
\be\la{eq:poisson}
\sum_{n} g(\theta + 2\pi n) = \frac{1}{2\pi}\sum_{m} \hat{g}(m)\,e^{im\theta}\,,
\ee
converts \eqref{eq:Z-theta-winding} into an electric-flux sum:
\be\la{eq:Z-theta-character}
Z(\beta,\theta) = \frac{1}{2\pi}\sum_{m \in \mathbb{Z}} e^{im\theta}\, e^{-t\,m^2/2}\,.
\ee
Equivalently, it expresses $Z(\beta,\theta)$ as a sum over $U(1)$ characters $\chi_m(\theta) = e^{im\theta}$.
The integer $m$ labels the irreducible representation of $U(1)$ with charge $m$, and $e^{-t\,m^2/2}$ is the Boltzmann weight for the quadratic Casimir $C_2(m) = m^2$.
The overall factor of $1/(2\pi) = 1/\mathrm{Vol}(U(1))$ is the standard group-volume normalization of the heat kernel, 
with the convention that the group measure is $d\theta$. 

To match with Eq.~\eqref{eq:JTYM-partition} in the main text, we go to the purely thermal case, $U = \mathbf{1}$.
Then, Eq.~\eqref{eq:Z-theta-character} gives
\be\la{eq:Zgauge-rep}
Z(\beta) = \frac{1}{2\pi} \sum_{m \in \mathbb{Z}} e^{-e^2\,m^2 \beta/2}\,.
\ee
Since,
for $U(1)$, every irreducible representation $R_m$ has $\dim R_m = 1$, $C_2(m) = m^2$, and $\chi_m(0) = 1$,
this result matches Eq.~\eqref{eq:JTYM-partition} up to the group-volume factor $1/(2\pi)$ which can be absorbed into a redefinition of the measure.

\subsection{Comparison of saddles in the flux and representation sums}
\label{sec:saddle-comparison}

In Section~\ref{sec:gauge}, we took the inverse Laplace transform
inside the sum over representations and obtained a well-defined
density of states.
While the flux sum \eqref{eq:Z-theta-winding} is equivalent to the
representation sum \eqref{eq:Z-theta-character}, their individual
summands have different saddle-point structures, as we now show.
Throughout, $e>0$ denotes the finite effective coupling defined
above, and we take $E>0$.

The term for the $n^{\text{th}}$ flux sector, including the gravity
contribution and setting $\phi_b=1/2$, reads
\begin{equation}
\label{eq:Z-theta-winding-2}
Z_n(\beta) \propto
\beta^{-3/2}e^{\pi^2/\beta}
\cdot\frac{1}{\sqrt{\beta}}
\exp\left[-\frac{(2\pi n/e)^2}{2\beta}\right],
\end{equation}
where the omitted normalization is common to all $n$.
We simplify this as
\begin{equation}
Z_n(\beta)\propto\beta^{-2}e^{A_n/\beta},
\qquad
A_n\equiv\pi^2\left(1-\frac{2n^2}{e^2}\right).
\end{equation}
Applying the identity \eqref{eq:bessel} with $\nu=1$ gives
\begin{equation}
\mathcal I_n(E)
\equiv
\int_{\mathrm{Br}}\frac{d\beta}{2\pi i}\,
\beta^{-2}e^{\beta E+A_n/\beta}
=
\frac{\sqrt E}{\sqrt{A_n}}\,
I_1\left(2\sqrt{A_n E}\right).
\end{equation}
Here the combination on the right is defined by analytic continuation
in $A_n$; in particular,
\begin{equation}
\label{eq:flux-transform-cases}
\mathcal I_n(E)=
\begin{cases}
\displaystyle
\sqrt{\frac{E}{A_n}}\,
I_1\left(2\sqrt{A_nE}\right),
& A_n>0,\\[2mm]
\displaystyle
\sqrt{\frac{E}{|A_n|}}\,
J_1\left(2\sqrt{|A_n|E}\right),
& A_n<0\,,
\end{cases}
\end{equation}
where the last line follows from $I_1(ix)=iJ_1(x)$.
Meanwhile, $A_n=0$ gives simply $\mathcal I_n(E)=E$.

The density of states is therefore
\begin{equation}
\label{eq:0f1}
\rho(E)\propto\sum_{n\in\mathbb Z}\mathcal I_n(E).
\end{equation}
This sum converges absolutely at each fixed $E>0$, and our choice to perform the inverse Laplace transform
term-by-term can be 
justified 
a posteriori by computing
the Laplace transform of the resulting sum.
Setting $e=1$, equality of the flux-lift and  representation
sums then gives the exact identity
\begin{equation}
\begin{aligned}
&\sum_{\substack{m\in\mathbb Z\\m^2/2<E}}
\sinh\left(2\pi\sqrt{E-\frac{m^2}{2}}\right)
=
\sqrt{2}\,\pi^2
\sum_{n\in\mathbb Z}\mathcal I_n(E).
\end{aligned}
\end{equation}

To understand the saddle structure of the flux terms, consider
their classical exponent
$f_n(\beta)=\beta E+A_n/\beta$.
Its stationary points satisfy
$f_n'(\beta)=E-A_n/\beta^2=0$.
There are therefore three distinct cases.

For $A_n>0$, the two stationary points are real:
\begin{equation}
\beta_{n,\pm}=\pm\sqrt{\frac{A_n}{E}}.
\end{equation}
The exact answer is the modified Bessel function in
Eq.~\eqref{eq:flux-transform-cases}.
Writing $x=2\sqrt{A_nE}$, its large-positive-$x$
expansion is
\begin{equation}
    I_1(xe^{\pm i0}) \sim \frac{e^{x}}{\sqrt{2\pi x}}(1+\dots)
    \mp i
    \frac{e^{-x}}{\sqrt{2\pi x}}(1+\dots)\,,
\end{equation}
where, as indicated, the separation of an exponentially small $e^{-x}$ contribution is subject to a Stokes
prescription. 
If we average the $\mp i\,e^{-x}/\sqrt{2\pi x}$ piece, it vanishes, rendering the status of the would-be inner horizon saddle at $\beta=-\sqrt{A_n/E}$ unclear.\footnote{We note that a related subleading-saddle interpretation of a Bessel-function answer has appeared in the study of $1/4$-BPS Wilson loops in $\mathcal N=4$ SYM \cite{Drukker:2006ga}.  In that case, the additional saddle is an unstable worldsheet solution, and it is matched to the exponentially
suppressed term in the large-argument asymptotic expansion of the Bessel function, whose imaginary prefactor
reflects the tachyonic fluctuation modes of that worldsheet.}

For $A_n=0$, the classical exponent is simply $\beta E$ and
has no finite stationary point. Nevertheless, the inverse
transform is well defined and equals $\mathcal I_n(E)=E$.

For $A_n<0$, the stationary points instead form an imaginary pair:
\begin{equation}
\beta_{n,\pm}
=
\pm i\sqrt{\frac{|A_n|}{E}},
\qquad
f_n(\beta_{n,\pm})
=
\pm2i\sqrt{|A_n|E}.
\end{equation}
Their oscillatory behavior is reflected in the ordinary Bessel
function $J_1$ in Eq.~\eqref{eq:flux-transform-cases}.
Indeed, for $x=2\sqrt{|A_n|E}\to+\infty$,
\begin{equation}
J_1(x)
=
\sqrt{\frac{2}{\pi x}}
\left[
\cos\left(x-\frac{3\pi}{4}\right)+O(x^{-1})
\right],
\end{equation}
whose two oscillatory exponentials correspond to the two
imaginary stationary points.

This differs from the representation expansion, where each
energetically allowed representation sector has the real saddle pair
\begin{equation}
\beta_{m,\pm}
=
\pm\frac{\pi}{\sqrt{E-e^2m^2/2}},
\qquad E>e^2m^2/2,
\end{equation}
and an unambiguous inner horizon saddle contribution.
There is no contradiction: 
The inner-horizon contributions that are manifest term by term
in the representation expansion are simply not organized
in the same way in the flux expansion.
However, it is still striking that the inner-horizon interpretation is not available term-by-term in the flux-lift representation.

Finally, let us comment on a curious feature of gauge fields on the inner horizon saddle.
Working in the metric Eq.~\eqref{eq:JT-metric},  we have 
\begin{equation}
    F=dA = f\,\text{vol}
=\pm f \sinh\rho \, d\rho\wedge d\theta\,,
\end{equation}
where the upper (lower) sign applies to the cigar (inner horizon) geometry, reflecting the relative sign of
$\sqrt{g}$ discussed in Section~\ref{sec:jt-setup}. Gauge-fixing $A_\rho=0$,
one finds
\begin{equation}
   A=\pm f(\cosh\rho-\cosh\rho_0) d\theta\,,
\end{equation}
where $\rho_0$ is a constant of integration.
Demanding that the gauge field is regular at the cap, where the angular Killing vector vanishes, we have, for the cigar geometry,
\be\label{eq:A-cigar}
A_{\text{cigar}} = f\,(\cosh\rho-1)\,d\theta\,,
\ee
and for the inner horizon geometry,
\be\label{eq:A-inner}
A_{\text{inner}} = -f\,(\cosh\rho+1)\,d\theta\,.
\ee
Curiously, the expression for $A$ generically takes complex values along the intermediate piece of
the contour, where the hyperbolic cosine inherits an imaginary part
from the contour deformation in $\rho$. 
We discuss the interpretation further in Section \ref{sec:discussion}.

\subsection{Comments on higher dimensions}
\label{sec:gaugeexh}

In this subsection, we illustrate the sector decomposition discussed at the end of Section \ref{sec:gauge} in a concrete higher-dimensional example: Maxwell theory on a four-dimensional background obtained by fibering an $S^{2}$ over the  AdS$_2$ background (which can be thought of as either the Euclidean cigar saddle or the quasi-Euclidean inner horizon saddle).
The example illustrates how, on a direct product AdS$_2\times S^2$ background with fixed sphere radius, the four-dimensional field decomposes into a two-dimensional zero-mode sector together with a non-topological Kaluza-Klein tower whose individual modes can be analyzed using the spectral KSW criterion.

We expand the four-dimensional gauge field into components along the sphere, $A_a$,
and components along the AdS$_2$, $A_{\mu}$.
The latter are scalars with respect to the sphere, and hence can be decomposed using ordinary scalar spherical harmonics:
\begin{equation}
    A_{\mu}(x,y) = \sum_{\ell,m}a_{\mu}^{\ell,m}(x) Y_{\ell m}(y)
\end{equation}
where $y$ denotes coordinates on the sphere, $x$ denotes coordinates on AdS$_2$, and $Y_{\ell m}(y)$ are spherical harmonics.
Hence, from the AdS$_2$ perspective, we are left with a tower of two-dimensional vectors $a_{\mu}^{\ell,m}(x)$.
Meanwhile, in a sector with zero magnetic flux through the sphere, $A_{a}$ admits a Hodge decomposition into longitudinal and transverse pieces:
\begin{equation}
A_a = \sum_{\ell,m} A_a^{\ell m}
\qquad
A_{a}^{\ell m}
=
b^{\ell m}(x)\nabla_{a}Y_{\ell m}(y)
+
c^{\ell m}(x)\epsilon_{a}{}^{b}\nabla_{b}Y_{\ell m}(y)\, ,
\end{equation}
so that it can be parameterized with scalar fields $b^{\ell m}(x)$ and $c^{\ell m}(x)$. 

First consider the $\ell=0$ harmonic.
Since $Y_{00}$ is a constant, the only nonzero component is the AdS$_2$ gauge field
\begin{equation}
    A_{\mu}\propto a_{\mu}^{0,0}(x)\,.
\end{equation}
From the two-dimensional perspective, this is precisely the Maxwell field analyzed in the previous subsections, whose path integral localizes onto electric-flux sectors labeled by the integer $m$ of Appendix~\ref{sec:rep-sum}.
In addition, the four-dimensional theory may be placed in a magnetic flux sector with flux $p$ through the sphere.
The full zero-mode sector is therefore labeled by discrete electric and magnetic fluxes $(m,p)$.

The $\ell\geq1$ harmonics, by contrast, provide nontrivial propagating degrees of freedom on the AdS$_2$, and moreover do not contribute to the total sphere flux.
The gauge-invariant combination
\begin{equation}
B_{\mu}^{\ell m}
=
a_{\mu}^{\ell m}-\nabla_{\mu}b^{\ell m}
\end{equation}
is a massive vector on AdS${}_{2}$ whose mass squared is set by the Laplacian eigenvalue on the sphere,
\begin{equation}
m_{\ell}^{2}=\frac{\ell(\ell+1)}{R_{S^{2}}^{2}}\,,
\end{equation}
and $c^{\ell m}$ is a massive scalar of the same mass. 
Since a massive vector in two dimensions
carries a single propagating degree of freedom, the contributions $c^{\ell m}$ and $B_{\mu}^{\ell m}$ combine to reconstruct the two physical polarizations of the four-dimensional photon.

Thus AdS${}_{2}\times S^{2}$ realizes the separation discussed at the end of Section \ref{sec:gauge}:
a sum over discrete  sectors,
and in each sector, Kaluza--Klein modes describing local fluctuations.

\section{On spectral cuts}
\label{app:agmon-details}

In this appendix, we provide extra examples in which the positioning of a spectral cut with respect to spectral accumulation regions influences the outcome of zeta regularization.
This supplements the chiral fermion example in Section \ref{sec:non-topological}.

We focus on the simple case of an abstract operator $\mathcal{O}$ with discrete spectrum $\lambda_n\in\mathbb{C}$.
Then, its zeta-regularized determinant with spectral cut $L_\theta$ is defined by
\begin{equation}
    \det \mathcal{O} = e^{-\zeta_\mathcal{O}'(0)}\,
    \qquad
    \zeta_{\cal O}(s) = \sum_n \lambda_n^{-s}\,, 
    \qquad \lambda_n^{-s}=|\lambda_n|^{-s}\,e^{-is\arg_\theta\lambda_n}\,.
\end{equation}
See~\cite{Seeley:1967ea,Shubin:2001pdo,Kontsevich:1994xe} for a more general analysis of the dependence of $\det\mathcal{O}$ on spectral cuts and the multiplicative anomaly in zeta function regularization. 

\subsection{Bad spectral cuts}

We begin with one of the simplest examples in which spectral cut placement heavily influences the outcome of zeta regularization.

Consider an operator with complex spectrum
\begin{equation}\label{eq:family-spectrum}
\lambda_n = n^2\, e^{i n\alpha}\,,\qquad n=1,2,3,\ldots\,,
\end{equation}
where $\alpha \in \mathbb{R}$ is a free parameter not equal to zero.
Let us choose a non-radial spectral cut $\Gamma$ which winds around in the complex plane and gives $\arg_\Gamma\lambda_n=n\alpha$.
Then, the spectral zeta function reads
\be\label{eq:dense-dirichlet}
\zeta_\mathcal{O}(s) \;=\; \sum_{n=1}^\infty n^{-2s}\, e^{-i s n\alpha}.
\ee
The series converges absolutely for real $s>1/2$ because $|\lambda_n^{-s}|=n^{-2\,\mathrm{Re}(s)}$, yielding a polylogarithm with $s$-dependent argument,
\be\label{eq:dense-polylog}
\zeta_\mathcal{O}(s) \;=\; \mathrm{Li}_{2s}\!\bigl(e^{-i s\alpha}\bigr),
\ee
however, it has  a singularity at $s=0$ and
the derivative $\zeta'_\mathcal{O}(0)$ does not exist for any choice of $\alpha$. 

This divergence is merely an artifact of how we chose the spectral cut.
Indeed, for $\alpha=2\pi$, the spectrum is purely real, and repeating the above computation with $\lambda_n =n^2$ gives the finite answer $\zeta_{\mathcal{O}}'(0)=2\zeta_R'(0)=-\log2\pi$.

To resolve this issue, one may introduce a spectral cut which does not wind around the origin.
For example, let us restrict to the case where
\begin{equation}
\label{eq:rational-spectrum}
    \alpha/(2\pi)= p/q
\end{equation}
is a rational number. 
Without loss of generality we take $p$ and $q$ to be relatively prime; then,  the spectrum accumulates along $q$ rays emanating radially from the origin.
Choosing a radial spectral cut $L_\theta$ which avoids these rays, we obtain
\begin{equation}
    \lambda_n^{-s} = 
    n^{-2s}
    e^{-i s\arg_{\theta}(e^{2\pi i n p/q})}\,.
\end{equation}
The exponential is invariant under $n\to n+q$,  so we may parameterize a general $n\in \mathbb{N}$ as $n=q\kappa + r$ with $r=1+((n-1)\bmod q)\in\{1,\ldots,q\}$, and $\kappa \ge 0$, so that
\begin{equation}
    \sum_n \lambda_n^{-s} = 
    \sum_{r=1}^q
    e^{-is\arg_{\theta}(e^{2\pi i r p/q})}
    \sum_{\kappa = 0}^{\infty}
    (q\kappa + r)^{-2s}\,.
\end{equation}
Using the Hurwitz zeta function,
\be\la{eq:hurwitz-def}
\zeta_H(s,a)=\sum_{k=0}^\infty (k+a)^{-s}\,,
\ee
this becomes
\begin{equation}\label{eq:zeta-rational-family}
    \sum_n \lambda_n^{-s} = 
    q^{-2s}
    \sum_{r=1}^q
    e^{-is\arg_{\theta}(e^{2\pi i r p/q})}
    \zeta_{H}(2s,r/q)\,.
\end{equation}
Finally, the derivative reads\footnote{Note that for $q=1$, Eq.~\eqref{eq:zeta-prime-rational-family} recovers the formula
\begin{equation}
    \zeta_{\theta}'(0)
    =
    -\log 2\pi+\frac{i}{2}\arg_{\theta}e^{i \alpha}
\end{equation}
for a single eigenvalue ray with fixed phase $\alpha$.
For $q>1$, Eq.~\eqref{eq:zeta-prime-rational-family} is not obtained by simply summing $q$ copies of this fixed-ray formula, because the eigenvalues on the $r$-th ray form the shifted subsequence $(q\kappa+r)^2$, whose zeta function is a Hurwitz zeta function rather than the unshifted Riemann zeta function.}
\be\label{eq:zeta-prime-rational-family}
\zeta'_\theta(0) =  -\log 2\pi
-i\sum_{r=1}^q\!\arg_\theta(e^{2\pi i r p/q})\,(\tfrac12-\tfrac{r}{q}) \,,
\ee
where we have used $\zeta_H(0,a)=\tfrac12-a$ and $\zeta_H'(0,a)=\log \Gamma(a)-\tfrac12 \log(2\pi)$.
Since this expression is manifestly finite, we conclude that replacing the winding spectral cut $\Gamma$ by the radial spectral cut $L_\theta$ has rendered the determinant well-defined.
Going forward, we accordingly restrict our attention to radial cuts.

\subsection{Spectral cut ambiguities}

We now discuss when spectral cut placement ambiguities can (or cannot) give rise to discrete ambiguities in the definition of $\det\cal{O}$.

Consider a general operator ${\cal O}$, and consider sweeping the spectral cut across an isolated eigenvalue $\lambda_n$.
As discussed in the main text, this does not modify the zeta-regularized quantity $\det\cal{O}$,  since increasing $\textrm{arg}_\theta \lambda_n$ by an integer multiple of $2\pi$ sends $e^{-\zeta_{\mathcal{O}}'(0)}$ to itself. 

What happens if we instead sweep the spectral cut across \textit{all} the eigenvalues $\lambda_n$ exactly once?
This has the effect of putting all of the eigenvalues $\lambda_n$ onto a new sheet of the $\arg$ function; that is, for each $n$, we have $\lambda_n^{-s}\to e^{-2\pi i s}\lambda_n^{-s}$.
This generates the following change:
\begin{equation}
   \zeta_{\cal O}(s)
   =
   \sum_n \lambda_n^{-s}
   \to
   \sum_n \lambda_n^{-s} e^{-2\pi i s} 
   =
   e^{-2\pi i s} \zeta_{\cal O}(s)\,,
\end{equation}
and accordingly
\begin{equation}
    \zeta_{\cal O}'(0)\to \zeta_{\cal O}'(0)-2\pi i\zeta_{\cal O}(0)
    \,.
\end{equation}
Since $\log\det \mathcal{O} =-\zeta_\mathcal{O}'(0)$,  we have
\begin{equation}
    \det\mathcal{O}\to e^{2\pi i \zeta_{\mathcal{O}}(0)}\det\mathcal{O}\,.
\end{equation}
Naively, one is led to conclude that the sweeping operation $L_\theta\to L_{\theta+2\pi}$ has generated a meaningful ambiguity in the definition of $\det\mathcal{O}$.
In particular, in the example of the previous subsection, we have $e^{2\pi i \zeta_{\cal O}(0)}=-1$ using the Riemann zeta function. 
However, in physical settings, we expect that 
such factors of $e^{\# \zeta_{\mathcal{O}}(0)}$ can be absorbed into a choice of local counterterm,\footnote{See, e.g. \cite{Vassilevich:2003xt} on the connection between zeta functions and integrals of local invariants.}
so that $\det\mathcal{O}$ and $e^{2\pi i \zeta_{\mathcal{O}}(0)}\det\mathcal{O}$ describe the same physics.

Next, we consider what happens when the spectral cut sweeps across an accumulation region of the spectrum.
We consider again the example of the previous subsection, and we let the spectral cut cross the accumulation ray at angle $2\pi r_* p/q$.
This shifts
$\arg_\theta(e^{2\pi i r_* p/q})$ by $2\pi$, and so we have
\be\label{eq:adjacent-ratio-family}
\zeta'_{\theta}(0) \to \zeta'_{\theta}(0)- 2\pi i\,\bigl(\tfrac12-\tfrac{r_*}{q}\bigr),
\qquad \det\nolimits_{\theta}\mathcal{O}\to -\,e^{-2\pi i r_*/q}\det\nolimits_{\theta}\mathcal{O}\,.
\ee
From this, we see there are up to $q$ inequivalent ways to define $\det\mathcal{O}$ --- up to the overall minus sign, which we can ignore following the discussion of local counterterms above.
Hence, in the presence of accumulation regions,  spectral cut placement can lead to genuine ambiguities in the definition of $\det\mathcal{O}$.
In QFT applications, we expect that this ambiguity can be fixed by physical considerations; for instance, analytic continuation from a reference background where the partition function is known.

More serious ambiguities arise when the spectral cut is forced to lie on an accumulation region of the spectrum.
Consider setting the spectral cut angle to $2\pi r_* p/q$, where the spectrum accumulates, and consider approaching the spectral cut from above or below $2\pi r_* p/q$.
By Eq.~\eqref{eq:adjacent-ratio-family}, these two limits can give different answers for the zeta-regularized determinant, hence rendering the zeta-regularized determinant for spectral cut angle $2\pi r_* p/q$ ill-defined.
In this particular example,  the issue can be resolved using a lip prescription or an $i\epsilon$ rotation which nudges the spectral cut away from the accumulation ray.
However, in other examples, the accumulation directions may be equidistributed on $S^1$, leading to serious ambiguities in the definition of $\det \mathcal{O}$.

For an explicit example of this, we set the quantity $\alpha/2\pi$ defined in Eq.~\eqref{eq:family-spectrum} to an irrational number.
In this case, the spectral accumulation directions are equidistributed on $S^1$.\footnote{By Weyl's
equidistribution theorem, the sequence
$\left\{\frac{n\alpha}{2\pi}\bmod 1
\right\}_{n\geq1}$
is equidistributed in $[0,1)$. Equivalently, the arguments
\begin{equation}
  \arg\lambda_n=n\alpha\bmod 2\pi
\end{equation}
are equidistributed in $[0,2\pi)$. Thus every open angular interval
contains infinitely many eigenvalues, with
$\lvert\lambda_n\rvert=n^2\to\infty$.}
To make sense of the determinant, we can try to approach $\alpha/2\pi$ with a sequence of rational numbers $p/q$, and for each rational number we can assign a spectral cut which avoids the accumulation regions.
We will show that the resulting zeta-regularized determinant is ambiguous in this limit.

\begin{figure}
    \centering
    \includegraphics[width=0.47\linewidth]{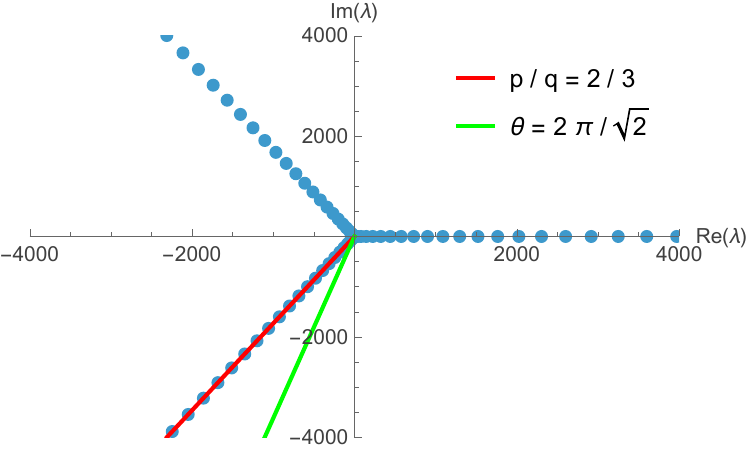}
    \hfill
\includegraphics[width=0.47\linewidth]{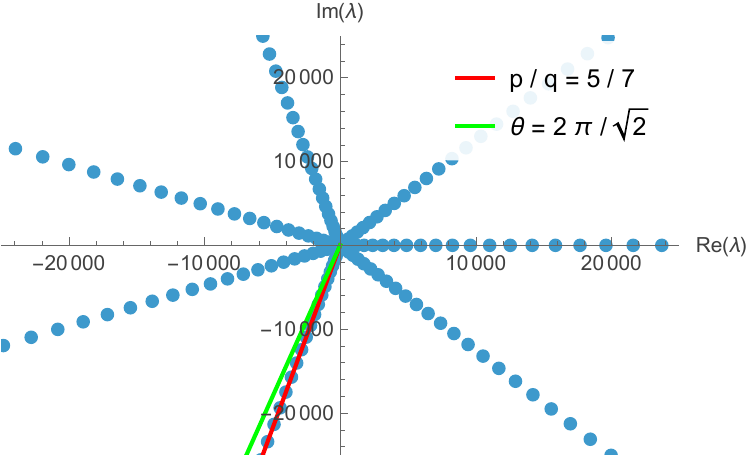}\caption{Depending on whether $p/q$ approaches $1/\sqrt{2}$ from above or below, the accumulation ray with angle $2\pi p/q$ will approach the spectral cut angle from the left or from the right.}
    \label{fig:rays}
\end{figure}

For concreteness, we take $\alpha/(2\pi)=1/\sqrt{2}=[0;1,2,2,2,\ldots]$, and we evaluate the imaginary\footnote{For every rational $p/q$, Eq.~\eqref{eq:zeta-prime-rational-family} has real part exactly $-\log(2\pi)$; only the imaginary part varies with the spectral data.} part of~\eqref{eq:zeta-prime-rational-family} numerically along the continued-fraction convergents $p_k/q_k$ of $\alpha/(2\pi)$, at the fixed spectral cut angle $\theta=2\pi/\sqrt{2}$:

\begin{center}
\begin{tabular}{rrr}
\hline
$p/q$ 
& $\mathrm{Im}\,\zeta'_\theta(0)$\\
\hline
$1/1$ & 
$3.14159$\\
$2/3$  & 
$2.79253$\\
$5/7$ & 
$4.48799$\\
$12/17$ & 
$2.03280$\\
$29/41$ & 
$4.82733$\\
$70/99$ & 
$1.88284$\\
$169/239$ & 
$4.88984$\\
$408/577$ & 
$1.85664$\\
$985/1393$ & 
$4.90070$\\
$2378/3363$ & 
$1.85214$\\
\hline
\end{tabular}
\end{center}
The sequence $\mathrm{Im}\,\zeta'_\theta(0)$ \textit{oscillates} between two limits: convergents that approximate $\alpha/(2\pi)$ from above give $\approx 4.90$, 
while convergents that approximate $\alpha/(2\pi)$ from below approach $\approx 1.85$.
One reason for this is that each rational spectrum contains an accumulation ray with angle $2\pi p/q$, and this ray closely approaches the spectral cut at angle $2\pi/\sqrt{2}$. 
See Fig.~\ref{fig:rays}.
Unlike the remainder of the accumulation rays, which are symmetrically arranged at some distance away, the ray with angle $2\pi p/q$  is highly sensitive to the location of the spectral cut.
The difference between having a $2\pi p/q$ ray which approaches $2\pi/\sqrt{2}$ from the left, versus a $2\pi p/q$ ray which approaches $2\pi/\sqrt{2}$ from the right, can be derived by looking at the $r=1$ term in Eq.~\eqref{eq:zeta-rational-family}:
\begin{equation}
    \zeta_{\cal O}(s) =\sum_n \lambda_n^{-s} \ni 
    q^{-2s}
    e^{-is\arg_{\theta}(e^{2\pi i  p/q})}
    \zeta_{H}(2s,1/q)\,.
\end{equation}
When $2\pi p/q$ crosses $\theta=2\pi/\sqrt{2}$, this term picks up a factor of $e^{-2\pi i s}$ and hence
\begin{equation}
    \zeta_{\cal O}'(0) \to \zeta_{\cal O}'(0) -2\pi i\zeta_H(0,1/q)
    \overset{q\to\infty}{\to}
    \zeta_{\cal O}'(0) - \pi i\,.
\end{equation}
This $\pi$ provides the dominant contribution to the separation between the limit points in the table above.

This example illustrates that zeta-regularized determinants can be highly sensitive to the placement of the spectral cut when the spectrum accumulates in all complex directions.
While a definite value for $\det \mathcal{O}$ can be assigned for a chosen cut,
an arbitrarily small deformation of that cut can cause accumulation regions of the spectrum to cross it,
leading to discontinuous jumps in $\det \mathcal{O}$.

\section{Deformed JT and multihorizon geometries}
\label{app:multihorizon}

In this appendix, we first explain how the results of Section \ref{sec:jt} generalize to the case of theories with richer horizon structure (subsection \ref{app:multihorizon-deform}).
Random matrix theory can be leveraged to determine whether these new inner horizon saddles contribute to the gravitational path integral, and we accordingly dedicate subsection \ref{app:joukowski} to outlining the relationship between a random matrix theory formula and the semiclassical picture discussed in the main text.

\subsection{Multihorizons from deformed JT}
\label{app:multihorizon-deform}

JT gravity, which is characterized by having the potential $U(\phi)=2\phi$ in Eq.~\eqref{eq:JT-action}, admits generalizations to certain nontrivial potentials $U(\phi)$ described in, e.g., Ref.~\cite{Kruthoff:2024gxc}.
For the classical discussion below, it suffices to characterize these potentials by the condition that $U(\phi)\to2\phi$ as $\phi\to\infty$, with deviations exponentially suppressed in $\phi$. 
We define the prepotential $W(\phi)$ by 
\begin{equation}
    W'(\phi)= U(\phi)\,,
\end{equation}
fixing its additive constant by the asymptotic convention $W(\phi)-\phi^2\to0$ as $\phi\to \infty$.

The relevant classical solutions to this theory read
\begin{equation}
ds^2 = f(r) d\tau^2 + \frac{dr^2}{f(r)}
 \qquad \phi(r) =  \phi_b r\,,
 \label{eq:classical-sol-W}
\end{equation}
where the equations of motion from Eq.~\eqref{eq:JT-action} give $f''(r) = U'(\phi(r))$ and $\phi_b f'=U(\phi)$,
hence
\begin{equation}
    f(r) = \frac{W(\phi(r))-W(\phi(r_0))}{\phi_b^2}\,,
\end{equation} 
where $r_0$ is chosen so that $f(r_0)=0$.
Depending on the pre-potential $W$, this geometry may have a rich horizon structure; for instance, if $f(r)$ vanishes at three distinct values of $r$, then the Lorentzian continuation will have three horizons.
Note that if the geometry caps off at one such horizon, call it $r_0$, then smoothness requires $\tau\sim \tau +  4\pi/f'(r_0)$, hence
\begin{equation}
    \beta = \frac{4\pi \phi_b}{W'(\phi(r_0))}\,.
    \label{eq:inverse-temp-W}
\end{equation}

Towards an understanding of the density of states, we now solve for $r_0$ subject to DN boundary conditions \eqref{eq:dn-bcs}. This gives
\begin{equation}
    r_b = \frac{1}{\epsilon}
    \qquad
    \frac{\phi_b}{\epsilon}\left(r_b-\sqrt{f(r_b)}\right)  = E \,.
\end{equation}
The second equation may be simplified using that
$W(\phi)\sim \phi^2$, with all corrections exponentially suppressed, when $\phi$ is large.
Then, the only term which is finite at $r_b\to\infty$ reads
\begin{equation}
E= \frac{W(\phi(r_0))}{2  \phi_b} \,,
\end{equation}
and therefore the solutions for $r_0$ are given by
\begin{equation}
\label{eq:r0-solutions-Dn-W}
    r_0 \in \left\{r_0 : E= \frac{W(\phi(r_0))}{2 \phi_b}\right\}\,.
\end{equation}
Following the discussion in the main text, we propose that each simple root of this equation corresponds to a classical geometry which caps off at $r_0$.

Indeed, as we now show, all of these classical solutions with DN boundary conditions have corresponding saddle points in the inverse Laplace transform of the classical partition function.
To evaluate $Z(\beta)$ in the classical approximation, we must evaluate the on-shell action of the solution in Eq.~\eqref{eq:classical-sol-W}, with $r_0$ fixed in terms of $\beta$ by Eq.~\eqref{eq:inverse-temp-W}.\footnote{Note that the value of $r_0$ solving Eq.~\eqref{eq:inverse-temp-W} need not be unique. 
Our analysis does not apply to this case and it would be interesting to understand the appropriate generalization.}
For the bulk action we note that for metrics of the form \eqref{eq:classical-sol-W}, we have $R=-f''(r)$, hence in the present case, $R=-W''(\phi(r))$. Then, following \cite{Witten:2020ert}, we have
\begin{equation}
\begin{aligned}
    I_{\text{bulk}} &= -\frac{1}{2}\int d^2x\sqrt{g}(\phi R + W'(\phi) )\\
    &=
    -\frac{\beta \phi_b}{2}\int_{r_0}^{\infty} dr (- r W'' + \frac{1}{\phi_b} W')\\
     &=
    -\frac{\beta}{2 \phi_b}\int_{r_0}^{\infty} dr (- r \partial_r^2 W +  \partial_r W)\,,
\end{aligned}
\end{equation}
which, using integration by parts, reduces to a pair of boundary terms:
\begin{equation}
\begin{aligned}
    I_{\text{bulk}} &=-\frac{\beta}{2\phi_b} (-r\partial_r W + 2W
    )\big|^{\infty}_{r_0}
    \,\\
    &=
    \frac{\beta}{2} \left(-r_0 W'(\phi(r_0)) +\frac{2}{\phi_b} W(\phi(r_0))\right)\,.
\end{aligned}
\end{equation}
The regulated GHY boundary term can straightforwardly be computed \cite{Witten:2020ert}, giving $I_{\text{bdy}}
=-\frac{\beta}{2 \phi_b}W(\phi(r_0))$, and hence, finally, the on-shell action reads
\begin{equation}
    I=\frac{\beta}{2} \left(-r_0 W'(\phi(r_0)) +\frac{1}{\phi_b} W(\phi(r_0))\right)
    \,.
    \label{eq:wittened}
\end{equation}
For JT gravity, $r_0 = 2\pi/\beta$ and $W(\phi)=\phi^2$; hence, this reduces to $I=-2\pi^2 \phi_b/\beta$, as expected.

Having computed the on-shell action, we may now compute the saddle points of the inverse-Laplace transform of $Z(\beta)$ in the classical approximation.
We retain the dilaton normalization $\phi(r)=\phi_b r$ used in the main text, and apply the following two identities coming from Eq.~\eqref{eq:inverse-temp-W}:
\begin{equation}
    \beta = \frac{4\pi\phi_b}{W'(\phi(r_0))}\,,
    \qquad
    -W''(\phi(r_0))\,\frac{d\phi(r_0)}{d\beta}
    = \frac{4\pi\phi_b}{\beta^2}\,.
\end{equation}
Using this, we find the saddle points of the integral are located where
\begin{equation}
    0=
    \partial_{\beta}\left(\beta E -I(\beta)\right)
    =
    E
    -
    \frac{W(\phi(r_0))}{2\phi_b}\,,
\end{equation}
that is, at every value of $\beta$ such that there is a corresponding classical solution with horizon value $r_0(\beta)$ given by Eq.~\eqref{eq:r0-solutions-Dn-W}.

As discussed in Section \ref{sec:two-saddles}, in order to determine whether these saddle points contribute with a steepest descent analysis, loop corrections are required.
Fortunately, for the class of deformed JT theories admitting the matrix-integral completion considered in Ref.~\cite{Kruthoff:2024gxc}, the disk-level density of states can be computed exactly, and it reads
\be\la{eq:pF-rho-double-W}
\rho(E) \;\propto\; \int_{E_{\rm edge}}^E \frac{dE'}{\sqrt{E-E'}}\int_{\mathcal{C}}\frac{d\phi_h}{2\pi i}\,\frac{e^{2\pi \phi_h}}{\sqrt{W(\phi_h) - 2\phi_b E'}}\,,
\ee 
where $E_{\rm edge}$ denotes the lower edge of the disk-level spectrum.
In the semiclassical limit, the critical points of the inner integral approach the branch points determined by  $W/(2\phi_b)=E'$, reproducing the classical horizon condition at leading order.
Moreover, in the triple-horizon example studied in Ref.~\cite{Kruthoff:2024gxc}, all three relevant real-horizon saddles are found to contribute, with the appropriate Stokes prescriptions.
As mentioned in Ref.~\cite{Kruthoff:2024gxc}, some solutions of Eq.~\eqref{eq:r0-solutions-Dn-W} are complex, and it would be interesting to understand their contributions further. 

\subsection{Random matrix theory and the inverse Laplace transform}
\label{app:joukowski}

The purpose of this appendix is to show that, in pure JT gravity, the random-matrix formula for the density of states in~\cite{Turiaci:2020fjj,Kruthoff:2024gxc} is equal, as an exact integral, to the inverse Laplace transform of the Schwarzian partition function $Z(\beta)$ used throughout the main text.
The equivalence is made explicit by a Joukowski-type substitution on the horizon-dilaton variable, and confirms that the $\beta$-contour prescription we rely on admits a first-principles matrix-integral origin.

We record the two presentations of $\rho(E)$.
On the one hand, the JT partition function $Z_{\mathrm{JT}}(\beta) = C\,\beta^{-3/2}\,e^{2\pi^2\phi_b/\beta}$ is one-loop exact~\cite{Stanford:2017thb,Saad:2019lba}, and its inverse Laplace transform along a Bromwich contour gives (suppressing the factor of $e^{S_0}$)
\be\la{eq:pF-ILT}
\rho(E) = C \int_{\mathrm{Br}} \frac{d\beta}{2\pi i}\,\beta^{-3/2}\,e^{\beta E + 2\pi^2\phi_b/\beta}.
\ee
On the other hand, the random-matrix presentation of~\cite{Turiaci:2020fjj,Kruthoff:2024gxc} expresses the same object as a double integral,
\be\la{eq:pF-rho-double}
\rho(E) \;\propto\; \int_0^E \frac{dE'}{\sqrt{E-E'}}\int_{\mathcal{C}}\frac{d\phi_h}{2\pi i}\,\frac{e^{2\pi\phi_h}}{\sqrt{\phi_h^2 - 2\phi_b E'}},
\ee
with $\mathcal{C}$ a vertical contour lying to the right of the two branch points of the square root.
From the matrix-integral point of view $\phi_h$ is merely an auxiliary variable, but, following~\cite{Kruthoff:2024gxc}, it is natural to identify it with the horizon dilaton value; its saddles at $\phi_h=\pm\sqrt{2\phi_b E'}$ reproduce the classical horizon condition.
In what follows we refer to the $E'$-integral as the \textit{$y$-integral}, with the understanding $y\equiv E'$; the notation emphasizes that, after an exchange in the order of integration, this integral is performed at fixed $\beta$ and plays the role of a one-dimensional transform in its own right.

The inner $\phi_h$-integral can be converted into
a $\beta$-integral by the Joukowski-type substitution
\be\la{eq:pF-joukowski}
\phi_h = \frac{1}{2}\left(
\frac{\beta E'}{\pi}+\frac{2\pi\phi_b}{\beta}
\right).
\ee
We use the inverse branch
\begin{equation}
\beta
=
\frac{\pi}{E'}
\left(\phi_h+\sqrt{\phi_h^2-2\phi_bE'}\right)\,,
\end{equation}
where
\begin{equation}
\frac{d\phi_h}{\sqrt{\phi_h^2-2\phi_bE'}}
=
\frac{d\beta}{\beta}.
\end{equation}
The vertical contour $\mathcal C$ maps to a curved contour
$\Gamma_{E'}$ in the right half of the $\beta$-plane, which
can be deformed to a vertical Bromwich contour without
changing the integral.\footnote{Write $\phi_h=d+it$, with $d>\sqrt{2\phi_bE'}$ fixed and $t\in \mathbb{R}$.
The chosen inverse branch has $\operatorname{Re}\beta>0$
and satisfies
\begin{equation}
\beta(d+it)
=
\frac{2\pi}{E'}(d+it)+O(t^{-1})
\qquad (t\to\pm\infty).
\end{equation}
Thus the image contour is asymptotic to a vertical line.
Moreover, the transformed integrand
$\beta^{-1}\exp(\beta E'+2\pi^2\phi_b/\beta)$ is
holomorphic for $\operatorname{Re}\beta>0$.
}
We obtain the exact identity
\be\la{eq:pF-phi-h-to-beta}
\int_{\mathcal{C}}\frac{d\phi_h}{2\pi i}\,
\frac{e^{2\pi\phi_h}}
{\sqrt{\phi_h^2 - 2\phi_b E'}}
=
\int_{\mathrm{Br}}\frac{d\beta}{2\pi i}\,
\frac{1}{\beta}\,
e^{\beta E' + 2\pi^2\phi_b/\beta}\,,
\ee
where the exponent on the right-hand side is the one-loop-exact Schwarzian answer $\log Z_{\mathrm{JT}}(\beta)$ up to the universal $\beta^{-3/2}$ measure and inverse-Laplace factor $\beta E'$.
As a consistency check, the saddles of the right-hand side sit at $\beta_\pm=\pm\sqrt{2\pi^2\phi_b/E'}$, in agreement with the semiclassical saddles $\phi_h=\pm\sqrt{2\phi_b E'}$ of the left-hand side via~\eqref{eq:pF-joukowski}. 

We now exchange the order of integration in~\eqref{eq:pF-rho-double}, using~\eqref{eq:pF-phi-h-to-beta} to trade the inner $\phi_h$-integral for a $\beta$-integral, and evaluate the $y$-integral at fixed $\beta$.
The relevant definite integral is
\be\la{eq:pF-y-integral}
\int_0^E \frac{e^{\beta E'}}{\sqrt{E-E'}}\, dE' = e^{\beta E}\sqrt{\pi/\beta}\,\bigl(1 - \mathrm{erfc}(\sqrt{\beta E})\bigr),
\ee
where $\mathrm{erfc}$ denotes the complementary error function and the branch of $\sqrt{\beta}$ is fixed by analytic continuation from the positive real axis.
This is the formula that connects the $y$-integral to the inverse Laplace transform; its two pieces combine with the $1/\beta$ of~\eqref{eq:pF-phi-h-to-beta} to yield a principal contribution proportional to $\beta^{-3/2}$: 
\be\la{eq:pF-rho-main}
\rho_{\mathrm{main}}(E) \propto \int_{\mathrm{Br}}\frac{d\beta}{2\pi i}\,\beta^{-3/2}\,e^{\beta E + 2\pi^2\phi_b/\beta},
\ee
and an erfc-weighted remainder, which integrates to zero by an application of Cauchy's theorem.

Hence, the $y$-integral in~\eqref{eq:pF-rho-double} collapses cleanly onto the inverse Laplace transform~\eqref{eq:pF-ILT}.
In pure JT gravity, the random-matrix presentation~\eqref{eq:pF-rho-double} and the inverse Laplace transform~\eqref{eq:pF-ILT} are therefore exactly equal.

\bibliographystyle{apsrev4-1long}
\bibliography{Universal/references}

\end{document}